\documentclass[aps,showpacs]{revtex4}
\usepackage{color}
\usepackage{graphicx}
\usepackage{graphics}
\usepackage{epsfig}
\usepackage{subfigure}
\usepackage{latexsym}
\usepackage{amsfonts}
\usepackage[english]{babel}
\usepackage[latin1]{inputenc}
\usepackage[all]{xy}
\usepackage{amsmath}
\usepackage{amssymb}
\newcommand{\be}{\begin{equation}}
\newcommand{\ee}{\end{equation}}
\newcommand{\ben}{\begin{eqnarray}}
\newcommand{\een}{\end{eqnarray}}
\newcommand{\bes}{\begin{subequations}}
\newcommand{\ees}{\end{subequations}}
\newcommand{\wt}{\widetilde}
\newcommand{\ov}{\overline}
\newcommand{\bb}{\bibitem}

\def \P{{\cal P}_{01}}

\def \p{\partial}
\def \pr{\prime}
\def \W{{\cal W}}

\def \dn{\mathop{\rm dn}\nolimits}
\def \sn{\mathop{\rm sn}\nolimits}

\def \bb{\bibitem}
\newcommand{\ba}[1]{\begin{array}{#1}}
\newcommand{\ea}{\end{array}}
\newcommand{\bea}[1]{\begin{equation}\left\{\begin{array}{#1}}
\newcommand{\eea}{\end{array}\right.\end{equation}}

\newcommand{\Ref}[1]{(\ref{#1})}
\begin{document}
\title{Construction of topological defect networks with complex scalar fields}
\author{V.I. Afonso$^{a}$, D. Bazeia$^{a}$, M.A. Gonzalez Le\'on$^{b}$, L. Losano$^{a}$, and J. Mateos Guilarte$^{c}$}
\affiliation{{$^{a}$Departamento de  Fisica, Universidade Federal da Para\'\i ba, BRAZIL}
\\{$^{b}$Departamento de Matematica Aplicada, Universidad de Salamanca, SPAIN}
\\{$^{c}$Departamento de Fisica and IUFFyM, Universidad de Salamanca, SPAIN}}

\begin{abstract}
This work deals with the construction of networks of topological defects in models described by a single complex scalar field. We take advantage of the deformation procedure recently used to describe kinklike defects in order to build networks of topological defects, which appear from complex field models with potentials that engender a finite number of isolated minima, both in the case where the minima present discrete symmetry, and in the non symmetric case. We show that the presence of symmetry guide us to the construction of regular networks, while the non symmetric case gives rise to irregular networks which spread throughout the complex field space. We also discuss bifurcation, a phenomenon that appear in the non symmetric case, but is washed out by the deformation procedure used in the present work.
\end{abstract}

\pacs{11.10.Lm, 11.27.+d}
\maketitle
\section{Introduction}

Defect structures have appeared in high energy physics almost fifty years ago, with some of the pioneer
results collected in Refs.~\cite{P1,P2,P3,P4}. Along the years, the subject has grown in importance,
accompanied with an increasing number of investigations on kinks in one spatial dimension, vortices in
two dimensions and monopoles in three dimensions, among other topological defects -- see, e.g. \cite{B}
for an extensive discussion of some of the most important results in the area.

The classical solutions which represent the defect structures can be of topological or non topological
nature, and here we will deal with perhaps the simplest topological structures, which appear in models
of scalar fields. To be specific, we will consider models of the Wess-Zumino type, described by a single
complex scalar field in the presence of discrete symmetry and in the more general case which engenders
no specific symmetry. Some of the models have been studied before in \cite{V,AT} -- see also \cite{H} for related issues --
with particular attention to the presence and stability of kinklike defects and junctions, and in \cite{SP}, where the kink orbits are written in
terms of real algebraic curves and the equations of motion are shown to be fully expressed in terms of first order differential
equations of the Bogomol'nyi-Prasad-Sommerfield (BPS) type \cite{P4}. 

The kinklike structures have been used in many different contexts, in $(1,1)$ and in higher space-time dimensions,
in particular in the form of junctions and networks of defects \cite{AT,JN1,JN2,JN3}. In $(3,1)$ dimensions they are
usually named domain walls, which can find applications in several distinct scenarios, in particular as seeds for the
formation of structures in the early Universe. In this context, although the standard scenario seems to show that the
presence of domain walls has little to contribute to the cosmic evolution, it has been suggested that domain walls
may perhaps be used as a source for the dark energy necessary to feed the current cosmic acceleration \cite{D1}.

Another line of research has recently appeared in gravity in higher dimensions \cite{DRS,RS}, with the hope to solve the hierarchy and
other problems in high energy physics. In $(4,1)$ dimensions, the braneworld model with warped geometry involving a single extra
dimension of infinite extent suggested in \cite{RS} has strongly impacted the subject. In this braneworld scenario, the inclusion of
scalar fields may contribute to smoothen the brane \cite{BR1}, to give rise to a diversity of situations of current interest, as one
can see, for instance, in the recent investigations \cite{BR2}.

The present study is a continuation of a former work \cite{abL}. Here we will focus mainly on the deformation procedure introduced in
\cite{DD}, and extended to other scenarios in \cite{DD1,SD}. Those investigations have led us to find a peculiar and very interesting
feature of the deformation procedure there implemented. The issue is that it is sometimes possible to deform a given model described by a
potential containing some minima, to get to another model, with the potential giving rise to a different set of minima, which may
increase periodically. This feature strongly suggests the possibility of using the deformation procedure to build lattices of
minima in the two-dimensional field space.

An interesting property of the deformation procedure is that it also constructs the kinklike solution of the deformed model in terms of
the kink solution of the original model. Thus, in the lattice of minima we can then nest a network of defects very naturally, that
is, as internal feature of the deformation itself. This is the idea underlying this paper, in which we apply the deformation
procedure to investigate the generation of networks of kinklike defects for the deformed models, which are
expanded networks. Although it is possible to start with the more general case, considering models with an arbitrary set of minima, we
shall firstly deal with the case involving $N$ minima in a $Z_N$ symmetric arrangement. We shall consider the symmetric $N=2$, $N=3$ and $N=4$ cases
explicitly, and later we relax the constraint to deal with three and four minima in the non symmetric case. The main reason for this is that we want
to keep the motivation set forward in our former work \cite{abL}, where we have investigated the construction of regular networks.
Moreover, as a pedagogical concern we believe that this route makes the problem easier to understand.

The idea of constructing networks of defects is not new, but the novelty here relies on the use of
the deformation procedure as a simple and natural way to generate networks. The mechanism is powerful and
suggestive, and fully motivates the present work. To make it short, direct, we have decided to consider models of
the Wess-Zumino type, driven by a single complex scalar field. These models are popular, of great importance and
easy to manipulate, and so they very much help us to highlight the idea to be explored below.

We start the investigation in Section~{\ref{sec:WZ}}, where we introduce the symmetric models and perform the
deformation procedure on general grounds. In Section~{\ref{sec:apply}} we illustrate the procedure with some
applications, considering two important cases, which engender three and four minima, forming an equilateral triangle and a square,
respectively. There we show how the deformed models tile the plane replicating the sets of minima in the entire field plane.
We then consider other possibilities in Section~{\ref{sec:nonsymm}}, and there we deal with more general models which 
three and four minima, engendering no symmetry anymore. We use the deformation procedure to get to many distinct and interesting patterns.
The more general case allows for a new phenomenon, bifurcation, and so in Section~{\ref{sec:bifur}} we deal with bifurcation, which concerns the
possibility of the system to allow for two or more distinct connections between two given minima. This is related to
the marginal stability curve, and has to do with the energy balance involving distinct orbits in field space, as
already investigated in \cite{AT}. We end the paper in Section~{\ref{sec:end}}, where we introduce some comments and
conclusions.

\section{Deformation of Wess-Zumino models}
\label{sec:WZ}

In this Section we start with a brief summary of the bosonic sector of the standard Wess-Zumino model engendering the $D_N$ symmetry. We
then propose a simple but very interesting way to deform the model, to generate an infinite family of new Wess-Zumino like models with
their defect solutions.

\subsection{The general case}

Let $\chi(x,t)=\chi_1(x,t)+i \chi_2(x,t)$ be a complex scalar field, written in terms of the two real partners $\chi_1(x,t)$ and
$\chi_2(x,t)$ in $(1,1)$ spacetime dimensions. The dynamics of the bosonic sector of Wess-Zumino models is governed by the Lagrange density
\be
{\cal L}=\frac12\partial_\mu\chi\partial^\mu\ov\chi-V(\chi,\ov\chi)
\ee
where the bar stands for complex conjugation. We refer to these systems as Wess-Zumino, or Landau-Ginzburg, models if the potential
energy density is determined from an holomorphic superpotential $W(\chi)$ such that the potential energy density of the scalar field
theory reads
\be
V(\chi,\ov{\chi})=\frac12 W'(\chi)\,{\ov{W'(\chi)}}
\ee 
The interest of these models lies in the fact that all of them admit a supersymmetric
version with ${\cal N}=2$ extended supersymmetry. It is also important to stress that there is a $U(1)$ ambiguity in the election
of the superpotential: $W_\alpha(\chi)=e^{-i\alpha}W(\chi)$, where $e^{i\alpha}\in U(1)$, produces the same dynamics.

We shall firstly consider polynomials of degree $N+1$ in $\chi$ with real coefficients as superpotentials, in this case
$\ov{W(\chi)}=W(\ov{\chi})$ is the same function of the conjugate complex field. The vacua manifold, the set of zeros of $V(\chi,\ov{\chi})$,
is given by the critical points of the superpotential, the $N$ roots of the polynomial $W^\prime(\chi)$
\be
\chi^{(j)}(x)=v^{(j)},\;\;\;\;\; W^\prime(v^{(j)})=0,\;\;\;\;\; j=1,2, \cdots , N 
\ee
The BPS static kinks satisfy the system of first-order ordinary differential equations
\be\label{eq:1a}
\frac{d\chi}{dx}=\,{\ov{W'_\alpha(\chi)}};\;\;\;\;\;\frac{d{\ov\chi}}{dx}=\,W'_\alpha(\chi)
\ee
We use these expressions to show that the field profiles and orbits should obey 
\be\label{eq:2}
dx=\frac{d\chi}{{\ov{W'_\alpha(\chi)}}}=\frac{d\overline{\chi}}{
W'_\alpha(\chi)}, \;\;\;\;\;W'_\alpha(\chi)\,d\chi-\,\overline{W'_\alpha(\chi)} \,d\overline{\chi} =0
\ee
Equation (\ref{eq:2}) (right) means that $d(W_\alpha-\overline{W}_\alpha)=0$ such that
\be\label{knkorb}
{\rm Im}W_{\alpha}(\chi(x))={\rm constant}
\ee
for the field orbits. These orbits are kink orbits if they connect two vacua
\be
{\rm Im}W_{\alpha^{(kj)}}(v^{(k)})={\rm
Im}W_{\alpha^{(kj)}}(v^{(j)})={\rm constant}
\ee
This criterion sets $\alpha^{(kj)}$ by requiring that $\left(W(v^{(k)})-W(v^{(j)}\right)e^{-i\alpha^{(kj)}})$ be real, or
\be
\alpha^{(kj)}={\rm arctan}\left[\frac{{\rm
Im}\left(W(v^{(k)})-W(v^{(j)})\right)}{{\rm
Re}\left(W(v^{(k)})-W(v^{(j)})\right)}\right] \quad {\rm mod}\, \pi,\;\;\;\;\; \alpha^{(jk)}=\alpha^{(kj)}+\pi
\label{eq:4}
\ee

Integration of (\ref{eq:2}) (left) gives
\be
x-x_0=\frac{1}{2}\int\,\left\{\frac{d\chi}{\ov{W_\alpha'}}+\frac{d\ov{\chi}}{W_\alpha'}\right\}
\ee
Interpretation of the meaning of this integral is reached from the identity between differential one-forms
\be\label{eq:5}
W_\alpha'(\chi)\, d\chi
+\overline{W_\alpha'(\chi)}\,d\overline{\chi}
=2\left|W^\prime(\chi)\right|^2dx \;\; \Rightarrow \;\;
d\,(W_\alpha+\overline{W}_\alpha)=2\left|W^\prime(\chi)\right|^2dx
\;\; \Rightarrow \;\; {\rm
Re}W_\alpha=\int\,\left|W^\prime(\chi)\right|^2dx =s
\ee
The kink profiles are then obtained by inverting these relations between the real part of the superpotential and the ``length" $s$ on the
kink orbits \Ref{knkorb} -- see, e.g. Ref.~{\cite{SP}}.

The energy of the static configurations can be written \`a la Bogomol'nyi in the form
\be
E=\frac{1}{2}\!\int dx
\left|\frac{d\chi}{dx}\!-\!\overline{W^\prime_\alpha(\chi)}
\right|^2+\frac12\left|\int\!d(W_\alpha+\overline{W}_\alpha)\right|\label{eq:bog}
\ee
which shows that the solutions of the first-order equations (\ref{eq:1a}) for the kink phases $\alpha^{(kj)}$ (\ref{eq:4}) have
energies given by
\be
M(kj)=\left|{\rm
Re}W_{\alpha^{(kj)}}(v^{(k)})-{\rm
Re}W_{\alpha^{(kj)}}(v^{(j)})\right|\nonumber
=\left|W(v^{(k)})-W(v^{(j)})\right| \label{eq:mass}
\ee

\subsection{Symmetric Wess-Zumino models}

The choice of the holomorphic superpotential in the form
\be
W(\chi)=\chi(x,t)-\frac{\chi^{N+1}(x,t)}{N+1},\;\;\;\;\; N\in\mathbb{N}
\ee
leads to the potential energy density
\be
V(\chi,\ov{\chi})=\frac{1}{2} \left(1-\chi^{N}(x,t)\right)
\left(1-{\ov\chi}^{N}(x,t)\right)
\ee
The dynamics is invariant with respect to the dihedral group $D_N=Z_2\times Z_N$, the symmetry group of a regular polygon of $N$ sides. In our case, the $Z_2$ sub-group is generated by the transformation $\chi\, \rightarrow\, \ov{\chi}$ and the elements of $Z_N$ are: $\chi\, \rightarrow e^{i\frac{2\pi}{N}(n-1)}\chi \, , \,n=1,2, \cdots , N$. Because the vacuum manifold is the set of the $N$th roots of the unity
\be\label{roots}
\chi^{(k)}(x,t)=v^{(k)}={\rm exp}(2\pi i(k-1)/N),\;\;\;\;\; k=1,2,\dots,N 
\ee
the $D_N$ symmetry is spontaneously broken to the complex conjugation $Z_2$ sub-group at every vacuum state.

The BPS static kinks satisfy the system of first-order ordinary differential equations
\be\label{eq:1}
\frac{d\chi}{dx}=e^{i\alpha}(1-{\ov\chi}^{N}(x));\;\;\;\;\;
\frac{d{\ov\chi}}{dx}=e^{-i\alpha}(1-\chi^{N}(x))
\ee
These equations can also be written as
\be\label{eq:2a}
dx=\frac{e^{-i\alpha}\,d\chi}{1-\bar{\chi}^{N}}
=\frac{e^{i\alpha}\,d{\ov{\chi}}}{1-\chi^{N}},\;\;\;\;\; e^{-i
\alpha}(1-\chi^{N})\, d\chi-e^{i
\alpha}(1-\overline{\chi}^{N})\,d\overline{\chi}=0
\ee
Thus, the real algebraic curves which solve (\ref{eq:2a}) (right)
\be\label{knkorba}
{\rm Im}\left[{e^{-i\alpha}}\left(\chi(x)-\frac{\chi^{N+1}(x)}{N+1}\right)\right]={\rm
constant}
\ee
are the orbits of the solutions. Kink orbits pass through two minima of the potential, so we have
\be
\label{knkcn}{\rm Im}\left[{e^{-i\alpha}}\left(v^{(k)}-\frac{(v^{(k)})^{N+1}}{N+1}\right)\right]=\frac{N}{(N+1)}
\sin(\frac{2\pi}{N}(k-1)-\alpha)={\rm constant}
\ee
and
\be
{\rm Im}\,W_{\alpha^{(kj)}}(v^{(k)})={\rm
Im}\,W_{\alpha^{(kj)}}(v^{(j)})\;\; \Leftrightarrow \;\;\left\{
\begin{array}{c} \alpha^{(kj)}=-{\rm arcsin}\left({\rm
cos}\left(\frac{\pi}{N}(k+j-2)\right)\right),\;\;\; k>j
\\ \alpha^{(kj)}=\;{\rm arcsin}\left({\rm
cos}\left(\frac{\pi}{N}(k+j-2)\right)\right),\;\;\;\;\;
k<j\end{array}\right. \label{eq:4a}
\ee
where $\alpha^{(jk)}=\alpha^{(kj)}+\pi$.

Integration of (\ref{eq:2a}) (left) gives
\be
x-x_0=\frac{1}{2}\!\int\left\{\frac{d\chi}{e^{i\alpha}(1-\ov{\chi}^{N}(x))}\!+\!
\frac{d\ov{\chi}}{e^{-i\alpha}(1-\chi^{N}(x))}\right\}
\label{knkpr}
\ee
This integral can be written in terms of the local parameter $\frac{ds}{dx}=\left|e^{-i\alpha}(1-\chi^{N}(x))\right|^2$ in order
to implicitly obtain the kink profiles
\be
{\rm Re}\left(e^{-i\alpha}(1-\chi^{N}(x))\right)=\int\,\left|e^{-i\alpha}(1-\chi^{N}(x))\right|^2dx=s
\ee
The kink energies are
\be
M(kj)=\frac{2N}{N+1}\left|{\rm sin}\left(\frac{\pi}{N}(k-j)\right)\right|\label{eq:mass2}
\ee

In \cite{V} it was shown that this superpotential in the ${\cal N}=2$ supersymmetric Landau-Ginzburg action is an integrable deformation of the ${\cal N}=2$ supersymmetric minimal $A_N$ series of conformal models. It was also suggested in \cite{V} the connection with the solitons of the affine Toda $A_N$ field theories -- see Ref.~\cite{H} for details -- which can be directly envisaged in the above expression \eqref{eq:mass2}.  
 
\subsection{The deformation procedure}

We now turn attention to the deformation procedure, which will allow us to obtain new Wess-Zumino like models. According to
Refs.~{\cite{DD,DD1}, we will express the deformed system in terms of a new complex field $\phi(x,t)=\phi_1(x,t)+i\phi_2(x,t),$ related
to the original one by means of the (a priori) holomorphic function $f(\phi)$ such that
\be
\chi=f(\phi)=f_1(\phi_1,\phi_2)+if_2(\phi_1,\phi_2)
\ee
This function has to obey
\be
\frac{\p f_1}{\p\phi_1}=\frac{\p
f_2}{\p\phi_2};\;\;\;\;\; \frac{\p f_1}{\p\phi_2} =-\frac{\p
f_2}{\p\phi_1}
\ee
The first-order equations become
\be\label{1eq}
\frac{d\phi}{dx}=e^{i\alpha}\frac{\overline{W^\prime(f(\phi))}}{f^\prime(\phi)};\;\;\;\;\;
\frac{d\ov{\phi}}{dx}=e^{-i\alpha}\frac{{W^\prime(f(\phi))}}{\ov{f^\prime(\phi)}}
\ee
that we choose to understand as determining the absolute energy minima associated to the ``deformed'' Lagrange density
\be
{\cal L}_D=\frac12\p_\mu\phi\p^\mu\ov{\!\phi}-
\frac{V(f(\phi),\ov{f(\phi)})}{f^\pr(\phi)\ov{f^\pr(\phi)}}
\label{eq:def}
\ee
The dynamics governed by ${\cal L}$ and ${\cal L}_D$ are different, but we can define ${\cal V}(\phi,\bar\phi)$ and
$\W(\phi)$ by
\be\label{eq:U}
{\cal V}(\phi,\bar\phi)=\frac{V(f(\phi),\overline{f(\phi)})}{\left|f'(\phi)\right|^2}
=\frac12\!\frac{W^\pr(f(\phi))}{\ov{f^\prime(\phi)}}
\frac{\overline{W^\pr(f(\phi))}}{f^\prime(\phi)}=\frac12\W^\pr(\phi)\,\overline{\W^\pr(\phi)}
\ee
such that the \lq\lq deformed" first-order equations \eqref{1eq} are
\be
\frac{d\phi}{dx}=\,e^{i \alpha}\,\overline{\W^\pr(\phi)};\;\;\;\;\;
\frac{d\bar{\phi}}{dx}=\,e^{-i\alpha}\,\W^\pr(\phi)
\label{eq:12}
\ee
The BPS kink solutions for this system are obtained from the solutions of (\ref{eq:1}) by simply taking the
inverse of the deformation function: $\phi^K(x)=f^{-1}(\chi^K(x))$. Thus, we can make the following relation between the deformed and
original equations: if $\chi^K(x)$ is a kinklike solution of the original model, we have that
\be\label{eq:defo1}
{\rm Im}\,W(\chi^K(x))={\rm constant}\, ;\;\;\;\;\;
{\rm Re} \,W(\chi^K(x))=s
\ee
and so we get that $\phi^K(x)=f^{-1}(\chi^K(x))$
is kinklike solution of the deformed model, obeying
\be\label{eq:defo2}
{\rm Im}\, \W(f^{-1}(\chi^K(x)))={\rm constant};\;\;\;\;\
{\rm Re} \,
\W(f^{-1}(\chi^K(x)))={\sigma}
\ee
where $\sigma$ is defined by
\be
{\sigma}=\int \, |\W^\pr\left(f^{-1}(\chi^K(x))\right)|^2\, dx
\ee

Alternatively, one could understand \eqref{1eq} as the first-order equations of the original model written in the form
\be
{\cal L}=\frac12f^\prime(\phi){\ov{f^\prime(\phi)}}\partial_\mu\phi\partial^\mu{\ov\phi}-V(f(\phi),{\ov{f(\phi)}})
\ee 
This interpretation means that the original system in the new variables appears as a nonlinear sigma model with target space a non-compact Riemannian manifold with metric
\be
G_{\phi\phi}(\phi,{\ov\phi})=0=G_{{\ov\phi}\,{\ov\phi}}(\phi,{\ov\phi});\;\;\;\;\;G_{\phi{\ov\phi}}(\phi,{\ov\phi})=
f^\prime(\phi){\ov{f^\prime(\phi)}}=G_{{\ov\phi}\phi}(\phi,{\ov\phi})
\ee

The merit of our approach is that we infer the kink solutions of one complicated but interesting field theoretical model from the well-known kinks of the associated simple system. The other point of view, in which one deforms the metric rather than the potential energy density, is sometimes also interesting.
Despite dealing with the same model, the use of appropriate coordinates in field space may lead to separation of variables in the first-order equations, 
sometimes reducing its integration to quadratures; see e.g., \cite{new1}. 

Inspired by former investigations on the deformation procedure, we select the deformation function as being equal to the new
superpotential, that is, we choose $f(\phi)=\W(\phi)$. This choice constrains the function $f(\phi)$ to obey the equation
\be
f'(\phi)
\overline{f'(\phi)}= \sqrt{2V(f(\phi),\overline{f(\phi)})}
\label{eq:sdsu}
\ee
A function $f$ satisfying this condition assures the relation \Ref{eq:U} to be fulfilled and presents the advantage of
providing a potential for the new model which is well defined (finite) at the critical points of $f(\phi),$
i.e. the zeros of $f^\prime(\phi)$. As a bonus, the procedure leads to a very simple expression for the deformed superpotential.

\section{Deformation of symmetric Wess-Zumino models}
\label{sec:apply}

To clarify the general considerations, let us now illustrate the above results with explicit examples. We will consider the cases $N=2$, $N=3$, and $N=4.$ The case $N=2$ is simpler, and it is very similar to the deformation used in the first work in \cite{DD1} to get to the sine-Gordon model. The cases $N=3$ and $N=4$ are harder. The deformation procedure leads to the formation of junctions of kink orbits from the original Wess-Zumino kinks. Since the original non deformed models engender sets of minima which depict equilateral triangles and squares, respectively, the deformation will then naturally tile the plane, with networks of defect orbits which we name expanded kink networks.

We will solve \eqref{eq:sdsu} for the Wess-Zumino model with solutions of $f^{\prime}(\phi)^2=(-1)^N(1-f^N(\phi)),$ and for $N=3,4$ these solutions are meromorphic functions. The issue here is that the deformation function $f(\phi)$ fails to be holomorphic in a discrete (infinite) set of points, $\Gamma$, and this induces the potential energy density to acquire a countably infinite set of poles (the metric in the target space in the second approach above acquires a countably infinite set of zeros). Analogous physical systems are described by the elliptic Calogero-Moser models (the elliptic tops in the second framework); see e.g., Ref.~\cite{new2}. The loss of holomorphicity can be avoided by restricting the new field to take values away from the set $\Gamma$, the lattice of poles of $f(\phi)$. We shall then take ${\mathbb C}/\Gamma$ as the $\phi$-field space.

\subsection{The case $N=2$}

In this case, we define the field $\chi$ as a function of the new field $\phi$ in the form $\chi=f(\phi)$. With this, we can use $f(\phi)$ to rewrite 
\be
W(\chi)=\chi-\frac13\chi^3;\;\;\;\;\;\;
V(\chi,\ov\chi)=\frac12(1-\chi^2)(1-\ov{\chi^2})
\ee
as
\ben
W(f)=f(\phi)-\frac13 f^3(\phi);\;\;\;\;\;\;
V(f)=\frac12(1-f^2(\phi))(1-\overline{f^2(\phi)})
\een
The deformation function $f(\phi)$ is the new superpotential if we impose 
\be\label{eq:dsupL}
f'(\phi) \overline{f'(\phi)}=\,
\sqrt{(1-f^2(\phi))(1-\overline{f^2(\phi)})}
\ee
The particular choice $f(\phi)=\sin(\phi)$ complies with \eqref{eq:dsupL} and leads to the deformed system defined by
\be
{\cal W}(\phi)=\cos(\phi)\, ;\;\;\;\;\;{\cal V}(\phi,\ov\phi)=\cos(\phi)\,\cos(\ov\phi)
\ee
which is the complex sine-Gordon model. Here the first-order equations are
\be
\frac{d\phi}{dx}=e^{i\alpha} \cos(\ov{\phi(x)});\;\;\;\;\;\;\frac{d\ov\phi}{dx}=e^{-i\alpha} \cos({\phi(x)})
\ee
The solutions for $\alpha=0,\pi$ are given by
\be
\phi^{K\pm}(x)=\pm{\rm gd}(x)+2n\pi=\pm\arcsin(\tanh(x))+2n\pi
\ee
where ${\rm gd}$ stands for the Gudermannian function \cite{Abr}, and $n$ is an integer. Here the kinks are analytic solutions and the superpotential is holomorphic.

\subsection{The case $N=3$}

In the $N=3$ case, we have
\be
W(\chi)=\chi-\frac14\chi^4\, ,\;\;\;\;\;
V(\chi,\ov\chi)=\frac12(1-\chi^3)(1-\ov\chi^3)
\ee
and putting $\chi=f(\phi)$, we rewrite this formula in the form
\be
W(f)=f(\phi)-\frac14{f^4(\phi)}\, ; \;\;\;\;\;
V(f)=\frac12(1-f^3(\phi))(1-\overline{f^3(\phi)})
\ee
As stated in \Ref{eq:sdsu}, the deformation function must then satisfy
\be\label{eq:dsup}
f'(\phi) \overline{f'(\phi)}=\,
\sqrt{(1-f^3(\phi))(1-\overline{f^3(\phi)})}
\ee
We choose in particular the holomorphic solution of \Ref{eq:dsup} which satisfies the separated equations
\be\label{eq:dsupb}
f'(\phi)^2=f(\phi)^3-1 \, ;\;\;\;\;\;
\overline{f'(\phi)^2}=\overline{f(\phi)^3}-1
\ee
The solution of \Ref{eq:dsupb}, henceforth a solution of \Ref{eq:dsup}, is the
equianharmonic case of the Weierstrass ${\cal P}$ function
\be
{\cal W}(\phi)=f(\phi)=4^{\frac{1}{3}}  \, {\cal P}(4^{-\frac{1}{3}}\phi; 0,1)
\ee
Recall that the Weierstrass ${\cal P}$ function -- see \cite{Abr} -- is defined as the solution of the ODE
\be
({\cal P}^\prime(z))^2=4{\cal P}^3(z)-g_2{\cal P}(z)-g_3
\ee
The Weierstrass ${\cal P}(z;g_2,g_3)$ elliptic function and its derivative that solve the differential equation above are doubly
periodic functions defined as the series
\bes\ben
{\cal P}(z)&=&\frac{1}{z^2}+\sum_{m,n}\biggl(\frac{1}{(z-2m\omega_1-2n\omega_2)^2}-\frac{1}{(2m\omega_1+2n\omega_2)^2}\biggr)
\\
{\cal P}^\prime(z)&=&\frac{-2}{z^3}-2\sum_{m,n}\frac{1}{(z-2m\omega_1-2n\omega_2)^3}
\een\ees
with $m,n\in\mathbb{Z}$ and ${m^2+n^2\neq 0}.$ Therefore, the deformation function is, up to a factor, the Weierstrass ${\cal
P}$ function with invariants $g_2=0$ and $g_3=1$, and we denote it by ${\cal P}_{01}(z)$. This function is meromorphic, with an
infinite number of poles congruent to the irreducible pole of order two in the fundamental period parallelogram (FPP). Thus, we suppose that the
$\phi$-field takes values away from the set of points $\Gamma_3$ in order to make the new superpotential holomorphic in the $N=3$ case.

A brief reminder of the essential properties of $\P$ and $\P^\pr$ is the following:

1. $\P(4^{-\frac{1}{3}}\phi)$ is a single-valued doubly periodic function with primitive periods:  $2 \omega_1$ and $2\omega_3$.
Defining $\omega_2=4^{\frac{1}{3}}{\Gamma^3({1}/{3})}/{4\pi}$, the primitive half-periods are
\be
\omega_1=\omega_2 \left(\frac{1}{2}-i\,\frac{\sqrt{3}}{2}\right);\;\;\;\;\;
\omega_3=\omega_2\left( \frac{1}{2}+i\, \frac{\sqrt{3}}{2}\right)
\ee
These periods determine the FPP. Note that only two of the half periods are irreducible: $\omega_2=\omega_1+\omega_3$.

2. $\P(4^{-\frac{1}{3}}\phi)$ has only one pole of order two at $\phi=0$ in the FPP.

3. The values of $\P(4^{-\frac{1}{3}}\phi)$ at the half-periods are
\be
\P(4^{-\frac{1}{3}}\omega_1)=4^{-\frac{1}{3}};\;\;\;\;\;
\P(4^{-\frac{1}{3}}\omega_2)=-4^{-\frac{1}{3}} \left(\frac{1}{2}+i\, \frac{\sqrt{3}}{2}\right);\;\;\;\;\;
\P(4^{-\frac{1}{3}}\omega_3)=-4^{-\frac{1}{3}} \left(\frac{1}{2}-i\, \frac{\sqrt{3}}{2}\right)
\ee

4. The zeros of the derivative $\P^\pr(4^{-\frac{1}{3}}\phi)$ in the FPP
are at the half-periods of $\P$
\be
\P^\pr(4^{-\frac{1}{3}}\omega_1)=\P^\pr(4^{-\frac{1}{3}}\omega_2)=\P^\pr(4^{-\frac{1}{3}}\omega_3)=0
\ee
and the origin is a third-order pole of $\P^\pr(z)$.

\begin{figure}[htbp]
\centerline{\epsfig{file=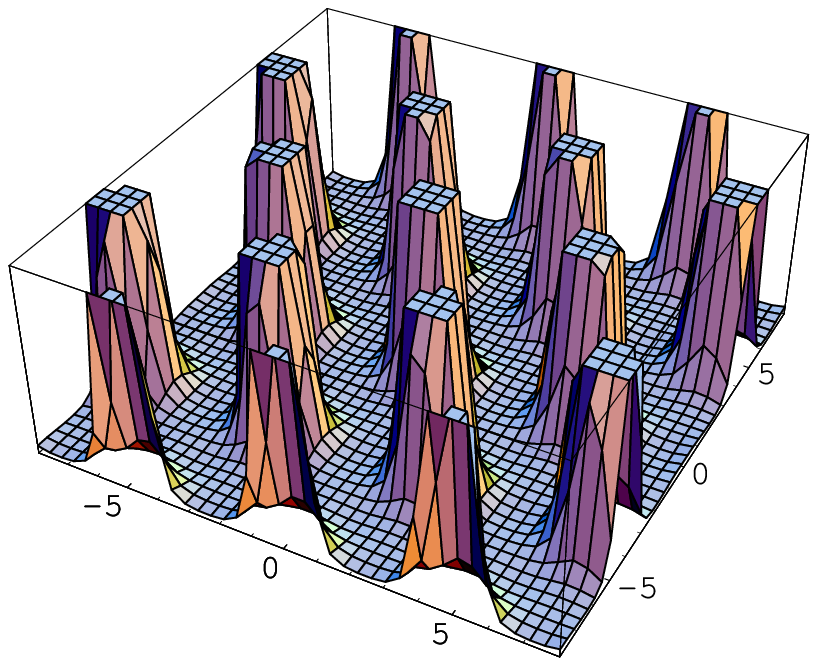,height=5cm}\hspace{1cm}\epsfig{file=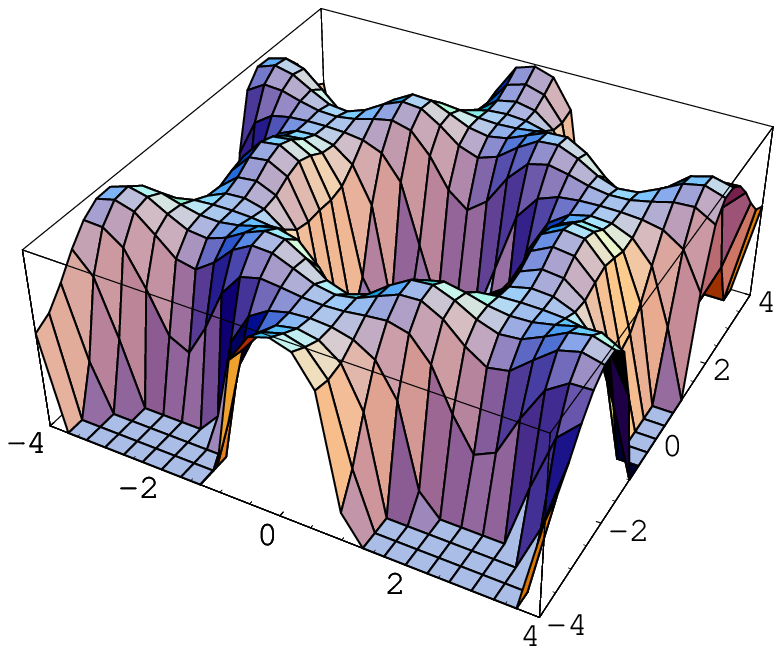,height=5cm}}
\caption{(Color online) The symmetric case $N=3$. 3D graphics of the potential ${\cal V}(\phi,\overline{\phi})$ (left panel),
and of $-{\cal V}(\phi,\overline{\phi})$ near a point in the lattice $\Gamma_3$ (right panel). Note that in the right panel the zeros are now maxima.}
\end{figure}

With these ingredients we write the deformed potential
\be
{\cal V}(\phi,\overline{\phi})=\frac{1}{2} \sqrt{
(1-4\,\P^3(4^{-\frac{1}{3}}\phi)) (1-4\,
\overline{\P^3}(4^{-\frac{1}{3}}\phi))} =\frac{1}{2}
\P^\pr(4^{-\frac{1}{3}}\phi)\, \overline{\P}\,^\pr(4^{-\frac13}\phi)
\ee
which is depicted in Fig.~1.

The new potential is doubly periodic with an structure inherited from the \lq\lq half-periods" of ${\cal P}$. The set of zeros of the
potential in the FPP (see Fig.~2) has three elements
\be
\phi^{(1)}=\omega_1=\omega_2\left(\frac12-i\frac{\sqrt{3}}{2}\right)\,;\;\;\;\;\;
\phi^{(2)}=\omega_3=\omega_2\left(\frac12+i\frac{\sqrt{3}}{2}\right)\,;\;\;\;\;\;
\phi^{(3)}=\omega_2=4^{\frac13}\frac{\Gamma^3(\frac13)}{4\pi}
\ee
The set of all the zeros of ${\cal V}$ form a lattice (see Fig.~3) which tile the entire configuration plane
\bes\ben
\phi^{(1)}_{(m,n)}&=&\omega_2
\left(m+n+\frac12+\sqrt{3}\,i(m-n-{\frac12})\right)
\\
\phi^{(2)}_{(m,n)}&=&\omega_2 \left( m+n+\frac12+\sqrt{3}\,i (m-n+\frac12)\right)
\\
\phi^{(3)}_{(m,n)}&=&\omega_2 \left(m+n+1 +\sqrt{3}\, i (m-n)\right)
\een\ees
The values of the superpotential at the minima are
\be
\W_\alpha(\phi^{(1)}_{(m,n)})= e^{-i\alpha}\,;\;\;\;\;\;
\W_\alpha(\phi^{(2)}_{(m,n)})=i
e^{-i(\alpha-\frac{\pi}{6})}\,;\;\;\;\;\;
\W_\alpha(\phi^{(3)}_{(m,n)})=-ie^{-i(\alpha+\frac{\pi}{6})}
\ee

\begin{figure}[htbp]
\centerline{\epsfig{file=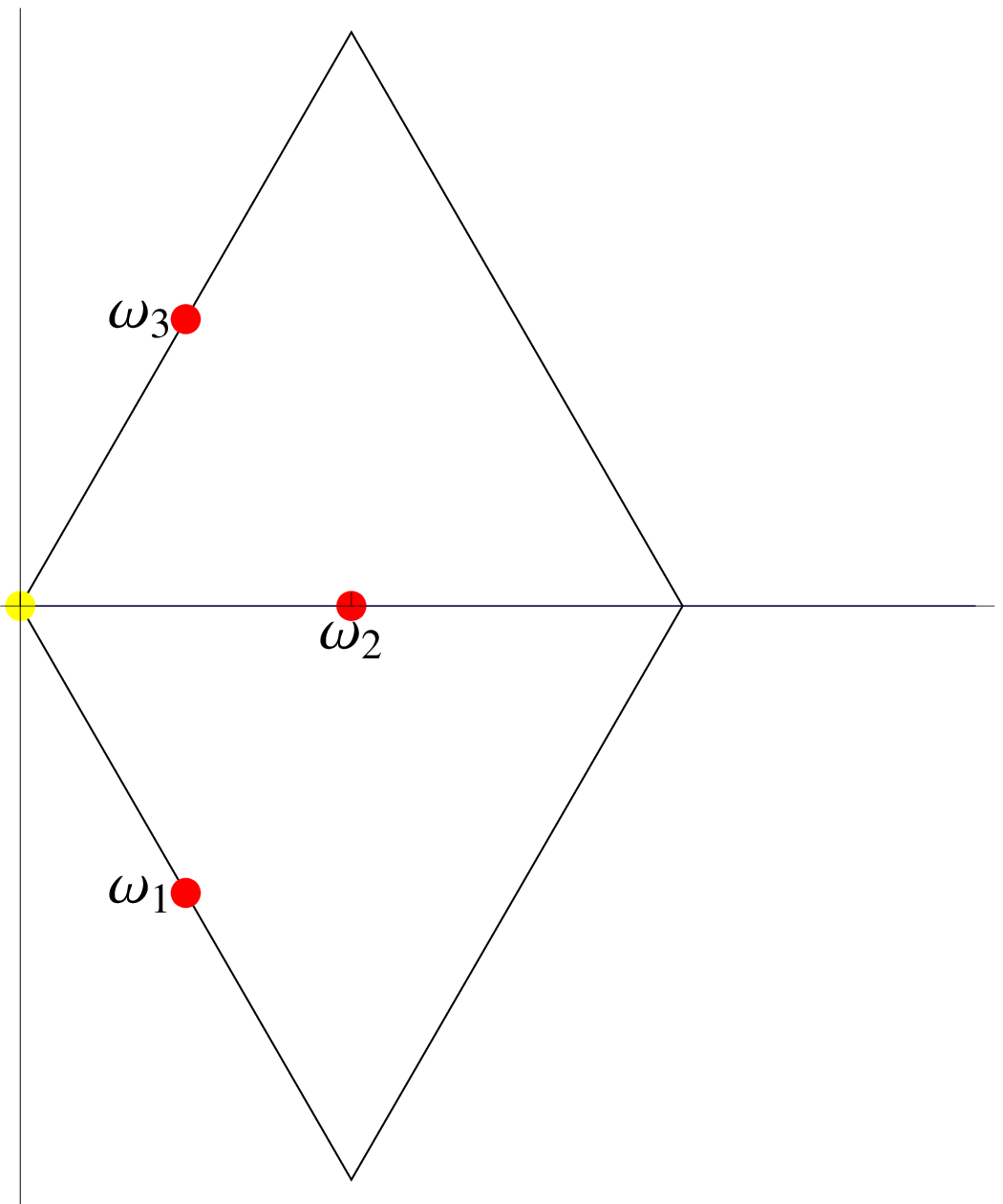,height=5cm}\hspace{1cm}\epsfig{file=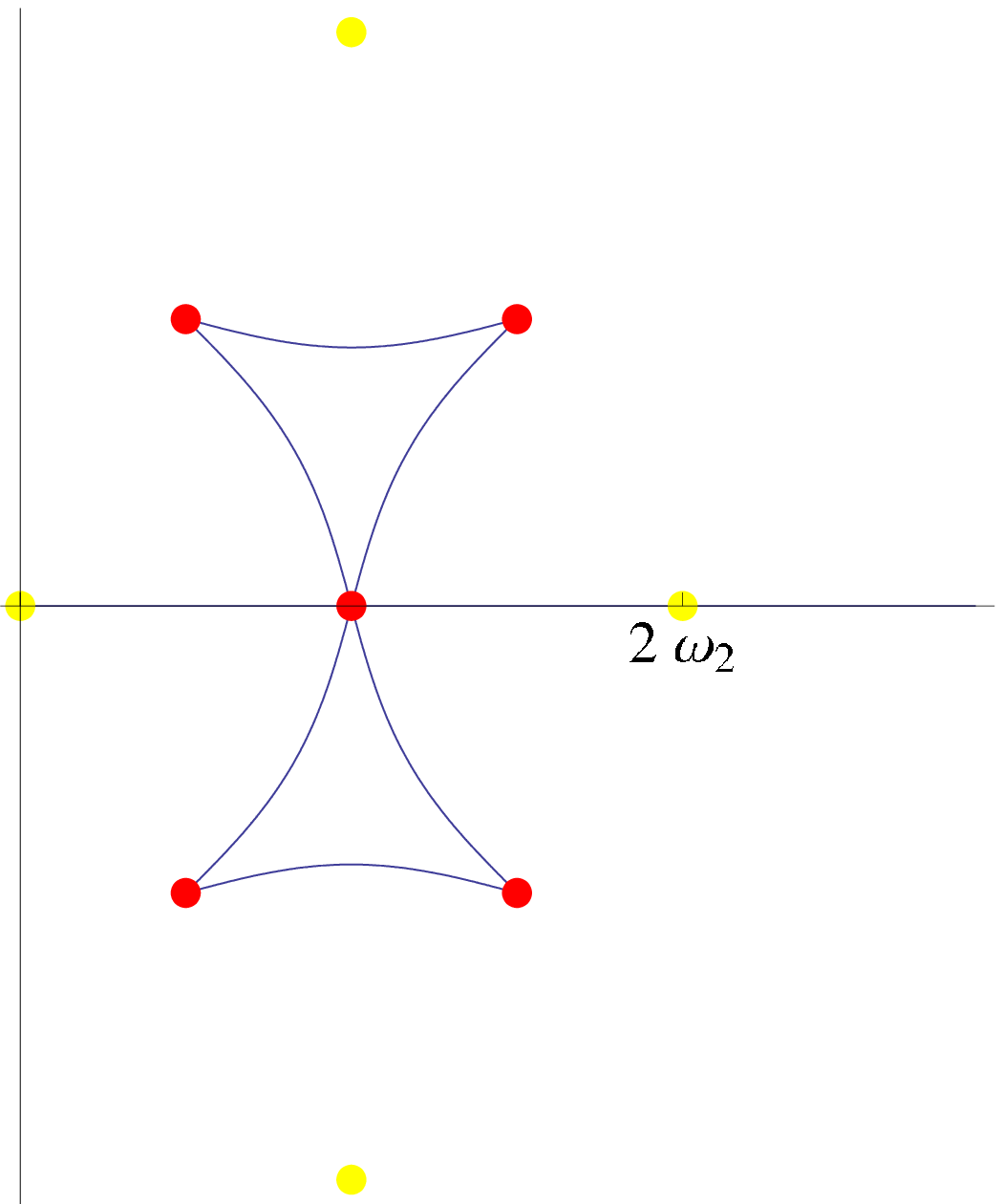,height=5cm}}
\caption{(Color online) The symmetric case $N=3.$ Zeros (red) of ${\cal V}$ and points (yellow) of the set $\Gamma_3$, and kink orbits (blue) connecting the zeros
of the potential (right panel).}
\end{figure}

In sum, the potential obtained from the deformation procedure has the same zeros in the FPP as the original model. Besides, one pole
arises at the origin due to the meromorphic structure of ${\cal P}^\prime_{01}$; see Fig.~2 and 3. However, this structure is
infinitely repeated in the deformed model, according to the two periods $\omega_1$ and $\omega_3$ determining the modular parameter
$\tau={{\omega_3}/{\omega_1}}=-{1}/{2}+i{\sqrt{3}}/{2}$ of the Riemann Surface of genus 1 associated with this ${\cal
P}$-Weierstrass function. As an aside, we note that the modular parameter ${\wt\tau}={1}/{2}+i{\sqrt{3}}/{2}$ gives the same
Riemann surface because $\wt\tau=(a\tau+b)/(c\tau+d)$ where
\[
\left(\begin{array}{cc} a & b \\ c & d\end{array}\right)=\left(\begin{array}{cc} 1 & 1 \\ 0 &1 \end{array}\right)
\]
is an element of the modular group $SL(2,{\mathbb Z})$.

Contrarily to the deformation function chosen in the case $N=2$, for $N=3$ the Weierstrass ${\cal P}$ function is not an entire function,
that is, it is not holomorphic in the whole complex plane ${\mathbb C}$. $\Gamma=\Omega_{mn}=2m\omega_1+2n\omega_2$ ($m,n \in {\mathbb Z}$)
is the lattice of points of ${\cal P}$ and ${\cal P}^\prime$ which are accordingly meromorphic functions. Thus, we suppose that the new field
take values in the space ${\cal C}/\Gamma_3$ to avoid the loss of holomorphicity. This point of view is very close to consider the genus 1
Riemann surface ${\mathbb C}/\!\!\sim{\!\!}\Gamma$ of modulus $\tau={\omega_3}/{\omega_2}$ minus the origin (the FPP with the edges identified pairwise minus
the origin) as the $\phi$-field space. Keeping, however, the infinite copies of this space contained in ${\cal C}/\Gamma_3$ gives a richer kink structure.

\subsubsection*{\bf B.1. Network of ${\cal P}$-kink orbits}

We shall compare the ${\cal P}$-kink orbits with the orbits of the original $N=3$ polynomial Wess-Zumino model. If $\chi^K(x)$ is a
$N=3$ solution of (\ref{knkorba}) and (\ref{knkpr}) then $\phi^K(x)=4^\frac13\P^{-1}(4^{-\frac13}\chi^K(x))$ solves
\be
{\rm Im}\;e^{-i\alpha}\,4^{\frac13}\,\P(4^{-\frac13}\phi^K(x))={\rm
constant}\, ; \;\;\;\;\;
{\rm Re}\;e^{-i\alpha}\,4^{\frac13}\,\P(4^{-\frac13}\phi^K(x))=\sigma
\label{eq:d6}
\ee
where
\be
\sigma=\int\,\left|\P^\pr(4^{-{\frac13}}\phi^K(x))\right|^2\, dx
\ee

\begin{figure}[htbp]
\centerline{\epsfig{file=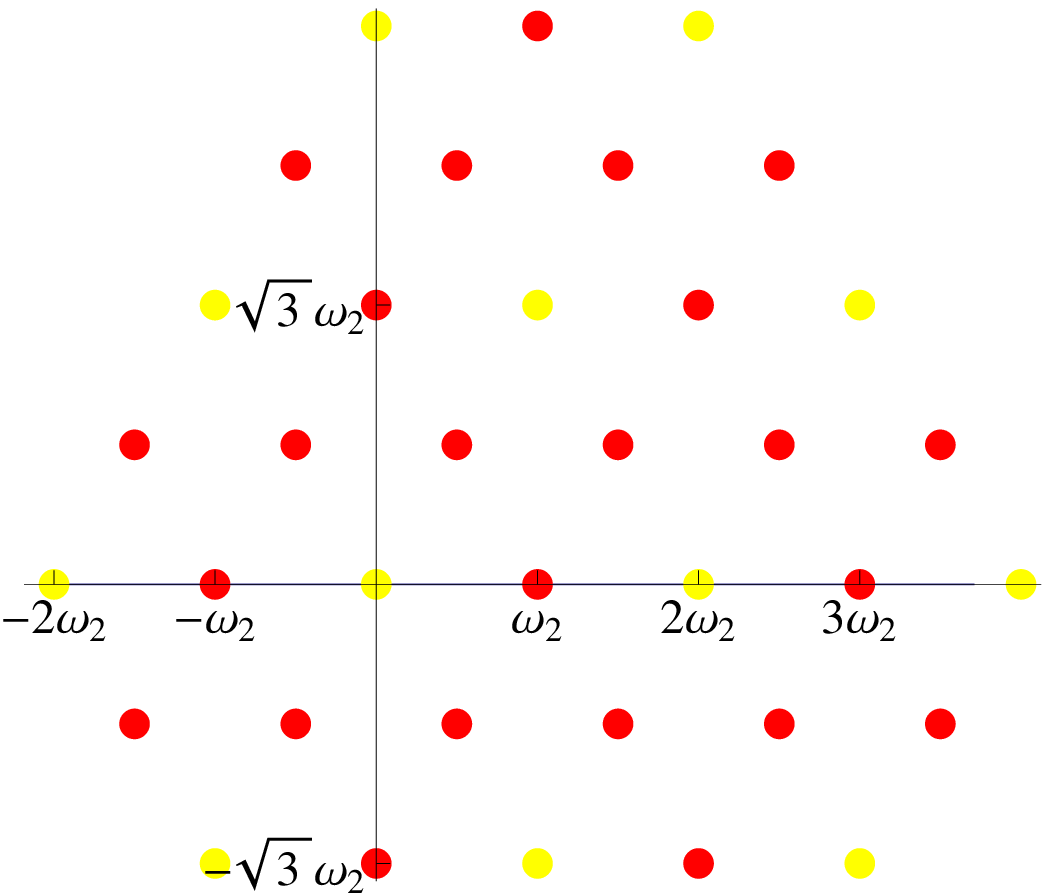,height=5cm}\hspace{1cm}\epsfig{file=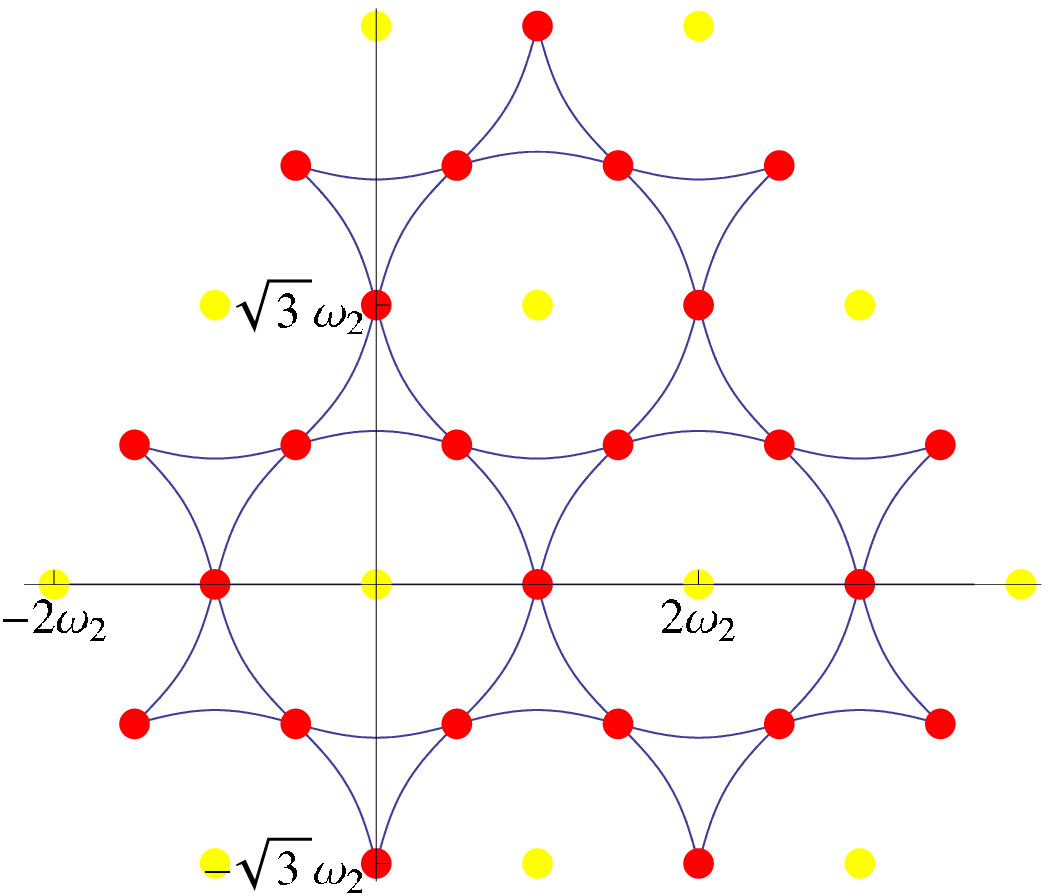,height=5cm}}
\caption{(Color online) The symmetric case $N=3.$ Lattice of zeros (red) of ${\cal V}(\phi,\ov{\phi})$ and points (yellow) of the set $\Gamma_3$, and the network of kink orbits (blue) in the lattice of minima (right panel).}
\end{figure}

Because of the relations
\be
W(\chi^{(k)})-W(\chi^{(j)})=\frac{3}{4}\left({\cal
W}(\phi^{(k)})-{\cal
W}(\phi^{(j)})\right)=\frac{3}{4}\left(e^{i\frac{2\pi}{3}(k-1)}-e^{i\frac{2\pi}{3}(j-1)}\right)
\ee
the same values of $\alpha$ as in the non deformed case,
\be
\alpha^{(kj)}={\rm arctan}\left[\frac{{\rm sin}\frac{2\pi}{3}(k-1)-{\rm sin}\frac{2\pi}{3}(j-1)}{{\rm
cos}\frac{2\pi}{3}(k-1)-{\rm cos}\frac{2\pi}{3}(j-1)}\right]
\ee
give the deformed kink orbits.

There are three types, which we show below.

{\subsubsection*{Type {\rm (13)}, non deformed}}

The condition $\,\mbox{Im}\,W_\alpha(\chi^{(3)})= \mbox{Im}W_\alpha(\chi^{(1)})$ is satisfied only for
$\alpha^{(31)}={\pi}/{6}$ (for antikinks) and $\alpha^{(13)}={7\pi}/{6}$ (for kinks). What is called kink and what is antikink is a
matter of convention. Our convention is that kink/antikink orbits run clock/anticlockwise in the $(W, {\ov W})$ plane. The orbits obey
\be
-\frac{3\sqrt{3}}{8}\leq{\rm Re} W_\frac{\pi}{6}(\chi^K)\leq
\frac{3\sqrt{3}}{8}\, \, \, ;\;\;\;{\rm Im}
W_\frac{\pi}{6}(\chi^K)=-\frac{3}{8} \;\;\; {\rm with} \;\;\;
{\rm Re}W_\frac{\pi}{6}(\chi^{(1)})=\frac{3\sqrt{3}}{8}\,;\;\;\;
{\rm Re} W_\frac{\pi}{6}(\chi^{(3)})=-\frac{3\sqrt{3}}{8}\nonumber
\ee
and
\be -\frac{3\sqrt{3}}{8}\leq{\rm Re}
W_\frac{7\pi}{6}(\chi^K)\leq\frac{3\sqrt{3}}{8}\, ;\;\;\;{\rm
Im} W_\frac{7\pi}{6}(\chi^K)=\frac{3}{8} \;\;\; {\rm with} \;\;\;
{\rm Re} W_\frac{7\pi}{6}(\chi^{(1)})=-\frac{3\sqrt{3}}{8}\,
;\;\;\;{\rm
Re}W_\frac{7\pi}{6}(\chi^{(3)})=\frac{3\sqrt{3}}{8}\nonumber \ee

{\subsubsection*{Type {\rm (13)}, deformed}}

As mentioned above, the condition Im${\cal W}_\alpha(\phi_{(m,n)}^{(3)})=$ Im${\cal W}_\alpha(\phi_{(m',n')}^{(1)})$ is again satisfied only for
$\alpha^{(31)}={\pi}/{6}$ and $\alpha^{(13)}={7\pi}/{6}$. The orbits obey
\be
-\frac{\sqrt{3}}{2}\leq{\rm Re}\W_\frac{\pi}{6}(\phi^K)\leq
\frac{\sqrt{3}}{2}\, ;\;\;\;{\rm Im}
\W_\frac{\pi}{6}(\phi^K)=-\frac12\nonumber \;\;\; {\rm with} \;\;\;
{\rm Re}
\W_\frac{\pi}{6}(\phi_{(m,n)}^{(3)})=-\frac{\sqrt{3}}{2}\,
;\;\;\;{\rm Re}
\W_\frac{\pi}{6}(\phi_{(m',n')}^{(1)})=\frac{\sqrt{3}}{2}
\ee
and
\be
-\frac{\sqrt{3}}{2}\leq{\rm Re}\W_\frac{7\pi}{6}(\phi^K)\leq
\frac{\sqrt{3}}{2}\, ;\;\;\;{\rm Im}
\W_\frac{7\pi}{6}(\phi^K)=\frac12\nonumber \;\;\; {\rm with} \;\;\;
{\rm Re}
\W_\frac{7\pi}{6}(\phi_{(m',n')}^{(1)})=-\frac{\sqrt{3}}{2}\,
;\;\;\;{\rm Re}
\W_\frac{7\pi}{6}(\phi_{(m,n)}^{(3)})=\frac{\sqrt{3}}{2}\nonumber
\ee

The nearest neighbor type (1) and type (3) minima are connected by the orbit following the sequence
$$
\phi_{(m,n)}^{(1)} \leftrightarrow\phi_{(m,n)}^{(3)} \leftrightarrow \phi_{(m+1,n)}^{(1)}
\leftrightarrow \phi_{(m+1,n)}^{(3)}\leftrightarrow\phi_{(m+2,n)}^{(1)}
$$
See Fig.~3.

The other two cases $(23)$ and (12) follow similarly. Here we just add that for the case $(23)$,
${(m,n)}$ and ${(m',n')}$ are restricted to link nearest neighbor type (3) and type (2) minima along the orbit. The sequence is
$$
\phi_{(m,n)}^{(2)} \leftrightarrow\phi_{(m,n)}^{(3)} \leftrightarrow \phi_{(m,n+1)}^{(2)}
\leftrightarrow \phi_{(m,n+1)}^{(3)}\leftrightarrow\phi_{(m,n+2)}^{(2)}
$$
Also, for the case $(12)$ we have that ${(m,n)}$ and ${(m',n')}$ must be chosen according to the following sequence
$$
\phi_{(m,n)}^{(1)}\!\!\leftrightarrow\!\phi_{(m,n+1)}^{(2)}\!\!\leftrightarrow\!\phi_{(m+1,n+1)}^{(1)}
\!\!\leftrightarrow\!\phi_{(m+1,n+2)}^{(2)}\!\!\leftrightarrow\!\phi^{(1)}_{(m+2,n+2)}
$$
in order to connect nearest neighbor minima of type (1) and (2).

We end the case $N=3$ collecting the corresponding energies. We have that the defect energies for the original non deformed
Wess-Zumino model are given by $M=3\sqrt{3}/4$, for kinks and anti-kinks for all the three sectors, with $(kj)=(12), (23),\, {\rm
and}\, (13)$. For the deformed model, the energies of the $\cal P$-defects are given by ${\cal M}=\sqrt{3},$ for the same cases.
\\

\subsection{The case $N=4$}

In the $N=4$ case we deal with
\be
W(\chi)=\chi-\frac15 \chi^5 \, , \;\;\;\;\;
V(\chi,\ov\chi)=\frac12(1-\chi^4)(1-\ov\chi^4)
\ee
Thus, putting $\chi=f(\phi)$, we have
\be
W(f)=f(\phi)-\frac15 f^5(\phi)\, , \;\;\;\;\;
V=\frac12(1-f^4(\phi))(1-\ov{f^4(\phi)})
\ee
The special deformation function must satisfy
\be\label{eq:dsup1}
f'(\phi) \ov{f'(\phi)}=\,\sqrt{(1-f^4(\phi))(1-\ov{f^4(\phi)})}
\ee
Arguing like in \Ref{eq:dsup}, we choose the holomorphic solution that satisfies the separated equations
\be\label{eq:dsup1b}
f'(\phi)^2=1-f(\phi)^4\, ,\;\;\;\overline{f'(\phi)^2}=1-\overline{f(\phi)^4}
\ee
The solution of \Ref{eq:dsup1b}, henceforth a solution of \Ref{eq:dsup1}, is the
elliptic sine of parameter $k^2=-1$, the Gauss's {\it sinus lemniscaticus}
\be
{\cal W}(\phi)=f(\phi)={\rm sn}(\phi,-1)
\ee
As the derivative of the Jacobi elliptic sine is ${\rm sn}\,u^\pr={\rm cn}\,u \;{\rm dn}\,u$, the identities
\be
{\rm cn}^2(\phi,-1)=1-{\rm sn}^2(\phi,-1);\;\;\;\;\; {\rm
dn}^2(\phi,-1)=1+{\rm sn}^2(\phi,-1)
\ee
show the solution of (\ref{eq:dsup1}) very directly.

The deformed potential reads
\be
{\cal V}(\phi,\ov{\phi})=\frac12\sqrt{ (1-{\rm sn}^4(\phi,-1)) (1-\ov{\rm
sn}^4(\phi,-1))}=\frac12|{\rm cn}(\phi,-1)|^2 |{\rm dn}(\phi,-1)|^2
\ee
which is depicted in Fig.~4. The superpotential is given by $ {\cal W}_\alpha(\phi)= e^{-i \alpha} \, {\rm sn}(\phi,-1)$

\begin{figure}[htbp]
\centerline{\epsfig{file=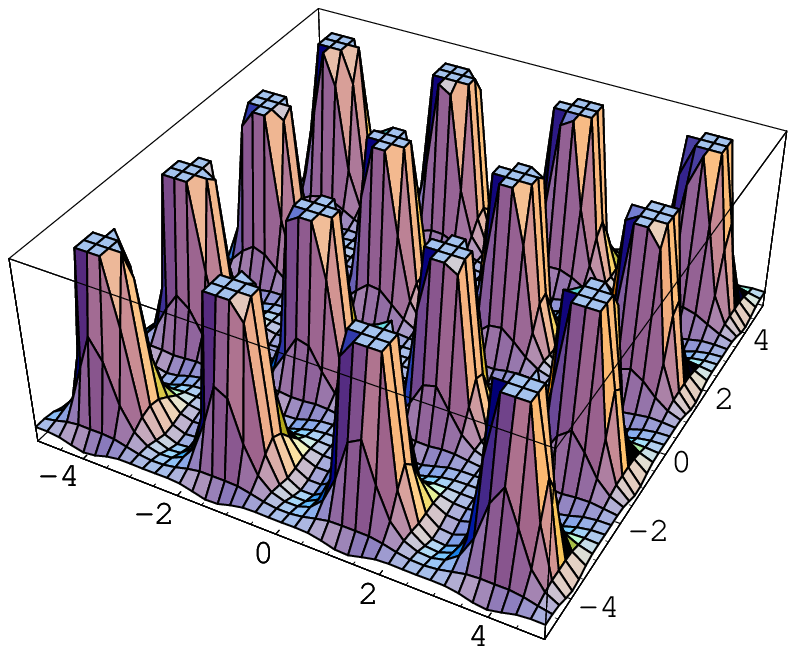,height=5cm}\hspace{1cm}\epsfig{file=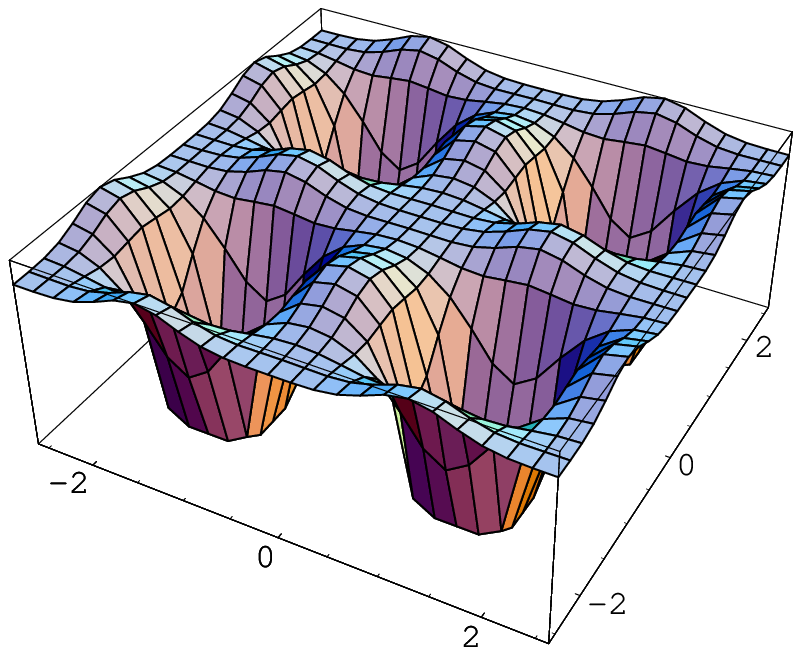,height=5cm}}
\caption{(Color online) The symmetric case $N=4.$ 3D graphics of the potential ${\cal V}(\phi,\ov{\phi})$ (left panel) and of $-{\cal V}(\phi,\ov{\phi})$
near {\it four points} of the set $\Gamma_4$ (right panel). Note that in the right panel the zeros are now maxima.}
\end{figure}

The new potential is doubly periodic with its structure inherited from the \lq\lq quarter-periods" $K(-1)={\omega_1/ 4}$ and $i
K(2)={\omega_2/ 4}$ of the twelve Jacobi elliptic functions; see Fig.~5. Here $K(-1)\approx 1.31103$ is the complete elliptic
integral of the first type, a quarter of the length of the lemniscate curve in field space:
$(\phi_1^2+\phi_2^2)^2=\phi_1^2-\phi_2^2.$ $K(2)\approx 1.31103-i1.31103$ is the complementary complete elliptic integral of $K(-1)$.

The set of zeros of the potential in the FPP are $\phi^{(1)}={\omega_1/4},\phi^{(2)}=i{\omega_1/4},
\phi^{(3)}=-{\omega_1/4},\phi^{(4)}=-i{\omega_1/ 4},$ whereas the set of all the zeros of ${\cal V}$ form a quadrangular lattice in the whole
configuration space. They are given by, explicitly
\bes\ben
\phi^{(1)}_{(m,n)}&=& {\frac{\omega_1}{4}}\left(2(2m+n)+1+ i
2n\right)\,;\;\;\;\;\;
\phi^{(2)}_{(m,n)}={\frac{\omega_1}{4}}\left(2(2m+n)+i(2n+1)\right)
\\
\phi^{(3)}_{(m,n)}&=&{\frac{\omega_1}{4}}\left(2(2m+n)-1+ i
2n\right)\,;\;\;\;\;\;
\phi^{(4)}_{(m,n)}={\frac{\omega_1}{4}}\left(2(2m+ n)+i(2n-1)\right)
\een\ees
because
\be
{\rm cn}(\phi^{(k)}_{(m,n)},-1)\cdot {\rm
dn}(\phi^{(k)}_{(m,n)},-1)=0 \, , \;\;\; k=1,2,3,4
\ee
Thus,
\be {\cal W}_\alpha(\phi^{(1)})=e^{-i\alpha} \, , \;\;\;
{\cal W}_\alpha(\phi^{(2)})=ie^{-i\alpha} \, , \;\;\;
{\cal W}_\alpha(\phi^{(3)})=-e^{-i\alpha} \, , \;\;\; {\cal
W}_\alpha(\phi^{(4)})=-ie^{-i\alpha}
\ee
since ${\rm sn}[\pm K(-1),-1]=\pm 1$ and ${\rm sn}[\pm iK(-1),-1]=\pm i$.

The modular parameter of the associated genus 1 Riemann surface is $\tau=i{K[2]}/{K[-1]}=1+i$. Identical Riemann surface is
associated to the lemniscatic case, $g_2=1$, $g_3=0$, of the Weierstrass ${\cal P}$ function. Like in the former case, however, the Jacobi elliptic
sine is not an entire function, and so we restrict the new field to live in ${\mathbb C}/\Gamma_4$ (where $\Gamma_4$ is the set of poles of $f(\phi)$
in this case) in order to make the superpotential holomorphic.

\subsubsection*{\bf C.1. Network of ${\rm sn}$-kink orbits}

We shall compare the ${\rm sn}$-kink orbits with the orbits of the original $N=4$ polynomial Wess-Zumino model. If $\chi^K(x)$ is a
solution of (\ref{eq:defo1}) then $\phi^K(x)={\rm sn}^{-1}(\chi^K(x),-1)$ solves
\be
{\rm Im}\, e^{-i\alpha}{\rm sn}(\phi^K(x),-1)={\rm constant}\, ; \;\;\;\;\;
{\rm Re}\, e^{-i\alpha}{\rm sn}(\phi^K(x),-1)=\sigma\label{eq:d61}
\ee
where
\be \sigma=\int \,\left|{\rm cn}[\phi^K(x),-1]\cdot {\rm
dn}[\phi^K(x),-1]\right|^2\, dx
\ee
The same values of $\alpha$ as in the non deformed case give the kink orbits
\be
W(\chi^{(k)})-W(\chi^{(j)})={\frac45}\left({\cal
W}(\phi^{(k)})-{\cal W}(\phi^{(j)})\right)={\frac45}\left(e^{i\frac{\pi}{2}(k-1)}-e^{i\frac{\pi}{2}(j-1)}\right)
\ee
selects
\be \alpha^{(kj)}={\rm arctan}\left[\frac{{\rm
sin}\frac{\pi}{2}(k-1)-{\rm sin}\frac{\pi}{2}(j-1)}{{\rm
cos}\frac{\pi}{2}(k-1)-{\rm cos}\frac{\pi}{2}(j-1)}\right]\, ,\;\;\; {\rm mod}\,\pi
\ee
as the angles for both the original and deformed kink orbits.

There are four types, which we show below.

{\subsubsection*{Type {\rm (12)/(34)}, non deformed}}

In this case we have Im$W_\alpha(\chi^{(1)})$=Im$W_\alpha(\chi^{(2)})$ and Im$W_\alpha(\chi^{(3)})$=Im$W_\alpha(\chi^{(4)})$ only for
$\alpha={3\pi}/{4}$ (kinks) or $\alpha={7\pi}/{4}$ (antikinks). The kink orbits obey
\be
-\frac{2\sqrt{2}}{5} \leq {\rm
Re}W_\frac{3\pi}{4}(\chi^K)\leq \frac{2\sqrt{2}}{5} \, ; \;\;\;
{\rm Im} W_\frac{3\pi}{4}(\chi^K)=-\frac{2\sqrt{2}}{5}\;\;\; {\rm
with} \;\;\;
\ba{c} {\rm Re}W_\frac{3\pi}{4}(\chi^{(1)})=-\frac{2\sqrt{2}}{5}\,
;\;\;\;{\rm Re}W_\frac{3\pi}{4}(\chi^{(2)})=\frac{2\sqrt{2}}{5} \\
{\rm or} \\
{\rm Re}W_\frac{3\pi}{4}(\chi^{(3)})=\frac{2\sqrt{2}}{5}\,;\;\;\;
{\rm Re}W_\frac{3\pi}{4}(\chi^{(4)})=-\frac{2\sqrt{2}}{5} \ea \nonumber
\ee
whereas for the antikinks
\be
-\frac{2\sqrt{2}}{5}
\leq {\rm Re}W_\frac{7\pi}{4}(\chi^K)\leq \frac{2\sqrt{2}}{5}\,
;\;\;\;{\rm Im} W_\frac{7\pi}{4}(\chi^K)=\frac{2\sqrt{2}}{5}\;\;\;
{\rm with} \;\;\;\ba{cc} {\rm Re}W_\frac{7\pi}{4}(\chi^{(1)})=\frac{2\sqrt{2}}{5}\,;\;\;\;
{\rm Re}W_\frac{7\pi}{4}(\chi^{(2)})=-\frac{2\sqrt{2}}{5}\\
{\rm or} \\ {\rm
Re}W_\frac{7\pi}{4}(\chi^{(3)})=-\frac{2\sqrt{2}}{5}\,\,\,
;\;\;\;{\rm Re}W_\frac{7\pi}{4}(\chi^{(4)})=\frac{2\sqrt{2}}{5} \ea\nonumber
\ee

{\subsubsection*{Type {\rm(12)/(34)}, deformed}}

Here we have Im${\cal W}_\alpha(\phi_{(m,n)}^{(1)})=$ Im${\cal W}_\alpha(\phi_{(m',n')}^{(2)})$ and Im${\cal
W}_\alpha(\phi_{(m,n)}^{(3)})=$ Im${\cal W}_\alpha(\phi_{(m',n')}^{(4)})$ only for $\alpha={3\pi}/{4}$
(kinks) or $\alpha={7\pi}/{4}$ (antikinks). The kink/antikink orbits obey
\be
-\frac{\sqrt{2}}{2}\leq{\rm Re}{\cal
W}_\frac{3\pi}{4}(\phi^K)\leq\frac{\sqrt{2}}{2}\,;\;\;\; {\rm
Im}{\cal W}_\frac{3\pi}{4}(\phi^K)=-\frac{\sqrt{2}}{2}\;\;\; {\rm
with} \;\;\; 
\ba{c} {\rm Re}{\cal W}_\frac{3\pi}{4}(\phi_{(m,n)}^{(1)})=-\frac{\sqrt{2}}{2}\,;\;\;\;
{\rm Re}{\cal W}_\frac{3\pi}{4}(\phi_{(m',n')}^{(2)})=\frac{\sqrt{2}}{2}\\
{\rm or} \\ 
{\rm Re}{\cal W}_\frac{3\pi}{4}(\phi_{(m',n')}^{(3)})=\frac{\sqrt{2}}{2}\,;\;\;\;
{\rm Re}{\cal W}_\frac{3\pi}{4}(\phi_{(m,n)}^{(4)})=-\frac{\sqrt{2}}{2} \ea
\nonumber
\ee
and
\be
-\frac{\sqrt{2}}{2} \leq{\rm Re}{\cal
W}_\frac{7\pi}{4}(\phi^K)\leq \frac{\sqrt{2}}{2} \, ;\;\;\;
{\rm Im}{\cal W}_\frac{7\pi}{4}(\phi^K)=\frac{\sqrt{2}}{2} \;\;\; {\rm
with} \;\;\; \ba{c} 
{\rm Re}{\cal W}_\frac{7\pi}{4}(\phi_{(m,n)}^{(1)})=\frac{\sqrt{2}}{2}\,;\;\;\;
{\rm Re}{\cal W}_\frac{7\pi}{4}(\phi_{(m',n')}^{(2)})=-\frac{\sqrt{2}}{2}
\\ {\rm or}
\\ {\rm Re}{\cal W}_\frac{7\pi}{4}(\phi_{(m',n')}^{(3)})=-\frac{\sqrt{2}}{2}\,\,\,
;\;\;\;{\rm Re}{\cal W}_\frac{7\pi}{4}(\phi_{(m,n)}^{(4)})=\frac{\sqrt{2}}{2}\ea
\nonumber
\ee 
where ${(m,n)}$ and ${(m',n')}$ are restricted to link nearest neighbor type (1) with type (2) and type (3) with type (4)
minima along the orbit. The sequences are
$$\phi_{(m,n)}^{(1)}\leftrightarrow\phi_{(m,n)}^{(2)}\leftrightarrow\!\phi_{(m-1,n+1)}^{(1)}
\leftrightarrow\phi_{(m-1,n+1)}^{(2)}\leftrightarrow\phi_{(m-2,n+2)}^{(1)}
$$
and
$$\phi_{(m,n)}^{(3)}\leftrightarrow\phi_{(m,n)}^{(4)}\leftrightarrow\phi_{(m-1,n+1)}^{(3)}
\leftrightarrow\phi_{(m-1,n+1)}^{(4)}\leftrightarrow\!\phi_{(m-2,n+2)}^{(3)}
$$
See Fig.~6.

\begin{figure}[htbp]
\centerline{\epsfig{file=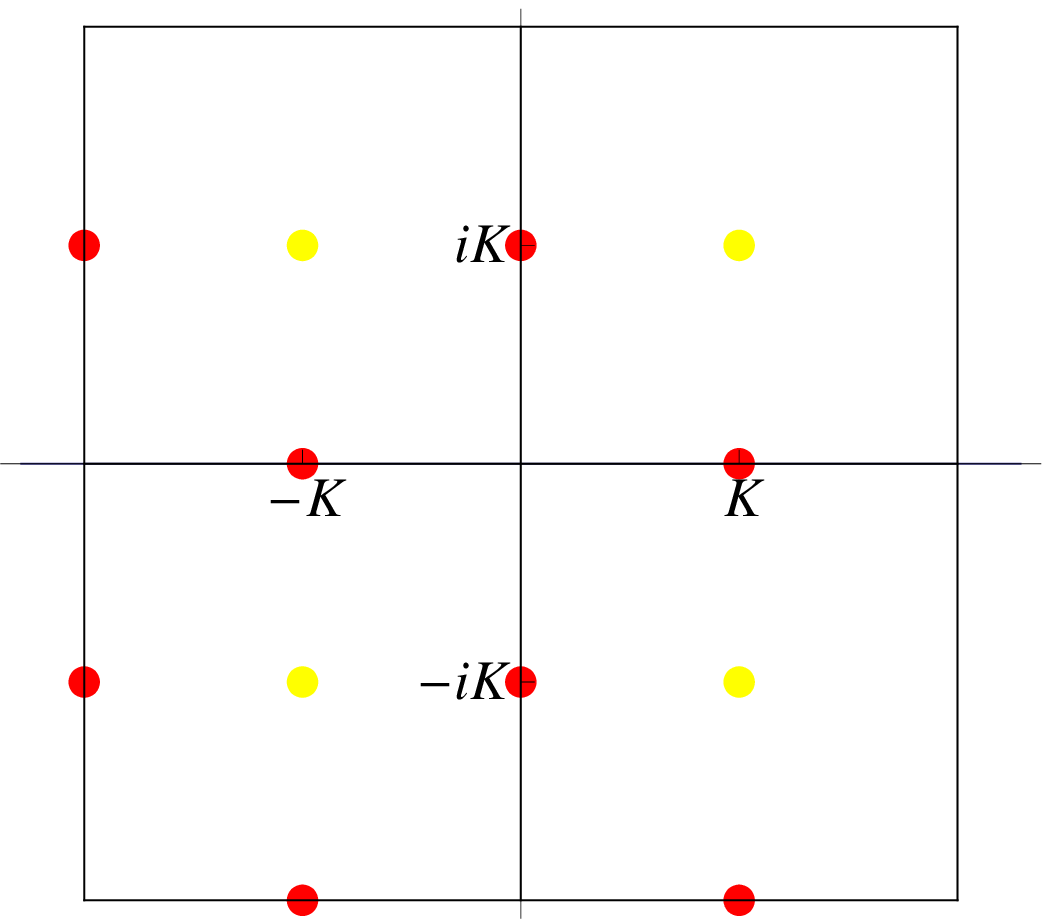,height=3.4cm}
\hspace{1cm}{\epsfig{file=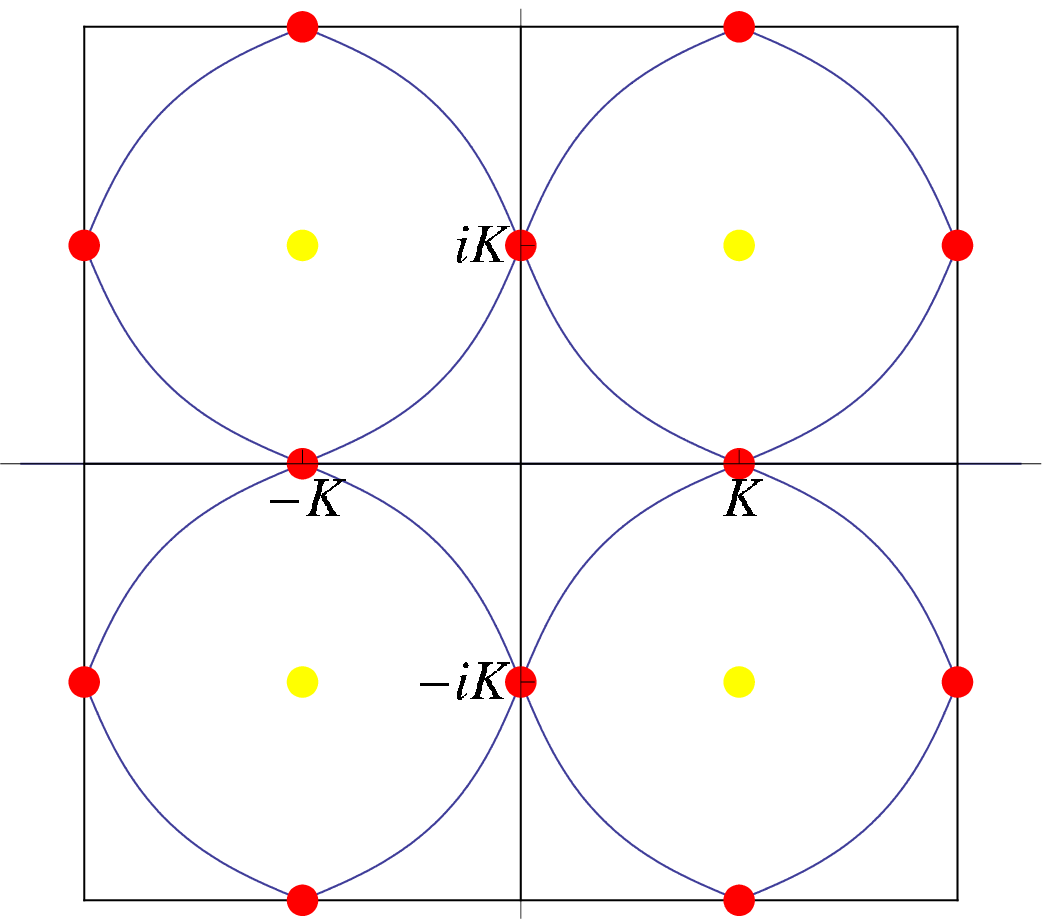,height=3.4cm}}}
\caption{(Color online) The symmetric case $N=4.$ Zeros (red) of ${\cal V}$ and points (yellow) of the set $\Gamma_4$, and the two (blue and black) possible orbits connecting the zeros of the potential.}
\end{figure}

\begin{figure}[htbp]
\centerline{\epsfig{file=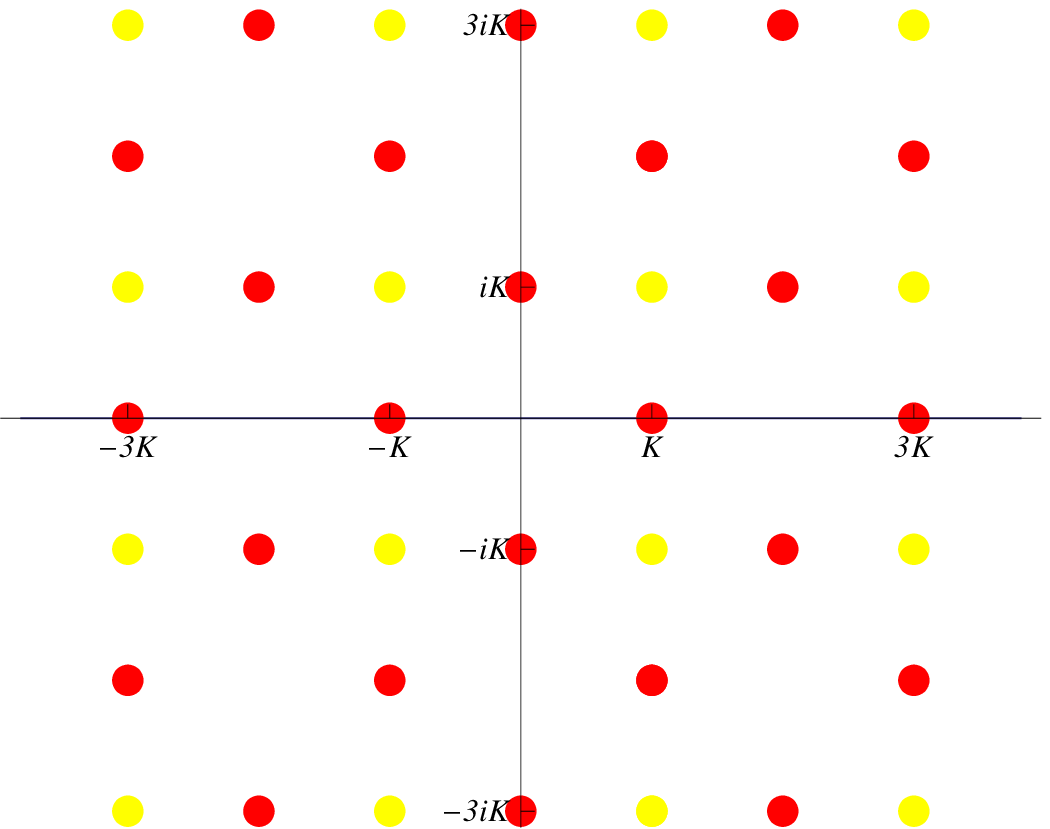,height=5cm}\hspace{1cm}{\epsfig{file=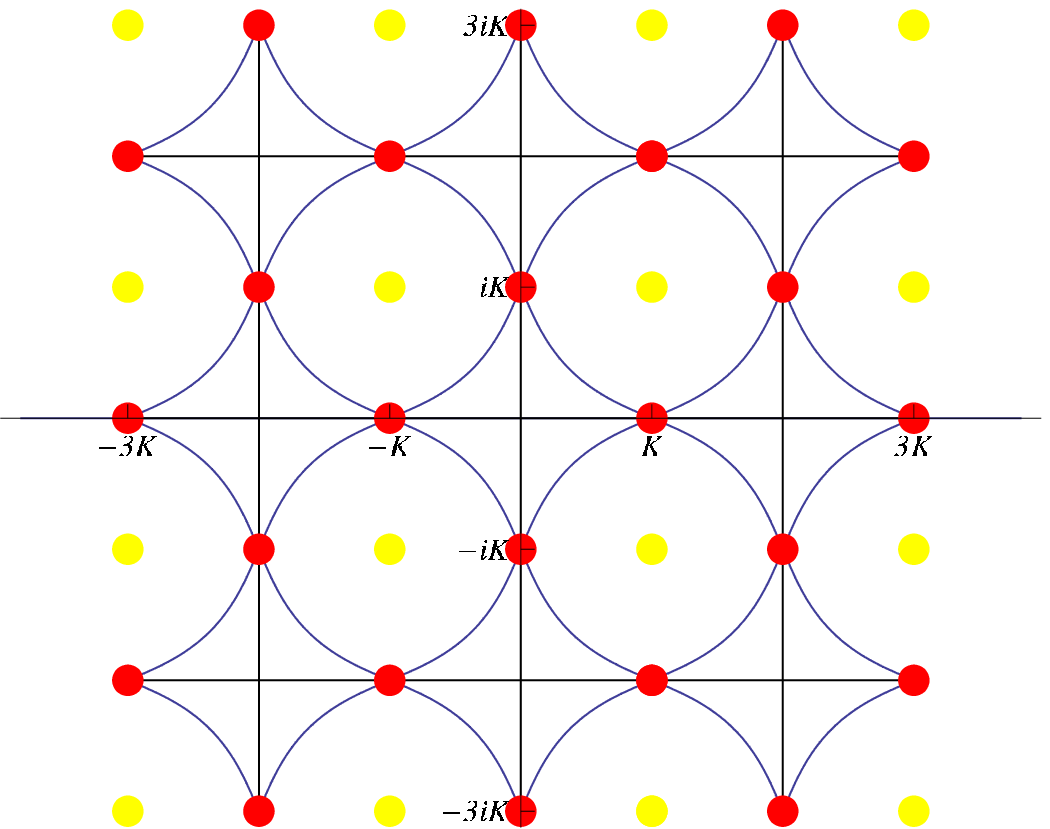,height=5cm}}}
\caption{(Color online) The symmetric case $N=4.$ Lattice of zeros (red) of ${\cal V}(\phi,\ov{\phi})$ and points (yellow) of the set $\Gamma_4$, and the networks of kink orbits (blue and black) in the lattice (right panel).}
\end{figure}

The other three cases, $(13)$, $(14)/(23)$, and $(24)$ follow similarly. Here we just add that in the case $(13)$ we have that ${(m,n)}$ and ${(m',n')}$ are restricted to link nearest neighbor type (1) and type (3) minima along the orbit, and so they must be chosen according to the following sequence
$$
\phi_{(m,n)}^{(3)} \leftrightarrow\phi_{(m,n)}^{(1)}\leftrightarrow\phi_{(m+1,n)}^{(3)}\leftrightarrow
\phi_{(m+1,n)}^{(1)}\leftrightarrow\phi_{(m+2,n)}^{(3)}
$$
In the case (14)/(23) we have that ${(m,n)}$ and ${(m',n')}$ are restricted to link nearest neighbor type (1) and type (2) minima respectively with type (4) and type (3) minima along the orbit. Therefore, the sequences are
$$
\phi_{(m,n)}^{(1)}\leftrightarrow\phi_{(m,n)}^{(4)}\leftrightarrow\phi_{(m,n+1)}^{(1)}\leftrightarrow
\phi_{(m,n+1)}^{(4)}\leftrightarrow
\phi_{(m,n+2)}^{(1)}
$$
and
$$
\phi_{(m,n)}^{(2)}\leftrightarrow\phi_{(m,n)}^{(3)}\leftrightarrow\phi_{(m,n+1)}^{(2)}
\leftrightarrow\phi_{(m,n+1)}^{(3)}\leftrightarrow
\phi_{(m,n+2)}^{(1)}
$$
In the case $(24)$ we have that ${(m,n)}$ and ${(m',n')}$ are restricted to link nearest neighbor type (2) and type (4) minima along the orbit.
Thus, they must be chosen according to the following sequence
$$
\phi_{(m,n)}^{(4)}\leftrightarrow\phi_{(m,n)}^{(2)}\leftrightarrow\phi_{(m-2,n+1)}^{(4)}
\leftrightarrow \phi_{(m-2,n+1)}^{(2)}
$$

We end the $N=4$ case collecting the energies of the defect structures. For the non deformed kinks we get
\be
M(13)=M(24)=\frac{8}{5}\,;\;\;\;\;\;
M(12)=M(34)=\frac{4\sqrt{2}}{5}\,;\;\;\;\;\;
M(14)=M(23)=\frac{4\sqrt{2}}{5}
\ee
whereas the energies of the deformed ${\rm sn}$-kinks read
\be
{\cal M}(13)={\cal M}(24)=2\,;\;\;\;\;\; {\cal M}(12)={\cal
M}(34)=\sqrt{2}\,;\;\;\;\;\; {\cal M}(14)={\cal
M}(23)=\sqrt{2}
\ee
with $M(kj)=M(jk)$ and ${\cal M}(kj)={\cal M}(jk)$. \vspace{1cm}

\section{Deformation of Abraham-Townsend models}
\label{sec:nonsymm}

Let us now move on to the case where the original model engenders no specific symmetry. This study is inspired on Ref.~\cite{AT}, in
which Abraham and Townsend consider some interesting situations, guided by more general superpotentials, which develop no specific symmetry.
Similar potentials were also considered in \cite{V}, but there the investigation was mainly on the symmetric case. In \cite{AT}, however,
the focus was on the non symmetric case, as a basic model underlying the study of intersecting extended objects in supersymmetric
field theories -- see also \cite{PT}, which deals with the forces between soliton states in the same model.

The absence of symmetry makes the investigation harder to follow, but it is still of current interest since it leads to more general
possibilities, bringing some new effects into the game and allowing for the presence of irregular network of defects. We follow as in
the former section, and below we consider models which develop three and four minima with no specific symmetry anymore, using 
${\wt\Gamma}_3$ and ${\wt\Gamma}_4$ as the set of poles of $f(\phi)$ as before, but now in the asymmetric cases, with $N=3$ and $N=4$,
respectively.

\subsection{Irregular network of ${\cal P}$-kink orbits}

Let us start with the simplest case, in which one considers a class of models that engenders three minima. Here we deal with the
superpotential
\be
W(\chi)=\delta\chi-\frac12\chi^2-\frac13\delta\chi^3+\frac14\chi^4
\ee
where $\delta=\delta_1+i\delta_2$ is a complex coupling constant which parametrizes the family of models. In this case, the potential
is given by
\be
V(\chi)=\frac12|\delta-\chi|^2 |1-\chi^2|^2
\ee
This choice leads to the following set of minima: two real minima which are fixed to be at $v_1=-1$ and $v_2=1,$ and a complex minimum
at $v_3=\delta$, which may move in the complex plane for different choices of the complex parameter $\delta.$ This is the most general
case with three arbitrary minima in the complex plane, since one can always choose the straight line joining two vacua as the abscissa axis,
crossing the perpendicular ordinate axis through the middle point between the vacua, setting the distance between them to be 2 by an
scale transformation. The values that the superpotential $W_\alpha(\chi)=e^{-i\alpha}W(\chi)$ takes now at the vacua are
\be
W_\alpha(\pm1)=\left(-\frac14\pm\frac23\delta\right)\,e^{-i\alpha}\, ; \;\;\;\;\;
W_\alpha(\delta)=\frac1{12}\delta^2(6-\delta^2)e^{-i\alpha}
\ee
From these expressions we obtain
\be
W_\alpha(1)-W_\alpha(-1)=\frac{4}{3}
\, \delta \, e^{-i\alpha}\, ,\;\;\;\;\; W_\alpha(\mp
1)-W_\alpha(\delta)=\frac{1}{12} (\delta\mp 3)(\delta\pm 1)^3 \,
e^{-i\alpha}
\ee
Therefore, the angles of the kink orbits are (mod $\pi$)
\be
\alpha^{(12)}=\arctan \left[\frac{{\rm Im}\delta}{{\rm Re}\delta}\right]\, , \;\;\;
\alpha^{(13)}=\arctan\left[\frac{{\rm Im} ((\delta-3)(\delta+1)^3)
}{{\rm Re} ((\delta-3)(\delta+1)^3 )}\right]\, , \;\;\;
\alpha^{(23)}=\arctan\left[\frac{{\rm Im} ((\delta+3)(\delta-1)^3)
}{{\rm Re} ((\delta+3)(\delta-1)^3 )}\right]  \nonumber
\ee
and now the kink energies are given by
\be
M(12)=\frac43(\delta_1^2+\delta_2^2)^{1/2}\, ,
\;\;\; \ba{c} M(23)=\frac1{12}\left(((\delta_1+3)^2+\delta_2^2)((\delta_1-1)^2+\delta_2^2)^3\right)^{1/2}\;\;
\\
M(31)=\frac1{12}\left(((\delta_1^2-3)^2+\delta_2^2)((\delta_1^2+1)^2+\delta_2^2)^3\right)^{1/2}\;\;
\ea\ee
with $M(kj)=M(jk)$. Note that the three masses $M(12)=M(13)=M(23)=(4/3)\sqrt{3}$ for $\delta=\pm i\sqrt{3}$, a
value of the parameter for which the vacua lie at the vertices of an equilateral triangle, leading us back to the symmetric case which engenders the $Z_3$ symmetry.

We now go to the deformation procedure, changing $\chi\to f(\phi),$ with $f(\phi)={\cal W}(\phi).$ In the present case, we get that
$f(\phi)$ should obey
\be
f^\prime(\phi)\overline{f^\prime(\phi)}=\sqrt{2V(f(\phi),\overline{f(\phi)})}
\ee
and this now gives, in the specially simple case which we have already considered in the former section,
\be
f^\prime(\phi)^2=\delta-f-\delta f^2+f^3
\ee
This is again the Weierstrass equation, and if we define $z=4^{-\frac13}\phi$ and $f=4^{\frac13}\psi+\delta/3$ we find
\be
\psi^\prime(z)^2=4\psi^3-g_2\psi-g_3 \label{cubic}
\ee
where $g_2=4^{\frac13}(1+\delta^2/3)$ and $g_3=-(2\delta/3)(1-\delta^2/9)$. The solution for $f(\phi)$ is then given by
\be
f(\phi)=\frac13\delta+4^{\frac13}{\cal P}(4^{-\frac13}\phi;g_2(\delta),g_3(\delta))
\ee
The half-periods $\omega_1$ and $\omega_3$ of the Weierstrass ${\cal P}$ function are obtained from
the invariants $g_2$ and $g_3$ by means of the equation
\be
g_2=60\sum_{(m,n)}\Omega_{mn}^{-4} \, ;  \;\;\;\;\; g_3=140\sum_{(m,n)}\Omega_{mn}^{-6}
\ee
where $\Omega_{mn}=2m\omega_3+2n\omega_1,$ with $m,n\in {\mathbb Z}$. We also have $\Delta=g_2^3-27g_3^2=4(\delta^2-1)^2$ and
$e_1=-4^{\frac13}(1+\delta/3),$ $e_2=4^{\frac13}(1-\delta/3),$ and $e_3=4^{\frac13}2\delta/3$ are respectively the discriminant and
roots of the cubic equation which appears from the right hand side of (\ref{cubic}).

It is interesting to note that the modular parameter $\tau(\delta)=\omega_3(\delta)/\omega_1(\delta)$ of the genus $1$ Riemann surface associated
to this ${\cal P}$-function depends on $\delta$. Thus, variations of $\delta$ correspond to motions in the Riemann surface moduli space.

The deformed model is governed by the potential
\be
{\cal V}(\phi)=\frac12|{\cal P}^\prime(4^{-\frac13}\phi;g_2(\delta)),g_3(\delta)|^2
\ee
It has zeros in the FPP at ${\cal P}^\prime(\omega_1)={\cal P}^\prime(\omega_3)={\cal P}^\prime(\omega_1+\omega_3)=0$. Thus,
the vacua of the deformed model are the constant field configurations
\be
\phi^{(1)}_{(m,n)}=4^{\frac13}(\omega_1+\Omega_{mn})\,;\;\;\;\;\;
\phi^{(2)}_{(m,n)}=4^{\frac13}(\omega_1+\omega_3+\Omega_{mn})\,;\;\;\;\;\;
\phi^{(3)}_{(m,n)}=4^{\frac13}(\omega_3+\Omega_{mn})
\ee
The values of the superpotential
\be
{\cal W}_\gamma(\phi)=\left(\frac{\delta}{3}+4^{\frac13}{\cal
P}(4^{- \frac13}\phi;g_2(\delta),g_3(\delta))\right)e^{-i\gamma}
\ee
at these vacua are
\bes\ben 
& &{\cal W}_\gamma(\phi^{(1)}_{(m,n)})=\left(\frac{\delta}{3}+2^{\frac23}{\cal P}_{g_2g_3}(\omega_1)\right)e^{-i\gamma}= \left(\frac{\delta}{3}+2^{\frac23}e_1\right)e^{-i\gamma}=-e^{-i\gamma} \nonumber\\ 
& &{\cal W}_\gamma(\phi^{(2)}_{(m,n)})=\left(\frac{\delta}{3}+2^{\frac23}{\cal P}_{g_2g_3}(\omega_2)\right)e^{-i\gamma}= \left(\frac{\delta}{3}+2^{\frac23}e_2\right)e^{-i\gamma}=e^{-i\gamma} \nonumber\\ 
& &{\cal W}_\gamma(\phi^{(3)}_{(m,n)})=\left(\frac{\delta}{3}+2^{\frac23}{\cal P}_{g_2g_3}(\omega_3)\right)e^{-i\gamma}= \left(\frac{\delta}{3}+2^{\frac23}e_3\right)e^{-i\gamma}=\delta e^{-i\gamma} \nonumber
\een\ees
Therefore the angles of the deformed kink orbits are (mod $\pi$)
\be
\gamma^{(12)}=0 \, , \;\;\;\;\; \gamma^{(13)}={\rm
arctan}\left[\frac{{\rm Im}\delta}{{\rm Re}(\delta+1)}\right] \,
, \;\;\;\;\; \gamma^{(23)}={\rm arctan}\left[\frac{{\rm Im}\delta}{{\rm Re}(\delta-1)}\right]
\ee
and the kink masses become
\be
{\cal M}(12)=2 \, , \;\;\;\;\;
{\cal M}(13)=\left((\delta_1+1)^2+\delta_2^2\right)^{\frac12} \, ,
\;\;\;\;\; {\cal M}(23)=\left((\delta_1-1)^2+\delta_2^2\right)^{\frac12}
\ee
Note that for $\delta=\pm i\sqrt{3}$ the three masses are equal, ${\cal M}(12)={\cal M}(13)={\cal M}(23)=2$, corresponding to
a regular triangular lattice of minima. The sequences of minima connected by the kink orbits in these families are
\bes\ben
\phi^{(2)}_{(m,n)}\;\;\; &\leftrightarrow& \;\;\; \phi^{(1)}_{(m, n)}
\;\;\;\;\; \leftrightarrow \;\;\;\;\; \phi^{(2)}_{(m-1,n+1)} \;\;\;
\leftrightarrow \;\;\; \phi^{(1)}_{(m-1,n+1)}\nonumber \\
\phi^{(3)}_{(m,n)}\;\;\; &\leftrightarrow& \;\;\; \phi^{(1)}_{(m,
n-1)} \;\; \leftrightarrow \;\;\;
\phi^{(3)}_{(m,n-1)} \;\;\;\;\;\;\; \leftrightarrow \;\;\; \phi^{(1)}_{(m,n-2)}\nonumber\\
\phi^{(3)}_{(m,n)}\;\;\; &\leftrightarrow& \;\;\; \phi^{(2)}_{(m, n)}
\;\;\;\;\; \leftrightarrow \;\;\;\;\; \phi^{(3)}_{(m+1,n)} \;\;\;\;\;
\leftrightarrow \;\;\;\;\; \phi^{(2)}_{(m+1,n)} \nonumber
\een\ees

In Fig.~7 and 8 we plot the potential, minima and network of kink orbits, respectively, for the specific value of the
complex parameter $\delta=1+i$. Comparison of these figures with figures 2 and 3 shows to what extent the complex
parameter induces irregularity when it differs from $\pm i\sqrt{3}$.

\begin{figure}[htbp]
\centerline{\epsfig{file=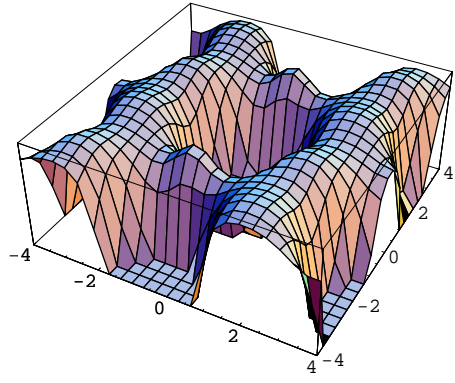,height=5cm}}
\caption{(Color online) The case of three asymmetric minima. 3D plot of the deformed potential $-{\cal V}(\phi,\bar\phi)$ for $\delta=1+i$ near a point of the set ${\wt\Gamma}_3$.}
\end{figure}

\begin{figure}[htbp]
\centerline{\epsfig{file=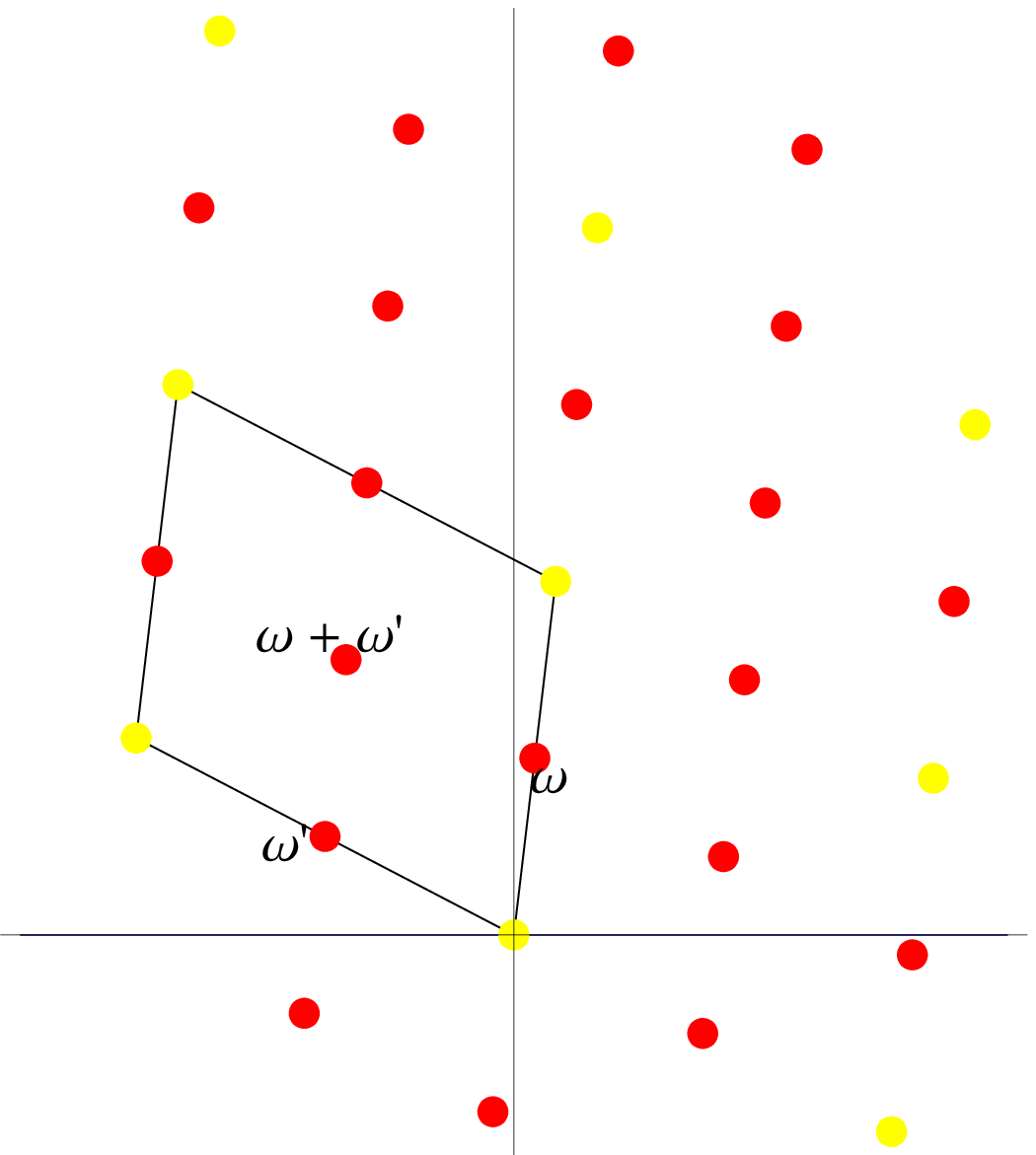,height=5cm}\hspace{1cm}\epsfig{file=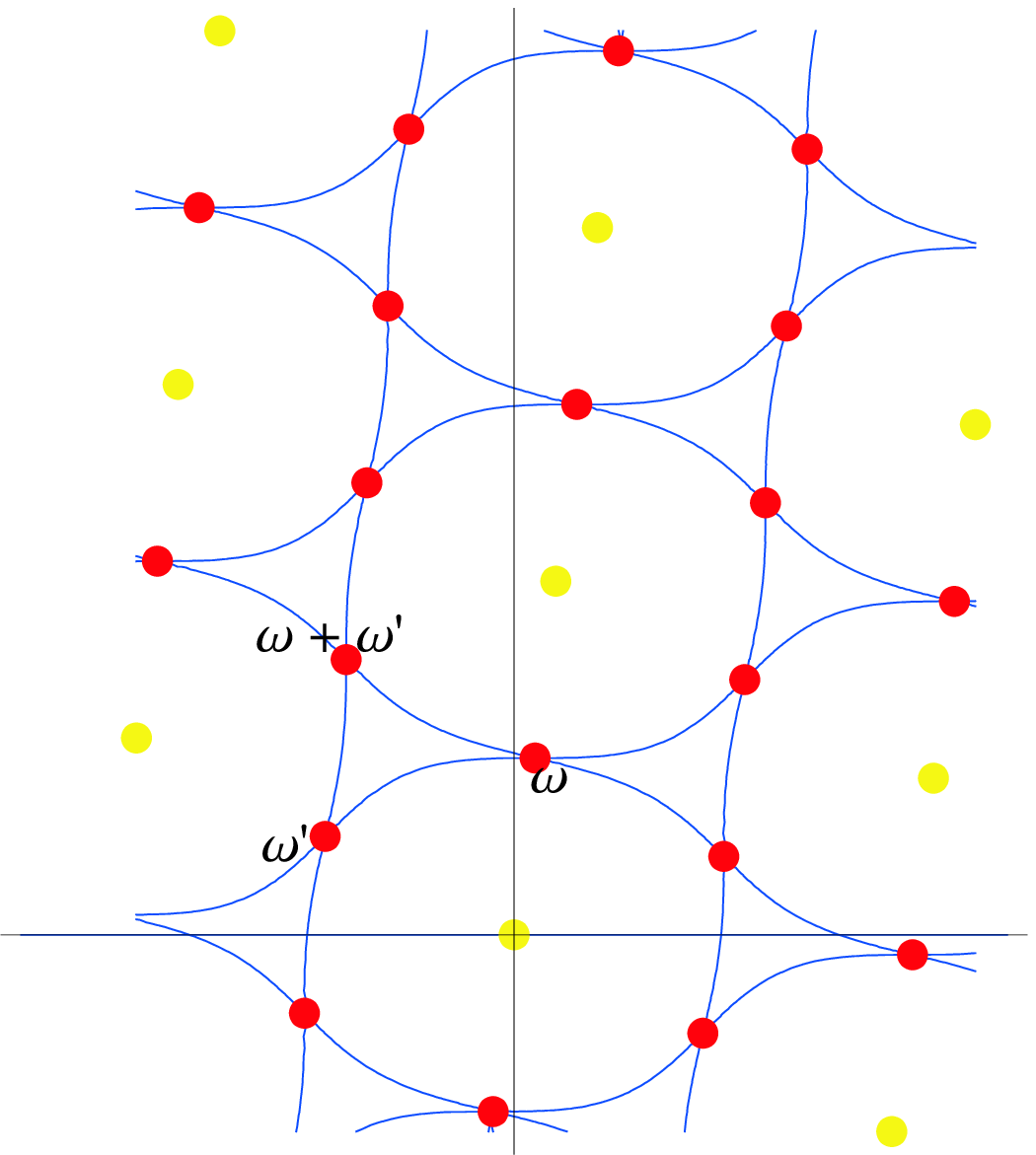,height=5cm}}
\caption{(Color online) The case of three asymmetric minima. Plots of the zeros (red) of the potential and points (yellow) of the set ${\wt\Gamma}_3$, and the networks of kink orbits (blue) connecting the zeros for $\delta=1+i$ (right panel).}
\end{figure}

\subsection{Irregular network of {\rm sn}-kink orbits}

We consider now the superpotential
\be
W(\chi)\!\!=\!-\delta\epsilon\chi\!+\frac12(\delta\!+\!\epsilon)\chi^2\!+\frac13(\delta\epsilon\!-\!1)\chi^3\!
-\frac14(\delta\!+\!\epsilon)\chi^4\!+\frac15\chi^5
\ee
where $\delta$ and $\epsilon$ are two complex coupling parameters which control the model. With this polynomial of fifth-order $W$ we get the potential
\be
V(\chi,\bar\chi)=\frac12 |1+\chi|^2 |1-\chi|^2 |\epsilon-\chi|^2 |\delta-\chi|^2
\ee
displaying the four minima $v_1=-1,\, v_2=\epsilon,\, v_3=1,$ and $v_4=\delta$. We notice that two of the
minima are again fixed at the values $\pm 1$, but this does not restrict generality of the procedure for the same reasons as before.

Again, we choose the deformation $f(\phi)={\cal W}(\phi)$ determined from the specially simple case
\be
f^\prime(\phi)^2=(f+1)(f-1)(f-\delta)(f-\epsilon)
\ee
This equation can be written as the elliptic sine equation $y^\prime=\sqrt{(1-y^2)(1-k^2y^2)}$ if we follow the usual procedure
\cite{ww}, in which we write the product $(f^2-1)(f-\delta)(f-\epsilon)$ as the product of the two factors $\left(A_+(f-\alpha_-)^2+A_-(f-\alpha_+)^2\right)$
and $\left(B_+(f-\alpha_-)^2+B_-(f-\alpha_+)^2\right),$ where
\be
A_\pm=\frac12\left(1\pm\frac{1+\delta\epsilon}{\sqrt{(1-\delta^2)(1-\epsilon^2)}}\right)\,
;\;\;\;
B_\pm=\frac12\left(1\pm\frac12\frac{2-\delta^2-\epsilon^2}{\sqrt{(1-\delta^2)(1-\epsilon^2)}}\right)
\, ; \;\;\;
\alpha_\pm=\frac{1+\delta\epsilon\pm\sqrt{(1-\delta^2)(1-\epsilon^2)}}{\delta+\epsilon}
\ee
We then make the homographic substitution $p=(f-\alpha_-)/(f-\alpha_+)$ to obtain
\be
\frac{dy}{d\phi}=\sqrt{(1-y^2)(1-k^2y^2)} \;\;\; \Rightarrow\;\;\;
y(\phi)={\rm sn}(\phi,k^2)
\ee
where $y=\sqrt{-A_+/A_-}\,p$, $\psi=(\alpha_--\alpha_+)\sqrt{-A_+B_-}\,\phi$, and $k^2=A_-B_+/A_+B_-$. Therefore, we finally find that
\be
f(\phi)=\frac{\alpha_+-\alpha_-\sqrt{-A_-/A_+}\;{\rm
sn}(\psi,k^2)}{1-\sqrt{-A_-/A_+}\;{\rm sn}(\psi,k^2)}
\ee
is the new superpotential from which we can construct the potential of the modified model in the form
\be
{\cal V}(\phi,\bar\phi)=\frac12\biggl|\frac{(\alpha_+-\alpha_-)^2\sqrt{A_-B_-}{\rm cn}(\psi,k^2)\dn(\psi,k^2)}{(\sqrt{-A_-/A_+}\sn(\psi,k^2)-1)^2}\biggr|
\ee

The above investigation is very general, and the presence of the two complex parameters $\delta$ and $\epsilon$ make the illustrations
awkward. For this reason, we shall restrict ourselves to the simpler case where $\epsilon=-\delta$ and only a single complex parameter is
free. Thus,
\be
W(\chi)=\delta^2\chi-\frac13(1+\delta^2)\chi^3+\frac15 \chi^5 \,
; \;\;\;\;\; V(\chi,\bar{\chi})=\frac12|1-\chi^2|^2|\delta^2-\chi^2|^2
\ee
are respectively the superpotential and potential of the original model. The four minima of $V$ are at the points
$v_1=-v_3=-1$, $v_2=-v_4=-\delta$ in the $\chi$-complex plane.

From the values of the superpotential at the minima
\be
W_\alpha(\pm
1)=\pm{\frac23}(\delta^2-{\frac15})e^{-i\alpha} \,,
\;\;\;\;\; W_\alpha(\pm\delta)=\pm{\frac23}\delta^3(1-\frac{\delta^2}{5})e^{-i\alpha}
\ee
we obtain
\bes\ben
W_\alpha(1)-W_\alpha(-1)&=&{\frac43}(\delta^2-{\frac15})e^{-i\alpha} \, ; \;\;\;\;\; W_\alpha(\pm
1)-W_\alpha(\pm\delta)=\pm{\frac23}\left(\frac{\delta^5-1}{5}-\delta^2(\delta-1)\right)e^{-i\alpha}
\\ W_\alpha(\delta)-W_\alpha(-\delta)&=&{\frac43}\delta^3(1-\frac{\delta^2}{5})e^{-i\alpha} \, ; \;\;\;\;\;
W_\alpha(\pm
1)-W_\alpha(\mp\delta)=\mp{\frac23}\left(\frac{\delta^5+1}{5}-\delta^2(\delta+1)\right)e^{-i\alpha}
\een \ees
The angles of the kink orbits are (mod $\pi$)
\bes \ben
\alpha^{(13)}&=&{\rm arctan}\left[\frac{{\rm Im}(1-5\delta^2)}{{\rm
Re}(1-5\delta^2)}\right] \, ; \;\;\;\;\;
\alpha^{(12)}=\alpha^{(34)}={\rm arctan}\left[\frac{{\rm
Im}(\delta^2-1-5\delta^2(\delta-1))}{{\rm Re}(\delta^2-1-5\delta^2(\delta-1))}\right]\\
\alpha^{(24)}&=&{\rm arctan}\left[\frac{{\rm
Im}(\delta^3(\delta^2-5))}{{\rm Re}(\delta^3(\delta^2-5))}\right]
\, ; \;\;\;\;\; \alpha^{(23)}=\alpha^{(14)}={\rm
arctan}\left[\frac{{\rm Im}(\delta^2+1-5\delta^2(\delta+1))}{{\rm
Re}(\delta^2+1-5\delta^2(\delta+1))}\right]
\een\ees

The corresponding energies are given by
\bes\ben
M(13)&=&\frac43|1/5-\delta^2|\, ;\;\;\;\;\;
M(12)=M(34)=\frac23|1/5+(\delta-1)\delta^2-\delta^5/5|
\\
M(24)&=&\frac43|(\delta^2/5-1)\delta^3|\, ;\;\;\;\;\;
M(23)=M(41)=\frac23|1/5-(\delta+1)\delta^2+\delta^5/5| 
\een\ees
plus $M(kj)=M(jk)$. Note that for $\delta=i,$ $M(13)=M(24)={\frac85}$, $M(12)=M(34)=M(23)=M(14)={\frac45}\sqrt{2}$ and we recover the case
of a regular square.

We now deform the model, choosing $f(\phi)={\cal W}(\phi)$. As before, we consider the specially simple case, which gives
\be
\label{elj}
f^\prime(\phi)^2=(1+f)(1-f)(\delta+f)(\delta-f)=\delta^2-(1+\delta^2)f(\phi)^2+f(\phi)^4
\ee
We compare this with the elliptic Jacobi sine equation to find $f(\phi)=\delta\; {\rm sn}(\phi,\delta^2)$ as the solution of equation
(\ref{elj}). Thus, the deformed superpotential and potential are given by
\be
{\cal W}_\gamma(\phi)=\delta\,{\rm
sn}(\phi,\delta^2)e^{-i\gamma}\,; \;\;\;\;\; {\cal
V}(\phi,\bar\phi)=\frac12 |\delta|^2 \,|{\rm
cn(\phi,\delta^2)}|^2 |{\rm dn}(\phi,\delta^2)|^2
\ee
The zeros of the potential form the lattice of vacua at the constant values of the field 
\bes\ben
\phi^{(1)}_{(m,n)}&=&-\frac14(\omega_1+\omega_2)+\Omega_{mn}\,;\;\;\;\;\;
\phi^{(2)}_{(m,n)}=-\frac14{\omega_1}+\Omega_{mn}
\\
\phi^{(3)}_{(m,n)}&=&\frac14(\omega_1+\omega_2)+\Omega_{mn}\,;\;\;\;\;\;
\;\;\phi^{(4)}_{(m,n)}=\frac14{\omega_1}+\Omega_{mn}
\een\ees
where the periodicity is determined by the quarter periods of the Jacobi elliptic sine
\be
\Omega_{mn}=n\omega_1+\frac12m\omega_2 \, , \;\;\;\;\;
\omega_1=4K(\delta^2) \, , \;\;\;\;\; \omega_2=4iK(1-\delta^2)
\ee
From the values of the superpotential at the minima
\be
{\cal W}_\gamma(\phi^{(3)}_{(m,n)})=e^{-i\gamma}=-{\cal
W}_\gamma(\phi^{(1)}_{(m,n)})\,; \;\;\;\;\; {\cal
W}_\gamma(\phi^{(4)}_{(m,n)})=\delta\,e^{-i\gamma}=-{\cal
W}_\gamma(\phi^{(2)}_{(m,n)})\ee we derive \bes \ben {\cal
W}_\gamma(\phi^{(3)}_{(m,n)})-{\cal
W}_\gamma(\phi^{(1)}_{(m',n')})&=&2\,e^{-i\gamma}\, ;\;\;\;\;\;
{\cal W}_\gamma(\phi^{(4)}_{(m,n)})-{\cal
W}_\gamma(\phi^{(2)}_{(m',n')})=2\delta\,e^{-i\gamma}
\\ {\cal W}_\gamma(\phi^{(2)}_{(m,n)})-{\cal
W}_\gamma(\phi^{(1)}_{(m',n')})&=&(1-\delta)e^{-i\gamma}={\cal
W}_\gamma(\phi^{(3)}_{(m,n)})-{\cal W}_\gamma(\phi^{(4)}_{(m',n')})
\\
{\cal W}_\gamma(\phi^{(3)}_{(m,n)})-{\cal
W}_\gamma(\phi^{(2)}_{(m',n')})&=&(1+\delta)e^{-i\gamma}={\cal
W}_\gamma(\phi^{(4)}_{(m,n)})-{\cal
W}_\gamma(\phi^{(1)}_{(m',n')})
\een\ees
The orbit angles and the energies of the deformed kinks are given by
\bes\ben
\gamma^{(13)}&=&0 \, ; \;\;\;{\cal M}(13)=2 \,
;\;\;\; \gamma^{(12)}=\gamma^{(34)}={\rm arctan}\left[\frac{{\rm
Im}(1-\delta)}{{\rm Re}(1-\delta)}\right] \, ; \;\;\; {\cal
M}(12)={\cal M}(34)=|1-\delta| \nonumber
\\
\gamma^{(24)}&=&{\rm arctan}\left[\frac{{\rm Im}\delta}{{\rm
Re}\delta}\right] \,; \;\;\; {\cal M}(24)=2|\delta|\,;\;\;\;
\gamma^{(23)}=\gamma^{(14)}={\rm arctan}\left[\frac{{\rm
Im}(1+\delta)}{{\rm Re}(1+\delta)}\right] \, ; \;\;\; {\cal
M}({41})={\cal M}(23)=|1+\delta| \nonumber
\een\ees
with $\gamma^{(kj)}=\gamma^{(jk)}+\pi$ and ${\cal M}(kj)={\cal M}(jk).$
The vacua connected by the kink orbits are organized according to the following sequences
\bes\ben
\phi^{(1)}_{(m,n)}\;\;\;
&\Leftrightarrow& \;\;\; \phi^{(3)}_{(m,n)} \;\;\; \Leftrightarrow
\;\;\; \phi^{(1)}_{(m+2,n+1)}\;\;\; \Leftrightarrow \;\;\;
\phi^{(3)}_{(m+2,n+1)} \nonumber\\
\phi^{(2)}_{(m,n)}\;\;\; &\Leftrightarrow& \;\;\; \phi^{(4)}_{(m,n)} \;\;\; \Leftrightarrow
\;\;\; \phi^{(2)}_{(m,n+1)}
\;\;\;\;\;\; \Leftrightarrow \;\;\; \phi^{(4)}_{(m,n+1)}\nonumber\\
\phi^{(1)}_{(m,n)}\;\;\; &\Leftrightarrow& \;\;\; \phi^{(2)}_{(m,n)}
\;\;\; \Leftrightarrow \;\;\; \phi^{(1)}_{(m+1,n)} \;\;\;\;\;\;
\Leftrightarrow \;\;\; \phi^{(2)}_{(m+1,n)} \nonumber\\
\phi^{(3)}_{(m,n)}\;\;\; &\Leftrightarrow& \;\;\; \phi^{(4)}_{(m,n)}
\;\;\; \Leftrightarrow \;\;\; \phi^{(3)}_{(m+1,n)}
\;\;\;\;\;\;\Leftrightarrow \;\;\; \phi^{(4)}_{(m+1,n)}\nonumber\\
\phi^{(1)}_{(m,n)}\;\;\; &\Leftrightarrow& \;\;\; \phi^{(4)}_{(m,n)}
\;\;\; \Leftrightarrow \;\;\; \phi^{(1)}_{(m+1,n+1)}\;\;\;
\Leftrightarrow \;\;\; \phi^{(4)}_{(m+1,n+1)}\nonumber\\
\phi^{(2)}_{(m,n)}\;\;\; &\Leftrightarrow& \;\;\; \phi^{(3)}_{(m,n)}
\;\;\; \Leftrightarrow \;\;\; \phi^{(2)}_{(m+1,n+1)}\;\;\;
\Leftrightarrow \;\;\; \phi^{(3)}_{(m+1,n+1)}\nonumber
\een\ees

To illustrate the investigations, in Fig.~9 and 10 we plot the potential, minima, points in the set ${\wt\Gamma}$, and orbits of the kinklike configurations.

\begin{figure}[htbp]
\centerline{\epsfig{file=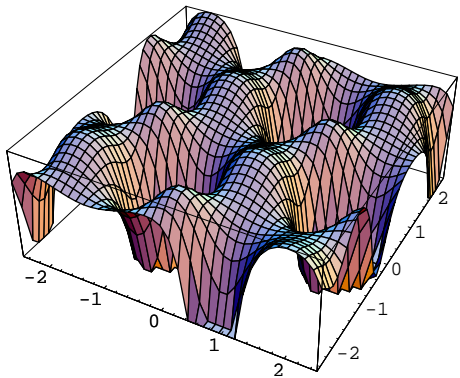,height=5cm}}
\caption{(Color online) The case of four asymmetric minima. 3D plot of the deformed potential $-{\cal V}(\phi,\bar\phi)$ near four points of the set ${\wt\Gamma}_4$ for $\delta=1+i$.}
\end{figure}

\begin{figure}[htbp]
\centerline{\epsfig{file=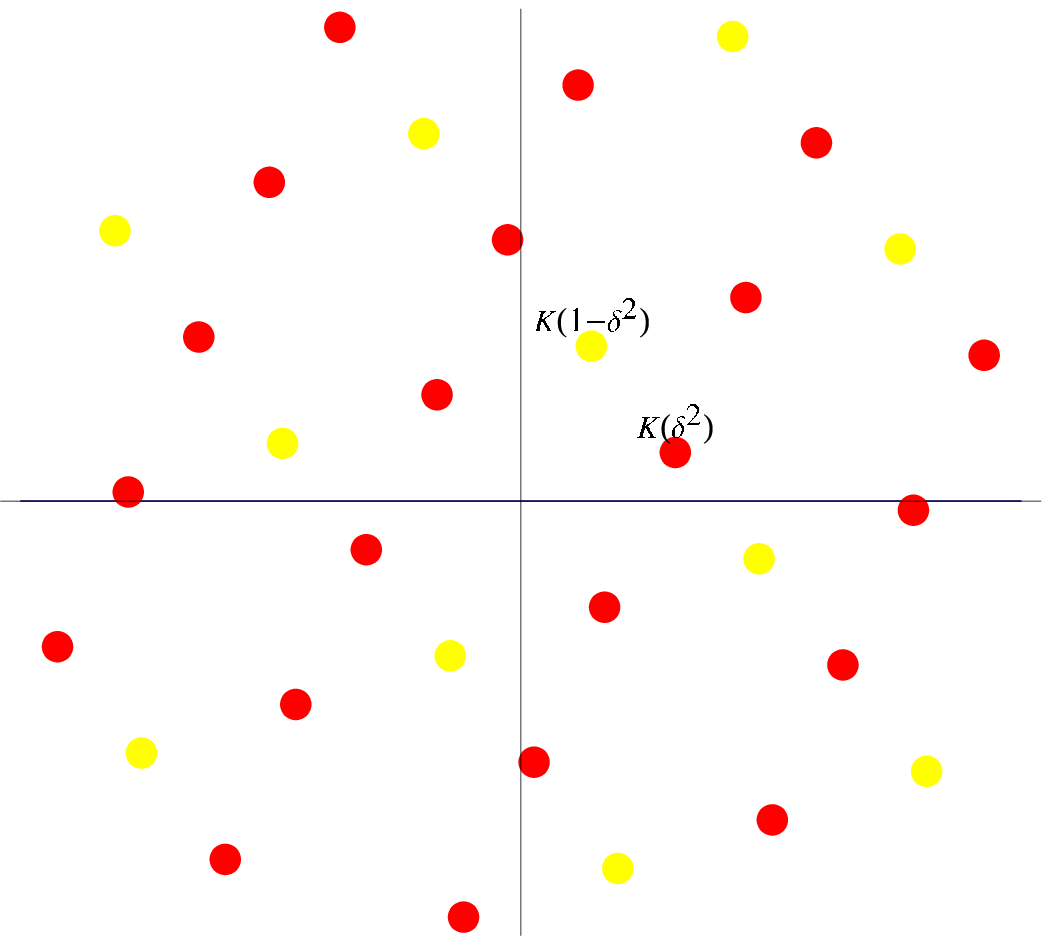,height=5cm}\hspace{1cm}\epsfig{file=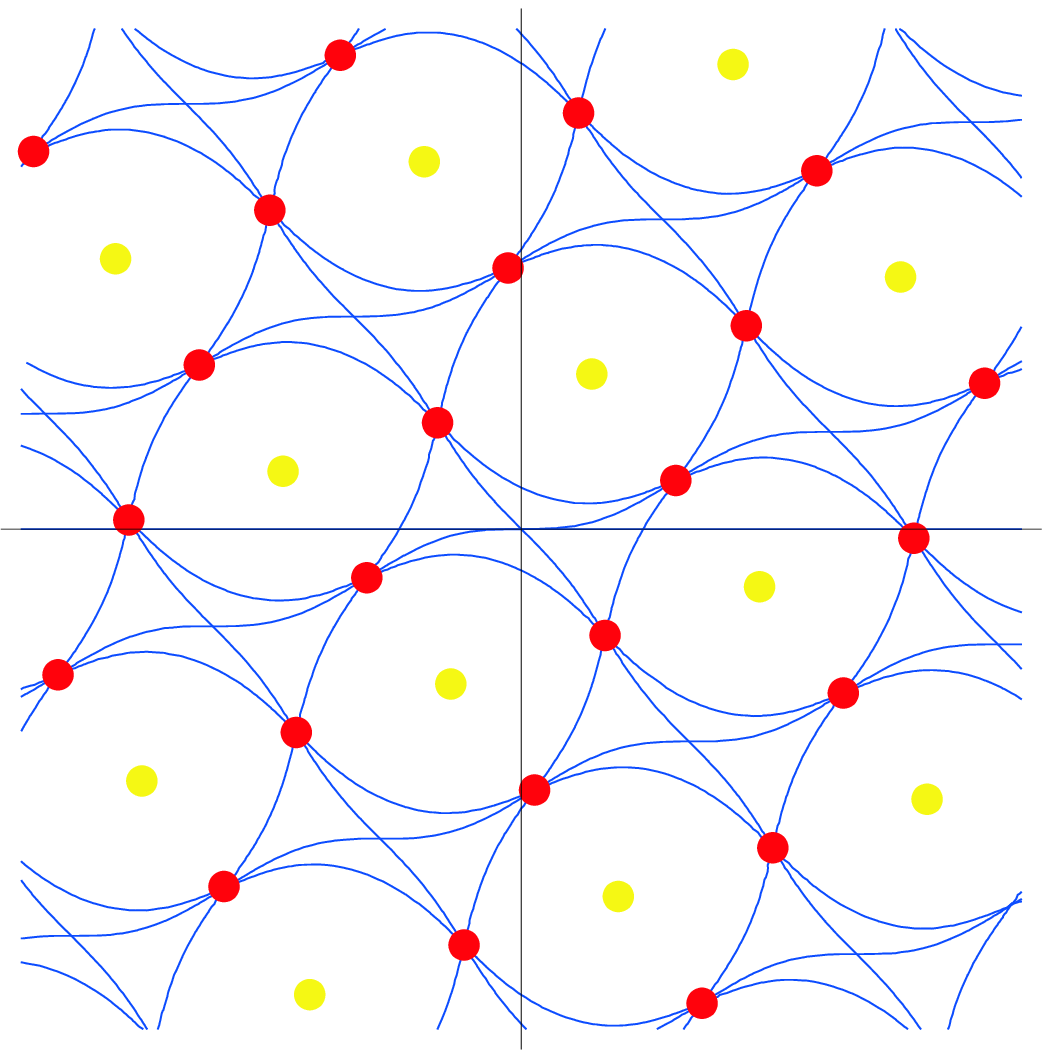,height=5cm}}
\caption{(Color online) The case of four asymmetric minima. Plots of the zeros (red) of the potential and points (yellow) of the set ${\wt\Gamma}_4$, and networks of kink orbits (blue) connecting the zeros for $\delta=1+i$ (right panel).}
\end{figure}

\section{Bifurcation}
\label{sec:bifur}

All the models which we have been studying so far present an interesting feature, which we now explore. It concerns the fact that
they have three or four minima. Thus, if we choose two minima arbitrarily, it may be possible that they are connected with two or
more distinct orbits. When this happens to be the case, we say that the system develops a bifurcation, since one can go from a given
vacua to another one, following two or more distinct kink orbits. This possibility is directly related to the balance of kink energies
providing an upper bound for the fusion of two of the kinks in a single kink of a third type. Such a process is energetically
possible -- and the outgoing kink stable -- if, given e.g. three minima $k,j,l$, the energy of the $(kj)$ kink is lower than the sum of the
other two kink masses
\be
M(kj)<M(kl)+M(lj)
\ee
All the kink masses depend on the $N-2$ complex parameters, indicating the arbitrary positions of
$N-2$ vacua, since the other two minima are fixed at the $\pm1$ points in the $\chi$-complex plane. 
A bifurcation occurs when the inequality becomes equality, since we can go from $k$ to $j$
following the direct $k\to j$ path, or then visiting $l$ through the path $k\to l\to j$ with the same energetic cost. In
the case $N=3$ it is shown in \cite{AT}, in the search for intersecting domain walls, that this possibility indeed happens -- see also Ref.~{\cite{PT}}.
In the generic case $N,$ there is a sub-manifold of real dimension $2N-5$ of the parameter space characterized by the equation
$M(kj)=M(kl)+M(lj)$. We shall refer to this sub-manifold as the marginal stability variety because some irreducible component of it
is a boundary between two regions of the $(N-2)$-dimensional complex parameter space; one region where $M(kj)<M(kl)+M(lj)$ (the $(kj)$ kink
is stable and cannot decay to $(kl)$ and $(lj)$ kinks), and the other region where $M(kj)>M(kl)+M(lj)$ (the $(kj)$ kink is unstable decaying
to the $(kl)$-$(lj)$ kink combination). In fact, things are slightly more complicated and a bifurcation occurs at the marginal stability variety,
with the kink orbit going to infinity beyond this point, with $M(kj)$ becoming divergent, breaking the direct connectivity between the vacua $k$ and $j.$

Since the energy in the topological sector is controlled by $W,$ we immediately see that bifurcation appears if and only if at least
three minima gets aligned in $W$ space, that is, iff $|W(k)-W(j)|=|W(k)-W(l)|+|W(l)-W(j)|$, with $W(l)$ in between $W(k)$ and $W(j)$. This is the general condition for bifurcation, and
below we use it to investigate the models introduced in the former section.

\subsection{The case of three minima}

Here we consider the case of three minima, with the model being described by the complex parameter $\delta.$ We notice that bifurcation
cannot appear in models which engender the $Z_3$ symmetry, since the symmetry implies that $M(12)=M(23)=M(31)$, henceforth
$M(12)<M(23)+M(31)$, etc. In the Abraham-Townsend model, the marginal stability curve is
characterized by the alignment of the three minima in the $(W, {\ov W})$ plane, i.e., it is the curve in this plane for which
$\alpha^{(12)}(\delta)=\alpha^{(23)}(\delta)=\alpha^{(13)}(\delta)\,\,\, {\rm mod}\, \pi$, or,
\be
\frac{{\rm Im}
( (\delta - 3)(\delta + 1)^3 )}{{\rm Re} ((\delta - 3)(\delta +1)^3)}
=\frac{{\rm Im}\delta}{{\rm Re}\delta}=\frac{{\rm Im}((\delta + 3)(\delta - 1)^3 )}{{\rm Re}((\delta + 3)(\delta - 1)^3 )}
\ee
These identities hold if
\be
(\delta-\bar\delta)(3+\delta\bar\delta(\delta^2+\bar\delta^2+\delta\bar\delta-6))=0
\ee
which is the algebraic equation that characterizes the marginal stability curve. We note that the curve has two irreducible components.

One component is the real axis in the $\delta$-plane: $\delta_2=0.$ For real values of the parameter it happens that all the three
minima are aligned in the real axis of the complex $\chi$-plane, and the system cannot develop bifurcation. In the two
very special cases with $\delta_1=\pm1,$ we have only two minima, and so only one kink orbit remains in the system. For $\delta_1<-1,$
there exist only two kink orbits, which we label $(31)$ and $12$, because the vacuum $v_1$ sits on the real axis in between $v_3$ and
$v_2$. For $-1<\delta_1<1,$ there exist the two kink orbits $(13)$ and $(32)$, and for $\delta_1>1,$ there exist the two kink orbits
$(12)$ and $(23)$ for similar reasons.

Things are more interesting for the other algebraically irreducible component. The quartic curve
\be
3-6\delta_1^2-6\delta_2^2+2\delta_1^2\delta_2^2+3\delta_1^4-\delta_2^4=
(\delta_2^2+3-\delta_1^2+2\sqrt{\delta_1^4-3\delta_1^2+3}\,)
(\delta_2^2+3-\delta_1^2-2\sqrt{\delta_1^4-3\delta_1^2+3}\,)=0
\ee
which allow for each one of the three possibilities, $M(12)=M(23)+M(31)$, or $M(23)=M(31)+M(12)$, or yet
$M(31)=M(12)+M(23)$, depending on the specific value of the complex parameter $\delta$, is the true boundary between regions in the
$\delta$ plane where there exist three or two kinks and the phenomenon of bifurcation takes place. This case is fully studied in
\cite{AT}, and below in Fig.~11 we plot the curves of marginal stability in the complex $\delta$ plane.

\begin{figure}[htbp]
\centerline{\epsfig{file=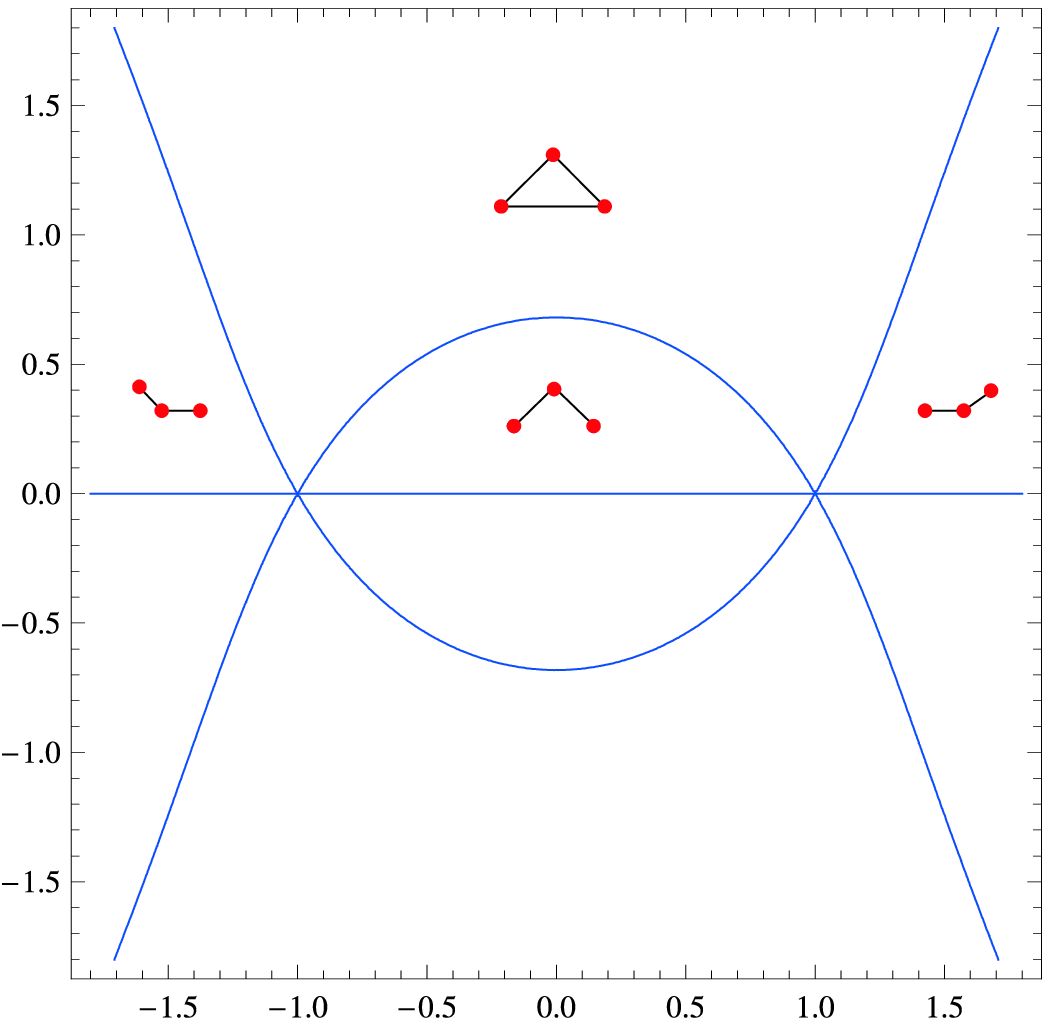,height=6cm}}
\caption{(Color online) The curves of marginal stability in the case of three minima. The set of (red) points illustrates the positions
of the minima and the kink orbits for distinct values of the parameter $\delta$ in the related regions.}
\end{figure}

\begin{figure}[htbp]
\begin{center}
\begin{tabular}{|c|c|c|c|}  \hline & & &  \\
& $\delta$ & Original kinks  & Deformed kinks
\\ & & &
\\ \hline {\small $\delta=0+i$}&
\parbox{2.5cm}{\epsfig{file=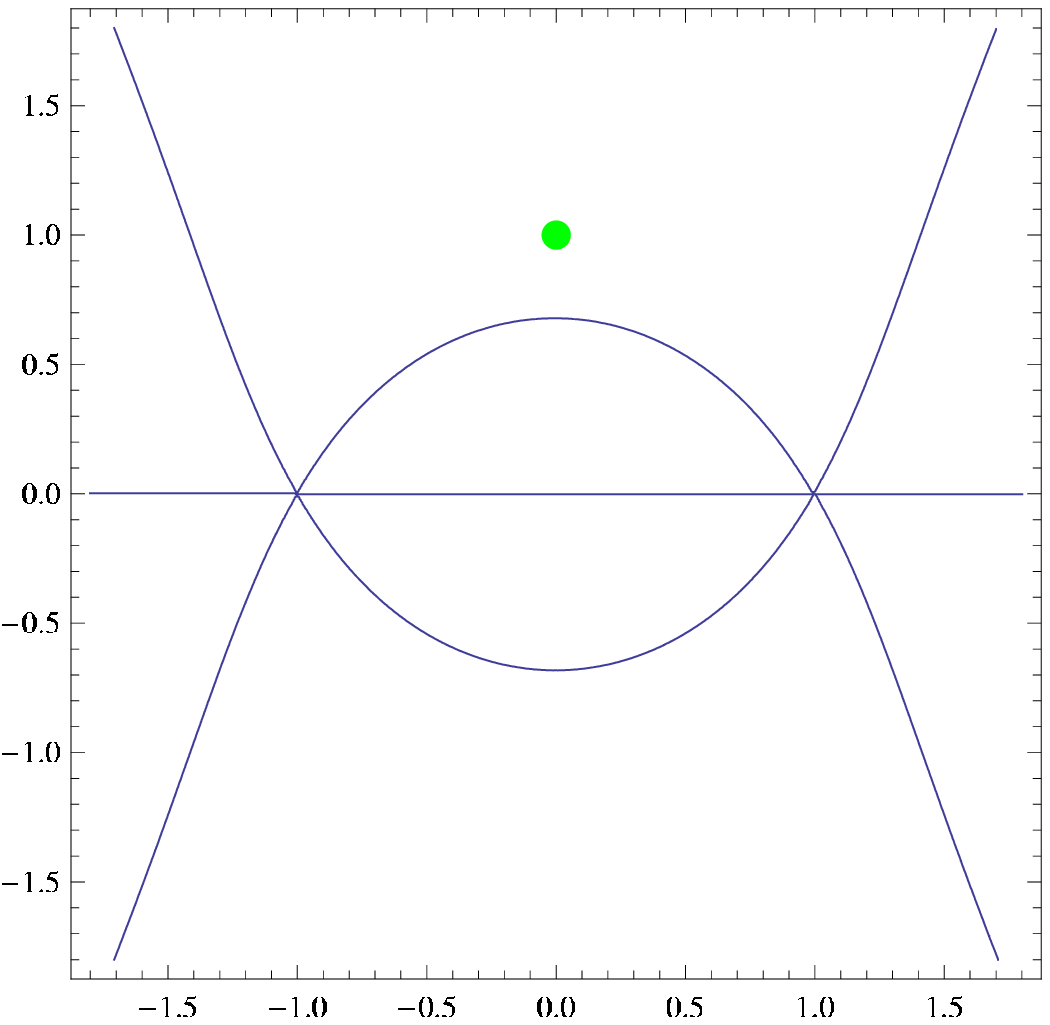,width=2.5cm}}&
\parbox{3.5cm}{\epsfig{file=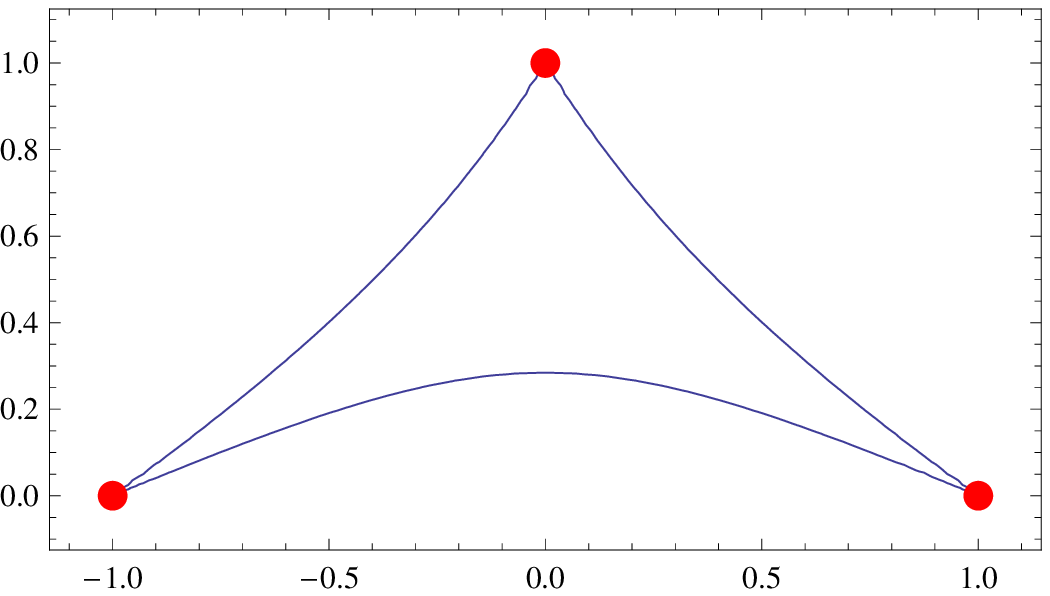,width=3.5cm}} & \parbox{3.5cm}{\epsfig{file=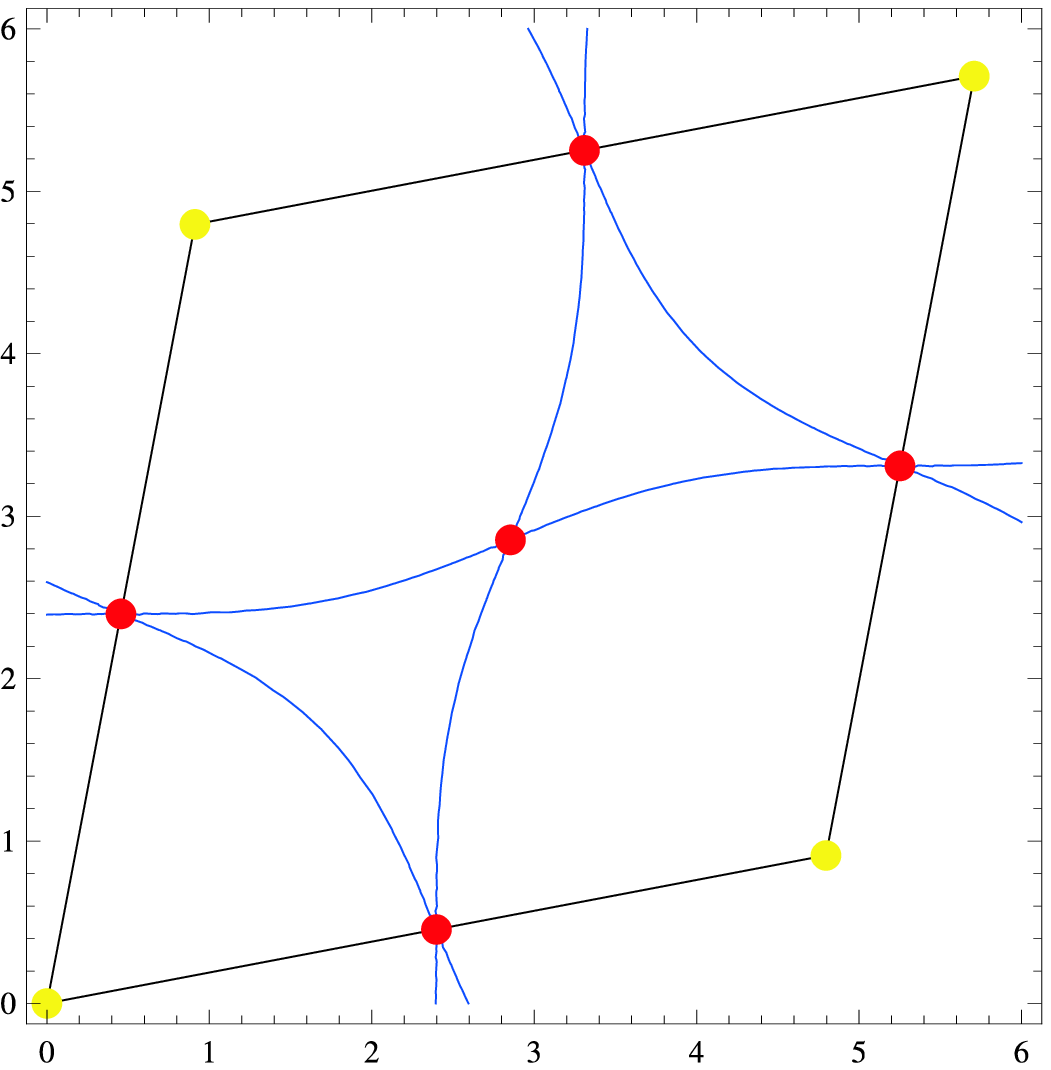,width=3.5cm}}
 \\ & & & \\
  \hline {\small $\delta=0+\frac{i}{2}$} & \parbox{2.5cm}{\epsfig{file=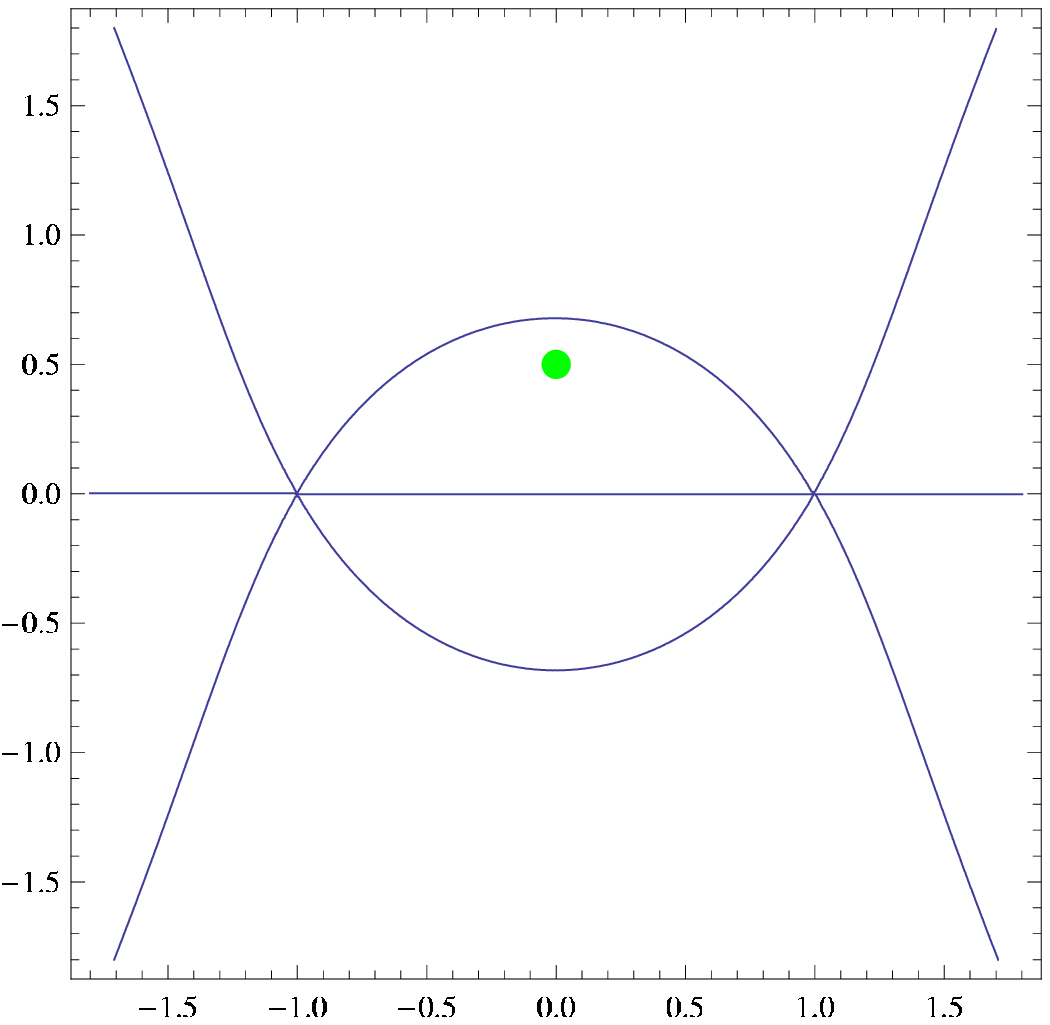,width=2.5cm}}&
\parbox{3.5cm}{\epsfig{file=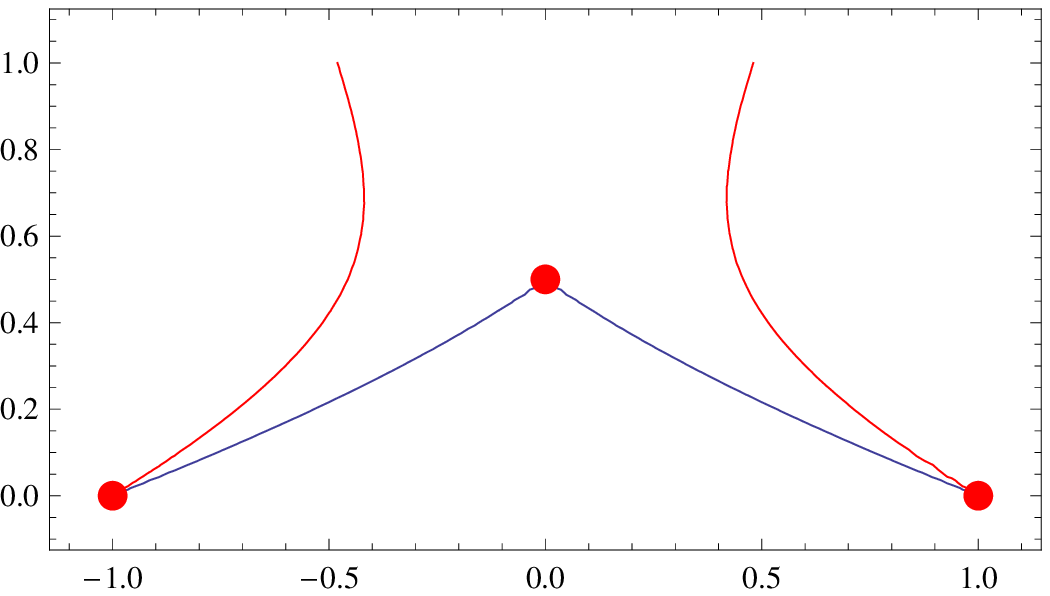,width=3.5cm}} &
\parbox{3.5cm}{\epsfig{file=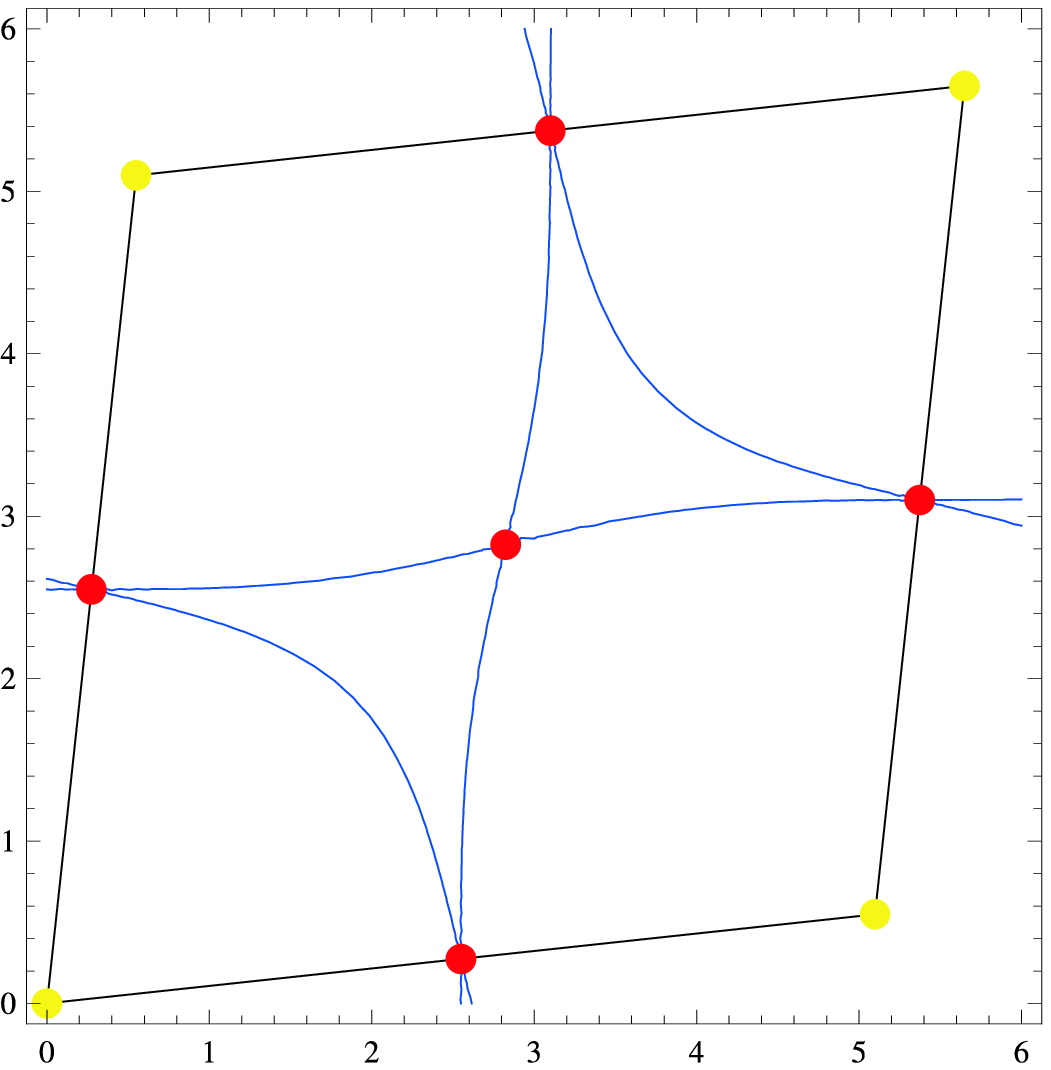,width=3.5cm}}
  \\ & & & \\
\hline  {\small $\delta=\frac{3}{2}+\frac{i}{2}$}&
\parbox{2.5cm}{\epsfig{file=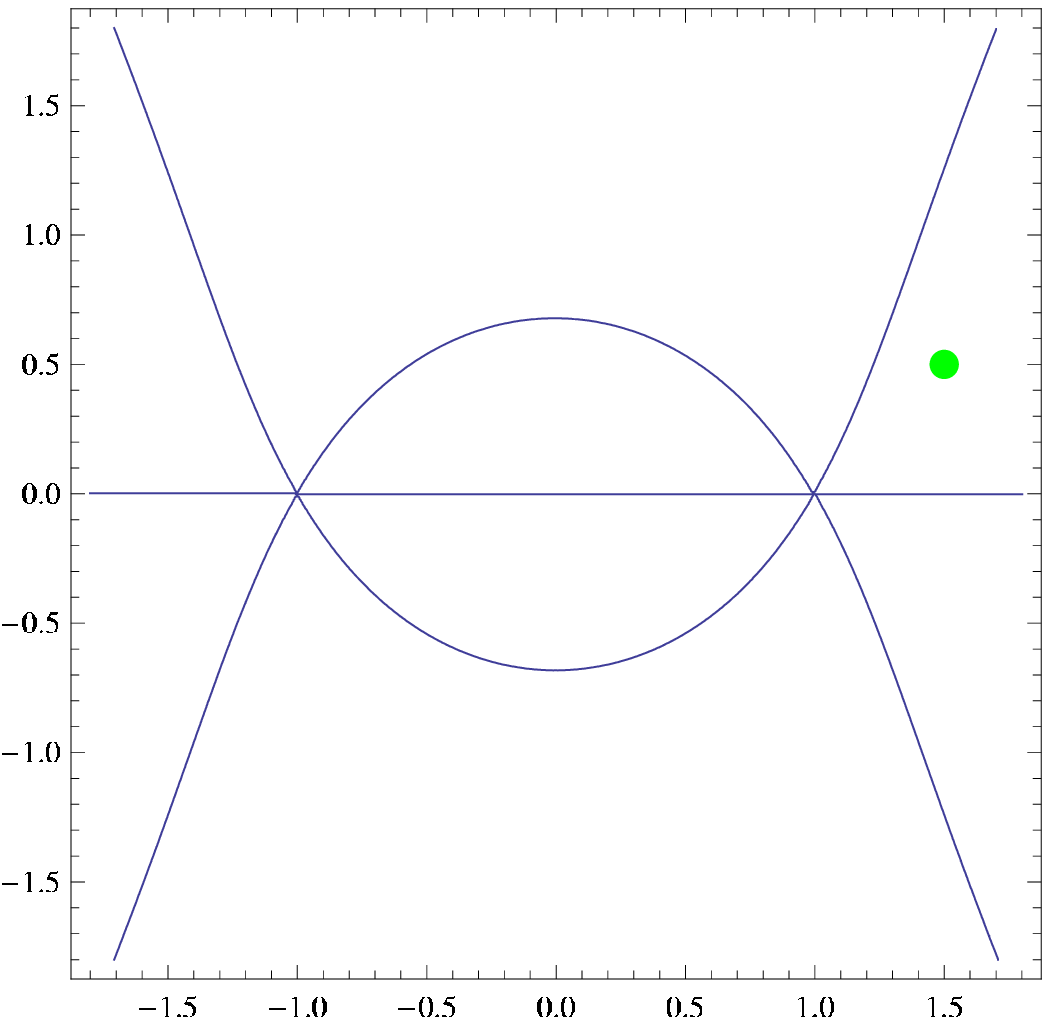,width=2.5cm}}&
\parbox{3.5cm}{\epsfig{file=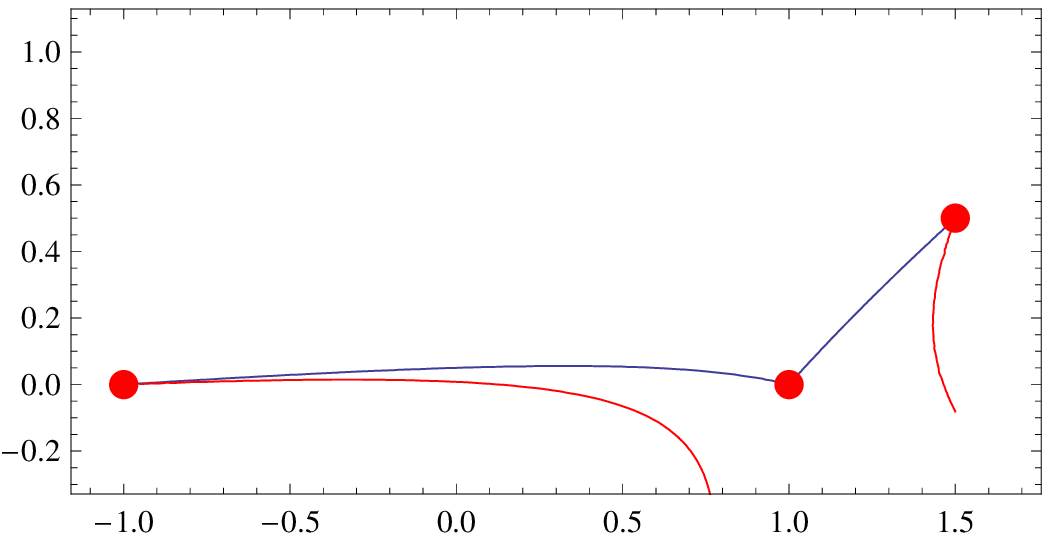,width=3.5cm}} &
\parbox{3.5cm} {\epsfig{file=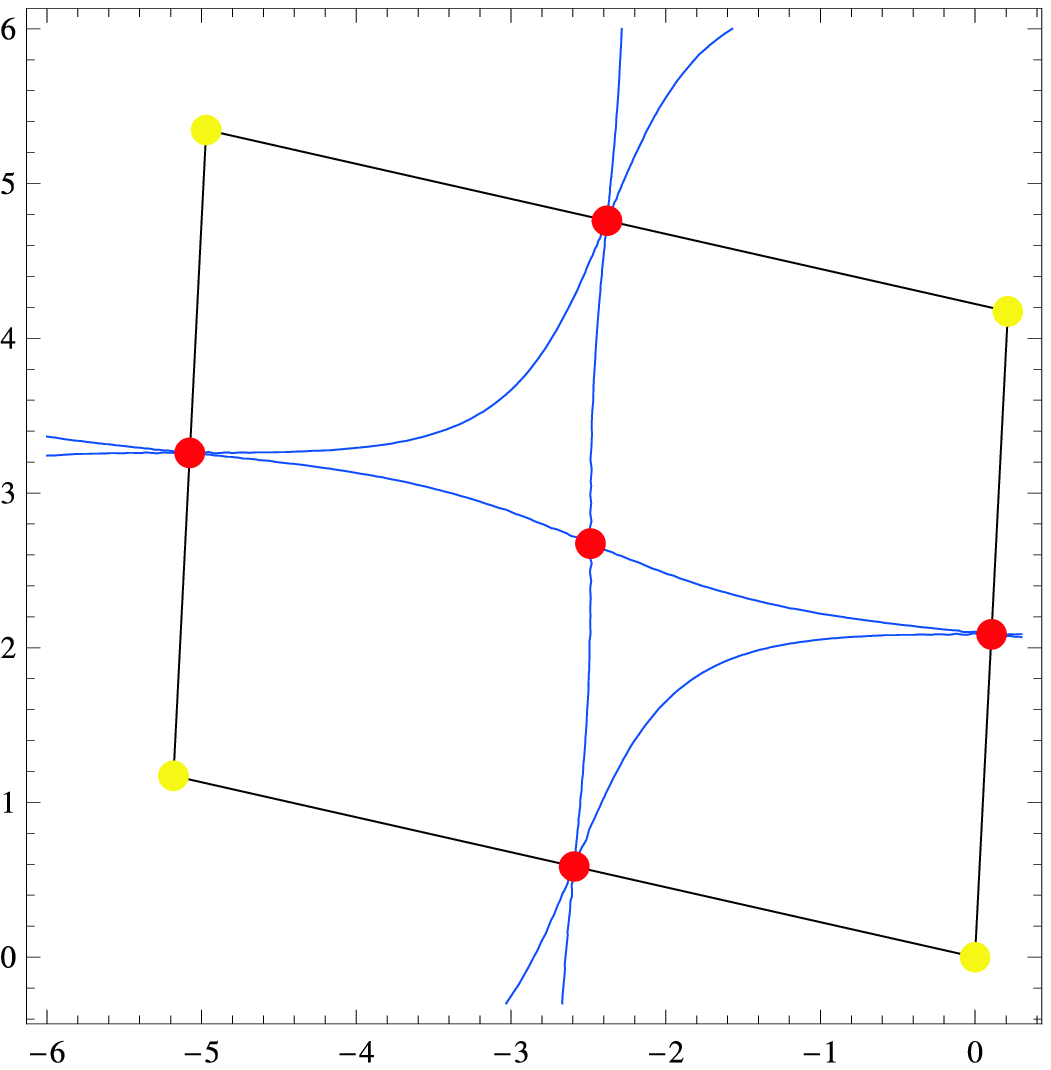,width=3.5cm}}  \\ & & & \\
 \hline
 \end{tabular}
\end{center}
\caption{(Color online) The bifurcation curves and some related illustrations with (green) points in the first column representing distinct values of the complex parameter $\delta$ in the case of three minima. The broken (red) lines in the second column indicate how the orbit goes to infinity, leading to divergent contributions which should be discarded.}
\end{figure}

The condition for alignment in the deformed model in the ${\cal W}$ plane $\gamma^{(12)}=\gamma^{(13)}=\gamma^{(23)}$ is
\be
0=\frac{{\rm Im}\delta}{{\rm Re}(\delta+1)}= \frac{{\rm Im}\delta}{{\rm Re}(\delta-1)}
\ee 
There is only one component, $\delta_2=0,$ but we have already seen that in this case there is no
bifurcation anymore. In fact on the real axis the kink masses are ${\cal M}(12)=2$, ${\cal M}(13)=|\delta_1+1|$, and  ${\cal
M}(13)=|\delta_1-1|$, such that: ${\cal M}(13)={\cal M}(12)+{\cal M}(23)$ if $\delta_1>1$, ${\cal M}(12)={\cal M}(13)+{\cal M}(23)$ if
$-1<\delta_1<1$, and ${\cal M}(23)={\cal M}(12)+{\cal M}(13)$ if $\delta_1<1$. If $\delta_2\neq 0$ the mass of one of the deformed
kinks is always lighter than the sum of the masses of the other two kinks because of the triangle inequality, the sum of the lengths of two
sides of a triangle is greater than the length of the third side. Thus, we notice that the deformation procedure wash out the marginal
stability curve, leaving no room for bifurcation in the deformed models which we are dealing with in this work. In Fig.~12 we illustrate
the $N=3$ case with several distinct possibilities for the complex parameter $\delta$. 

\subsection{The case of four minima}

Let us now consider the case of four minima, with the model being driven by the two complex parameters $\delta$ and $\epsilon$. The general
case is very complicate, so we move to the simpler case in which we use $\epsilon=-\delta,$ leading to a single complex parameter $\delta$.

The condition for alignment of all the minima in the $W$ plane now is $\gamma^{(12)}=\gamma^{(13)}=\gamma^{(23)}=\gamma^{(24)},\;
{\rm mod}\, \pi $ (note that because the vacua in the pairs $(v_1,v_3)=(1,-1)$ and $(v_2,v_4)=(-\delta,\delta)$ are always
aligned in the $W$-plane, there cannot be alignments of three non vanishing vacua). The identities between kink angles hold if
\be \frac{{\rm
Im}[1-5\delta^2+\delta^3(5-\delta^2)]}{{\rm
Re}[1-5\delta^2+\delta^3(5-\delta^2)]}= \frac{{\rm
Im}[\delta^3(5-\delta^2)]}{{\rm Re}[\delta^3(5-\delta^2)]}=
\frac{{\rm Im}[1-5\delta^2]}{{\rm Re}[1-5\delta^2]}=\frac{{\rm
Im}[1-5\delta^2-\delta^3(5-\delta^2)]}{{\rm
Re}[1-5\delta^2-\delta^3(5-\delta^2)]}
\ee
which is satisfied if
\be
(\delta-\bar\delta)\left(5(\delta^2+\bar\delta^2)-\delta^4-\bar\delta^4+
\delta\bar\delta\left(5(\delta^2+\bar\delta^2+\delta\bar\delta)-26\right)\right)=0\ee

One irreducible component is again the abscissa axis $\delta_2=0.$ The system does not support bifurcation since all the minima are now
in the real axis of the complex $\chi$-plane. There are three special values: $\delta_1=\pm1$ and $\delta_1=0;$ for
$\delta_1=\pm1,$ there exist only two minima and one kink orbit (12) and for $\delta_1=0$ there exist three minima, with the two distinct
kink orbits $(12)$ and $(23).$ Other possible values are: for $\delta_1<-1,$ there exist four minima, but only the three kink
orbits $(21)$, $(13)$, and $(34);$ for $-1<\delta_1<0,$ there exist four minima, but only the three kink orbits $(12)$, $(24)$, and
$(41);$ for $0<\delta_1<1,$ there exist four minima, but only the three kink orbits $(14)$, $(42)$, and $(23);$ for $\delta_1>1,$
there exist four minima, but only the three kink orbits $(41)$, $(13)$, and $(32)$.

The other algebraically irreducible component is given by
\be
15\delta_1^2-5\delta_2^2-30\delta_1^4-26\delta_2^4-40\delta_1^2\delta_2^2+15\delta_1^6
-5\delta_2^6+25\delta_1^4\delta_2^2+5\delta_1^2\delta_2^4=0
\ee

In Fig.~13 we plot the marginal stability curves which follow from the above expression, and we illustrate how the minima behave
in the complex $\delta$ plane. The two minima $\pm1$ and any other pair of asymmetrically positioned minima on the curves introduces
two distinct kink orbit possibilities, and so it illustrates the bifurcation phenomenon once again.

\begin{figure}[htbp]
\centerline{\epsfig{file=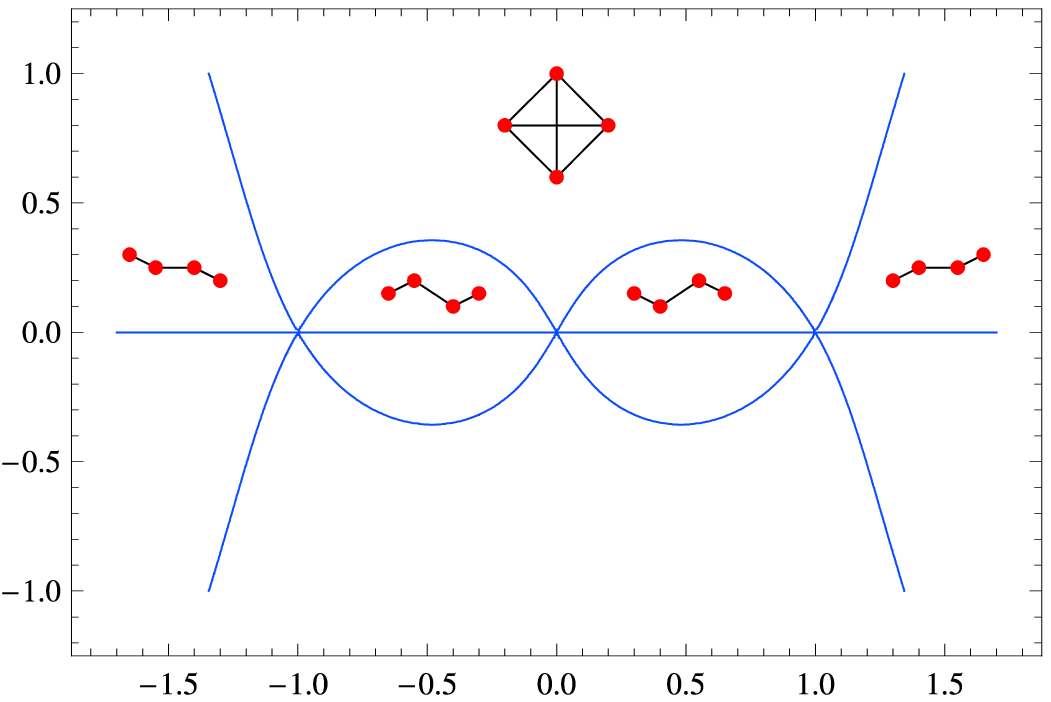,height=6cm}}
\caption{(Color online) The curves of marginal stability in the case of four minima, for $\epsilon=-\delta$. The set of (red) points illustrates the positions of the minima and the kink orbits for distinct values of the parameter $\delta$ in the related regions.}
\end{figure}

\begin{figure}[htbp]
\begin{center}
\begin{tabular}{|c|c|c|c|}  \hline & & &  \\
& $\delta$ & Original kinks  &  Deformed kinks
\\ & & &
\\ \hline {\small $\delta=0+i$}&
\parbox{3.7cm}{\epsfig{file=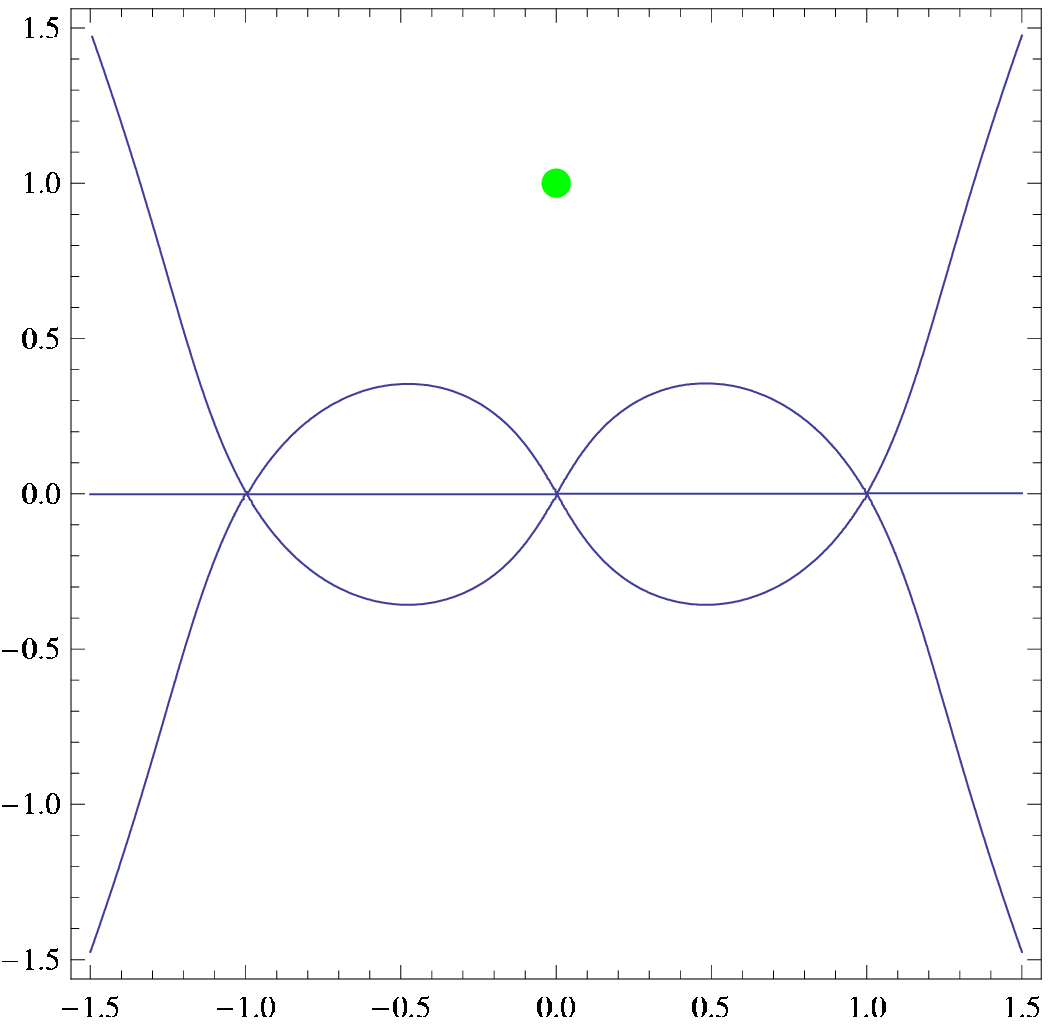,width=3.0cm}}&
\parbox{3.7cm}{\epsfig{file=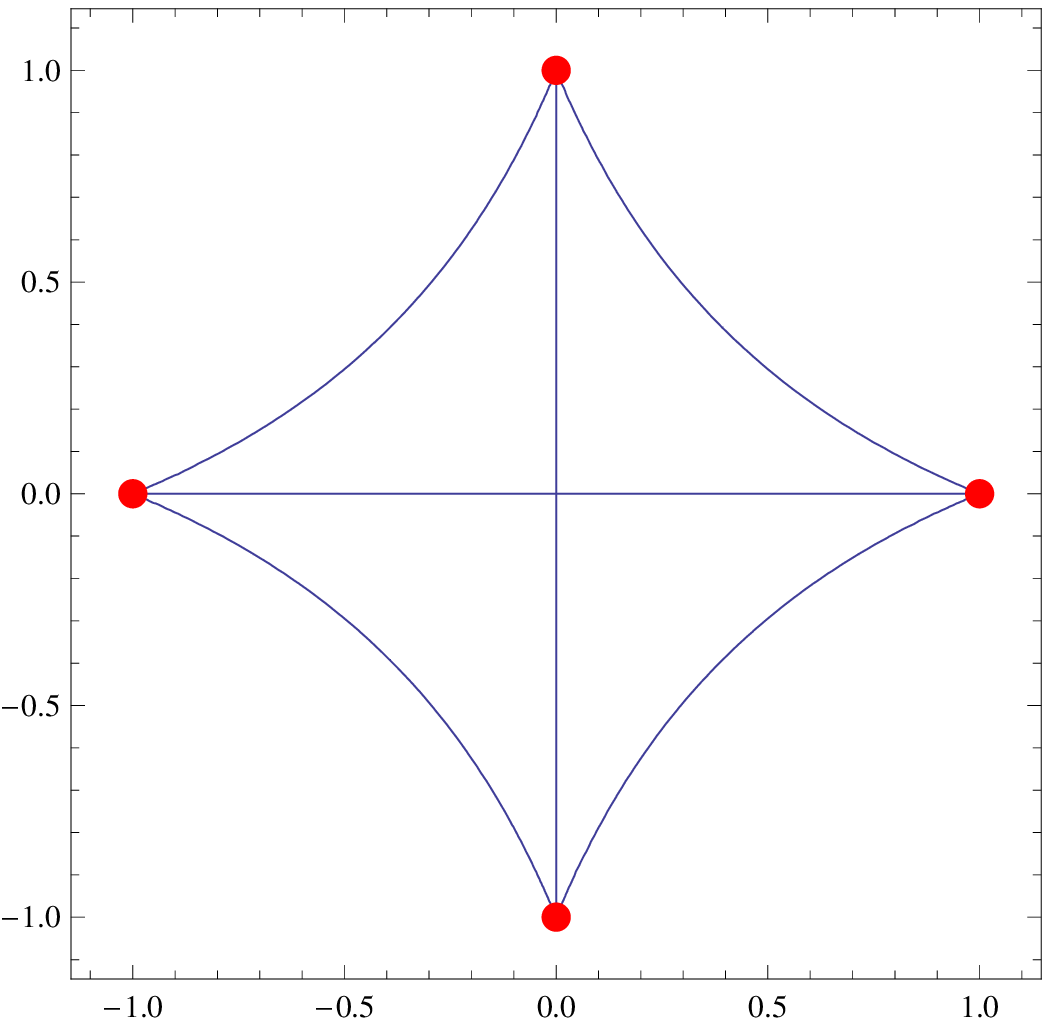,width=3.2cm}}&
\parbox{3.7cm}{\epsfig{file=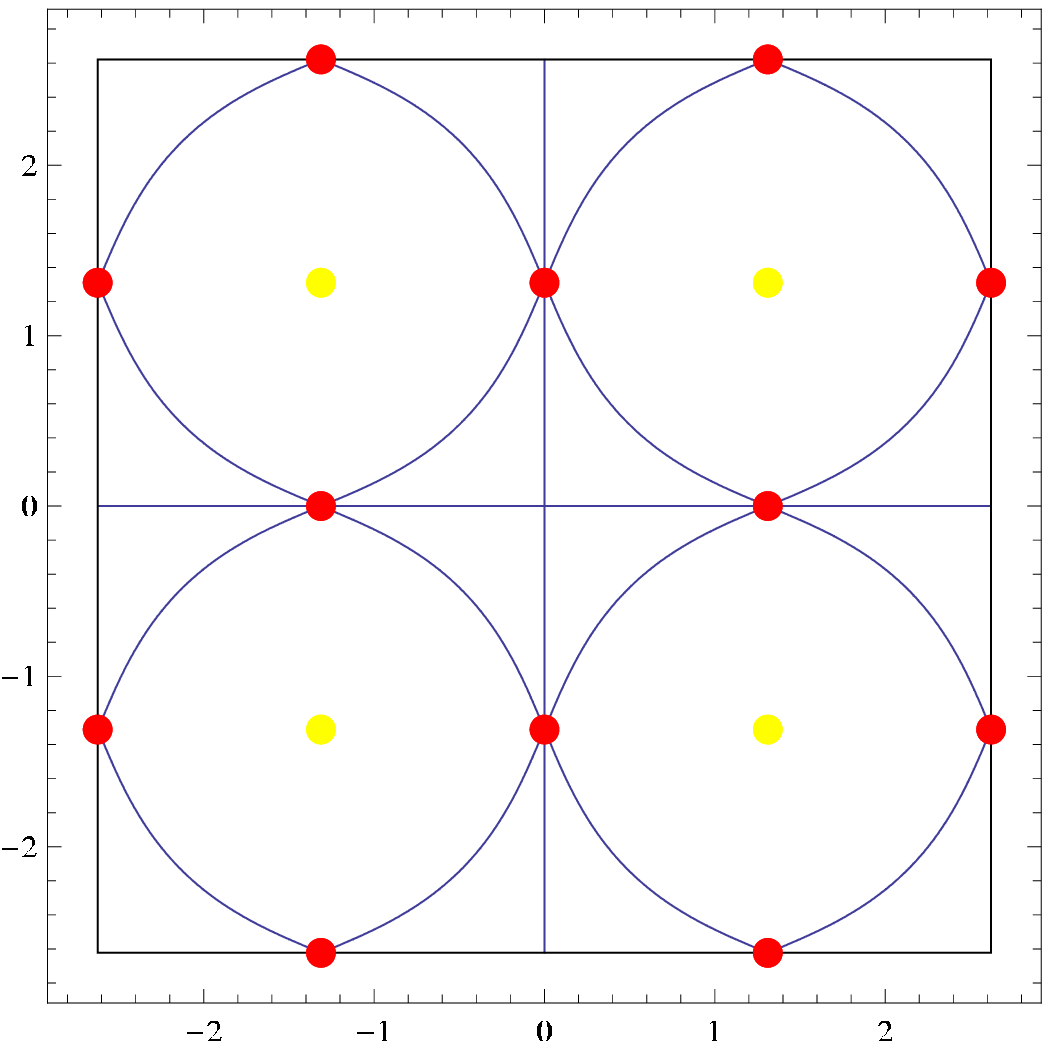,width=3.5cm}}
 \\ & & & \\
 \hline {\small $\delta=1+i$} & 
\parbox{3.7cm}{\epsfig{file=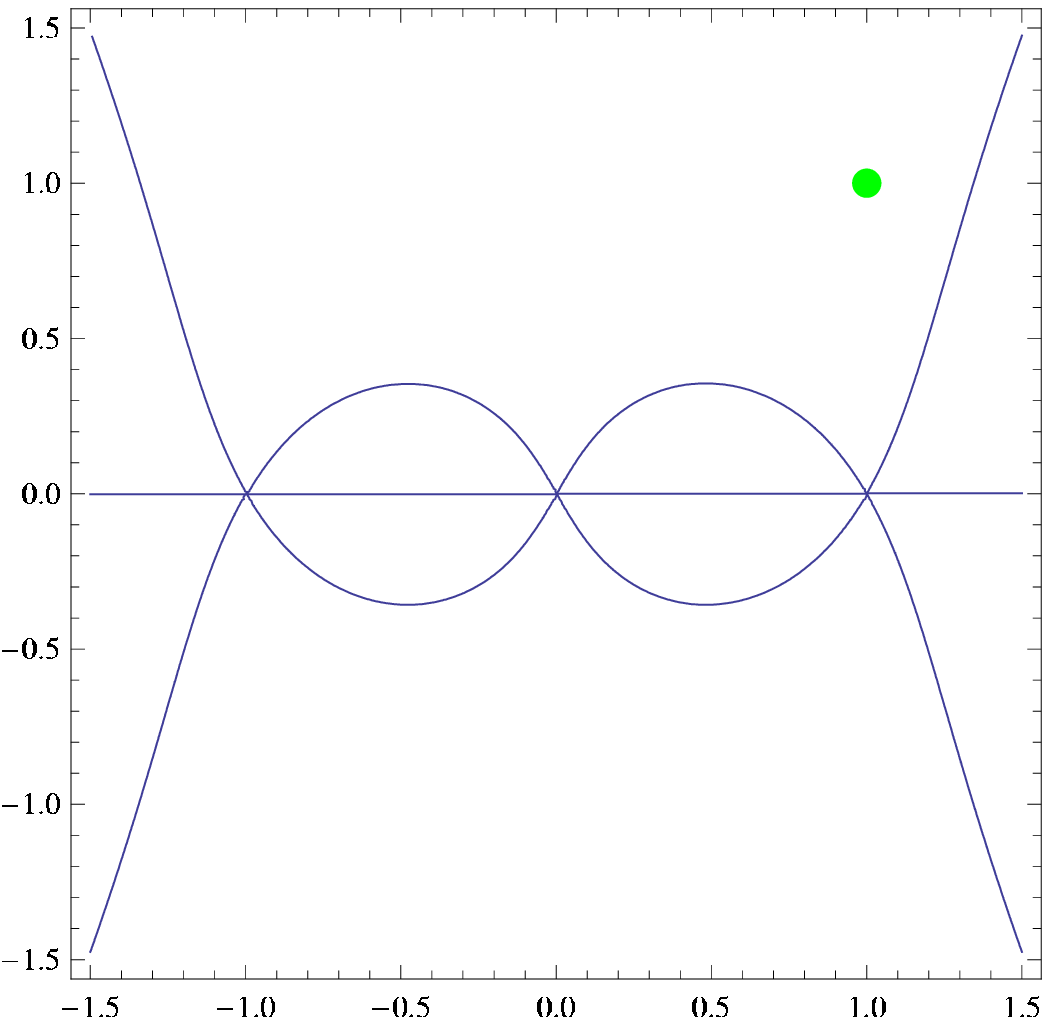,width=3.0cm}}&
\parbox{3.7cm}{\epsfig{file=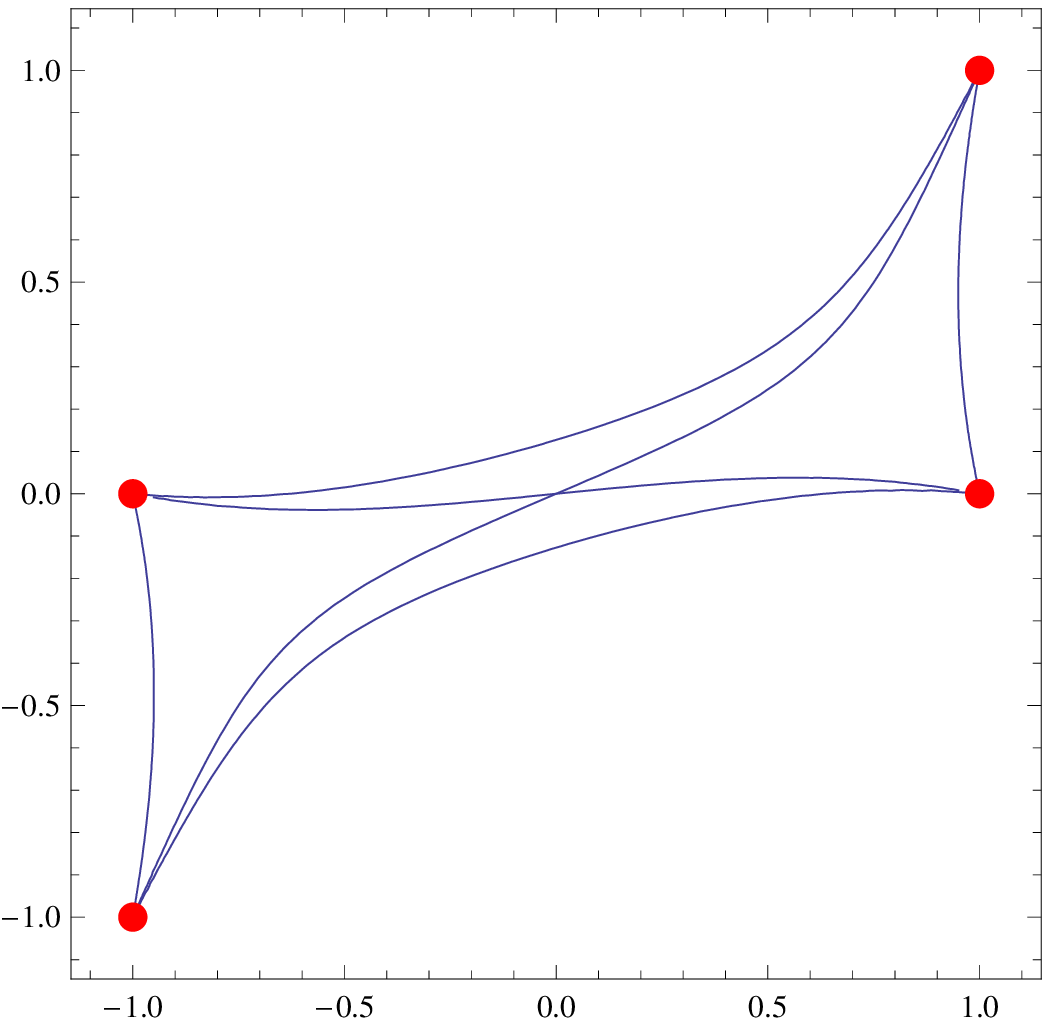,width=3.2cm}} &
\parbox{3.7cm}{\epsfig{file=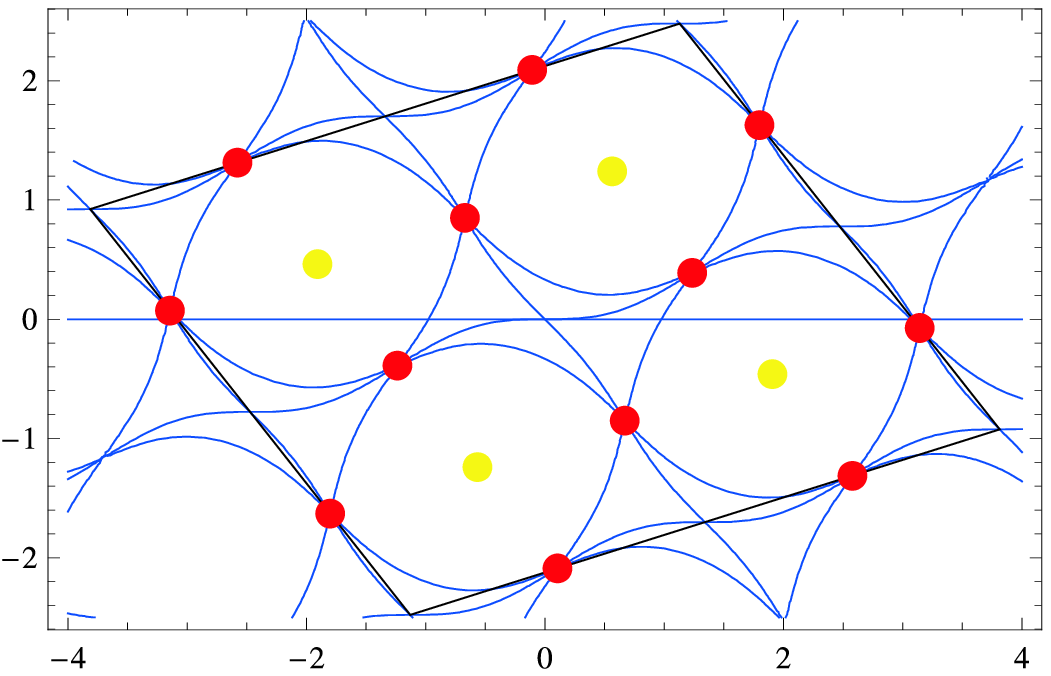,width=3.5cm}}
 \\ & & & \\
\hline  {\small $\delta=\frac{1}{2}+\frac{i}{5}$}&
\parbox{3.7cm}{\epsfig{file=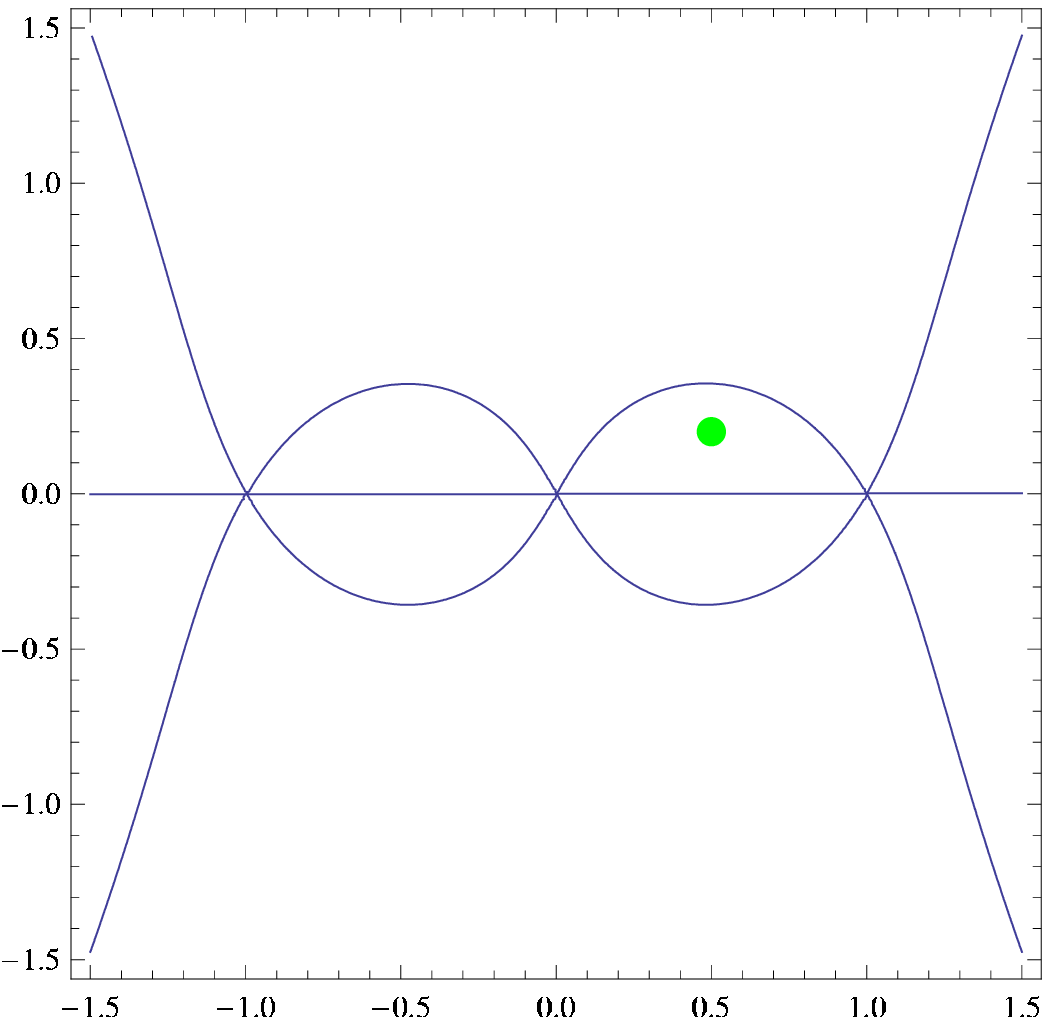,width=3.0cm}}&
\parbox{3.7cm}{\epsfig{file=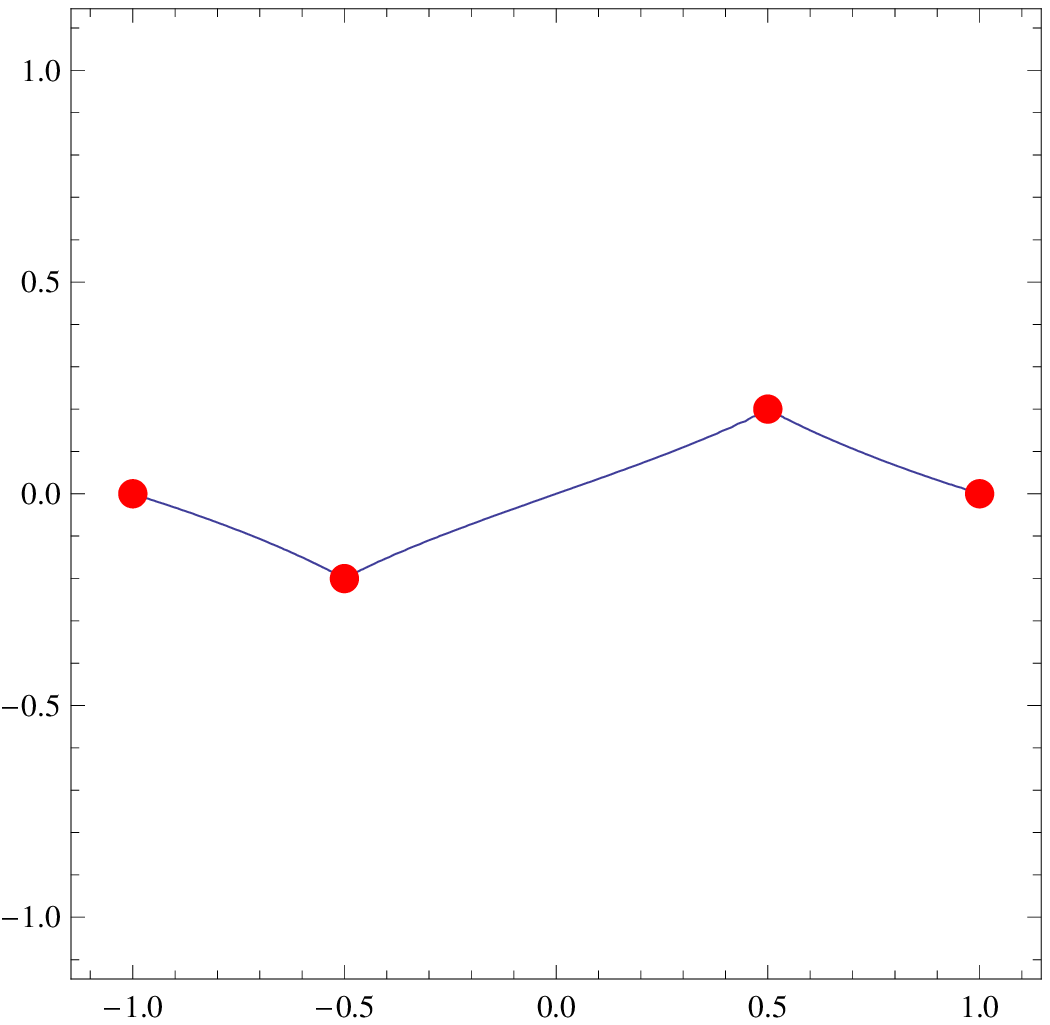,width=3.2cm}} &
\parbox{3.7cm} {\epsfig{file=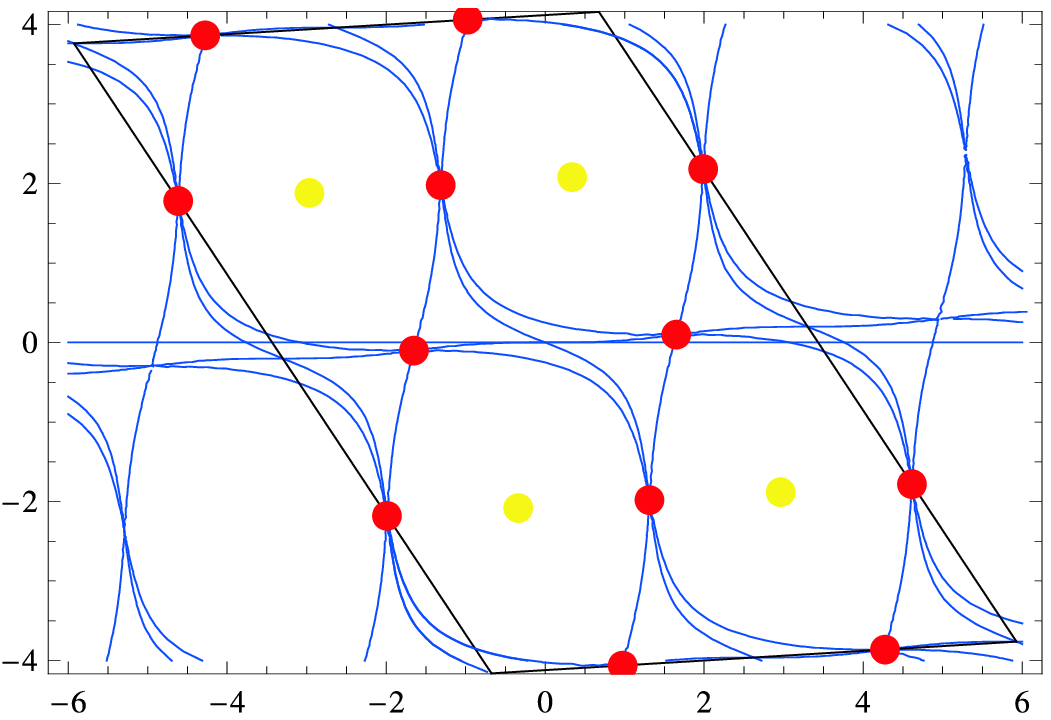,width=3.5cm}}
\\ & & & \\
\hline  {\small $\delta=2+i$}&
\parbox{3.7cm}{\epsfig{file=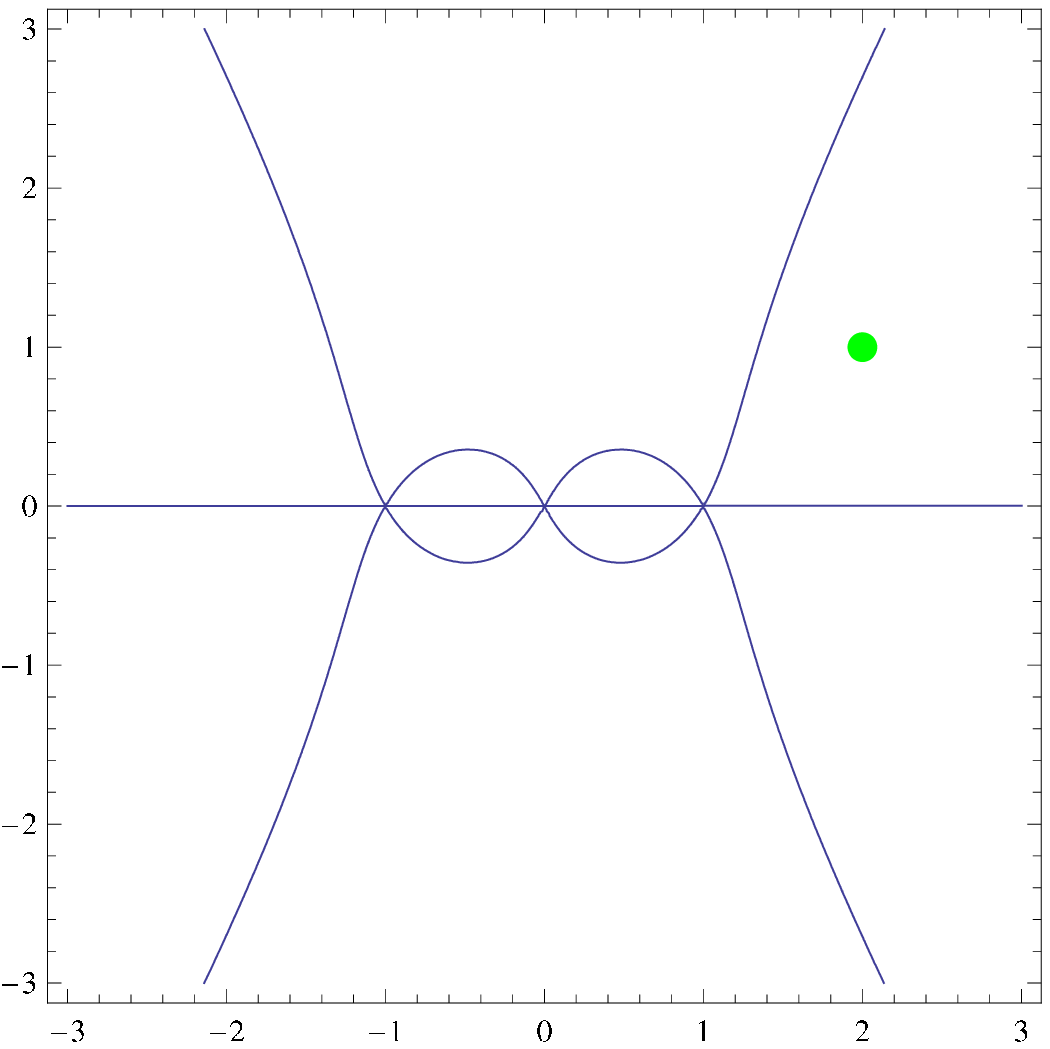,width=3.0cm}}&
\parbox{3.7cm}{\epsfig{file=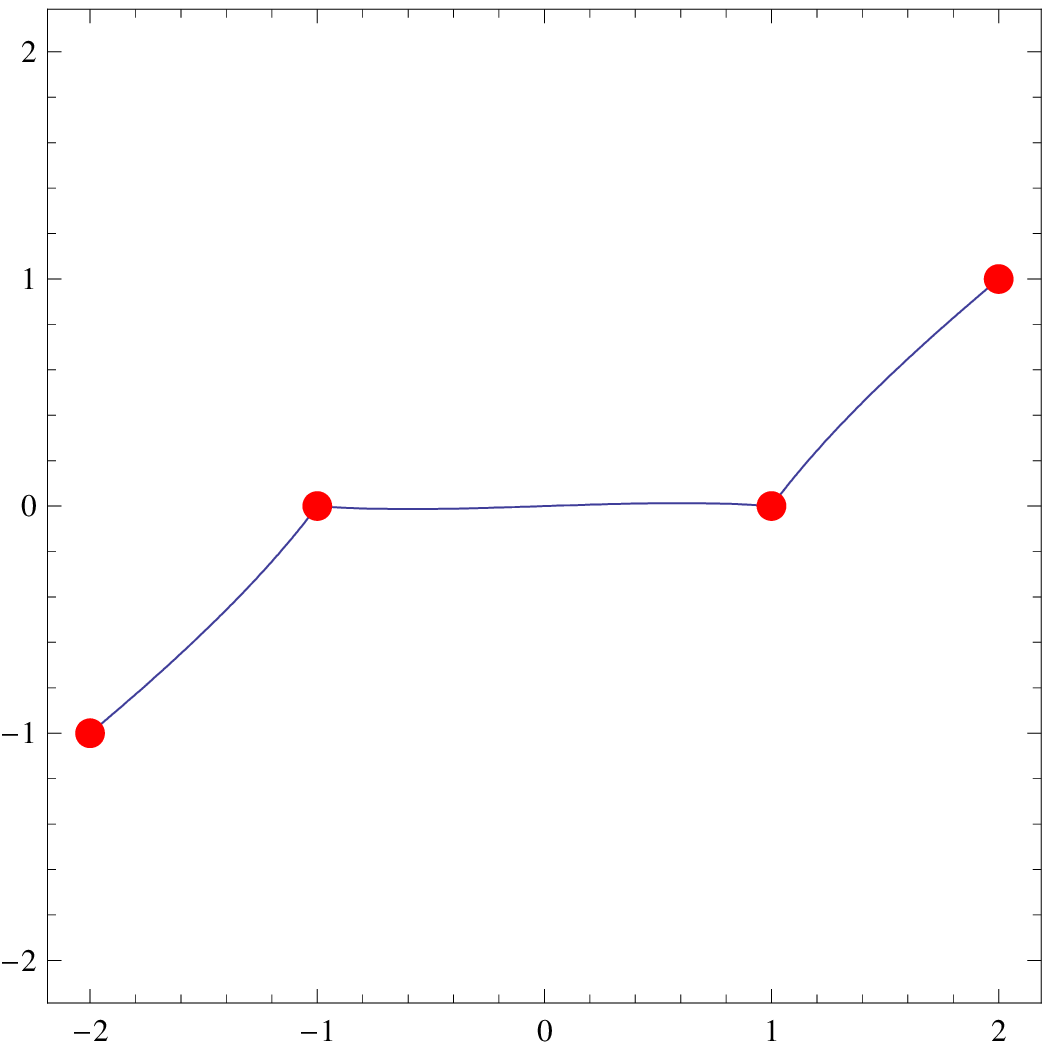,width=3.2cm}} &
\parbox{3.7cm} {\epsfig{file=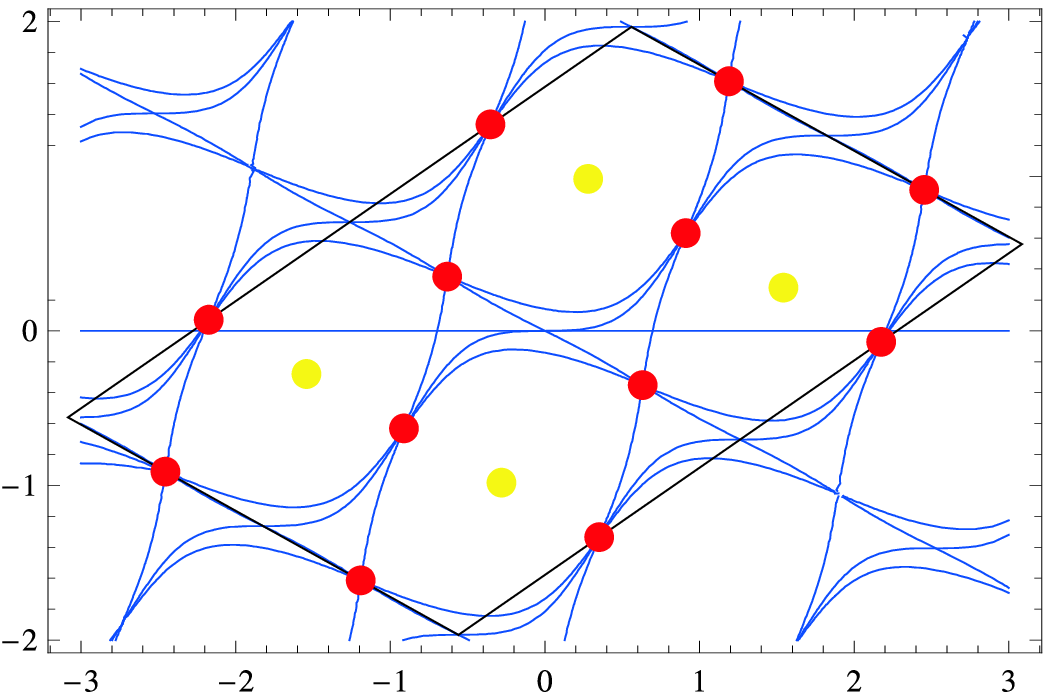,width=3.5cm}}
\\ & & & \\
 \hline
 \end{tabular}
\end{center}
\caption{(Color online) The bifurcation curves and some related illustrations with (green) points in the first column representing distinct values of the complex parameter $\delta$ in the case of four minima.}
\end{figure}

It is interesting to notice that the condition for alignment in the deformed model in the ${\cal W}$ plane, leads to $\delta_2=0$ again.
But we have already seen that in this case there is no bifurcation anymore. Thus, once again we notice that the deformation procedure
wash out the marginal stability curve, leaving no room for bifurcation also in this case. In Fig.~14 we illustrate the case $N=4$
with several distinct possibilities, driven by the complex parameter $\delta.$

\section{Final comments}
\label{sec:end}

In this work we have firstly dealt with the standard Wess-Zumino model driven by a complex scalar field engendering discrete $Z_N$ symmetry. The model is defined in terms of a superpotential, a holomorphic function of the complex field which contains $N$ minima, the vacua manifold which represents a set of points with the very same $Z_N$ symmetry of the model. The minima determine several topological sectors, which can be represented by algebraic curves and solved by first-order differential equations of the BPS type, as shown in Ref.~\cite{SP}.

The main idea of this work concerns the deformation procedure developed in \cite{DD,DD1}, which was used to deform the model in a
way such that the set of $N$ minima could be replicated in the entire configuration space, the plane described by the complex
field. In this way, the algebraic orbits in field (target) space of the original Wess-Zumino model are also replicated in the
entire plane, naturally leading to a network of defects, a spread network of kinklike orbits in field space. As we have seen,
the idea was implemented very efficiently, and we have illustrated the procedure with three interesting cases, involving the $Z_2$, $Z_3$ and
$Z_4$ symmetries, with the sets of minima forming an equilateral triangle and a square in the last two cases, respectively.

We have then investigated the more general case, with models engendering $N$ non symmetric minima, with $N=3$ and $N=4$.
The case $N=3$ is driven by a single complex parameter $\delta,$ and we have illustrated the results with two distinct values of $\delta,$
showing how it modifies the regular structure that we have obtained in the symmetric case. The case $N=4$ is more complicated, since it is controlled by two complex parameters, $\delta$ and $\epsilon$. We have then made an interesting simplification, reducing the model to a single complex parameter, with $\epsilon=-\delta$. In this case, we have also illustrated the results with two distinct values of the parameter, to show how it changes the regular structure of the symmetric case. In the non symmetric case, we have also examined bifurcation and the marginal stability curve for both $n=3,$ and $n=4,$ in the last case after introducing the simplification which leads to models described by a single complex parameter. In both cases, the complex parameter gives rise to a diversity of very nice possibilities, but the deformation washes out bifurcation from the deformed models.

The present investigation poses several issues, one of them concerning the $N=5$ and $N=6$ cases, which follows the natural course of this work. In the asymmetric case of four minima, we could also consider other relations between the two parameters $\delta$ and $\epsilon$. Another issue concerns the tiling of the plane with regular polygons, which is constrained to appear with regular hexagons, squares and equilateral triangles, in this order of decreasing efficiency. It would be interesting to search for possible connection among these regular tilings, the respective basic polygons, and the deformation procedure. It is also interesting to consider the issue studied in Ref.~{\cite{PT}}, in which one deals with the problem of marginal stability in terms of forces between soliton states in the case of three minima. A natural extension to the case of four minima seems desirable, and could also include investigations on how the forces between solitons would behave under the deformation procedure used in the present work.

Another route of interest is related to a topological change in the field plane
itself: if we let the fields to live in a genus 0 or genus 1 Riemann surface, that is, if we consider the target space to
be the two-sphere ${\mathbb S}^2$ or the two-torus ${\mathbb T}^2={\mathbb S}^1\times{\mathbb S}^1,$ instead of the plane
${\mathbb R}^2,$ we could be able to tile the surface ${\mathbb S}^2$ or ${\mathbb T}^2$ with other patterns, nesting
distinct networks of defects. This last case may possibly lead us to an issue of current interest: the conflict between geometry and
topology, related to the geometric features of the tiling with regular polygons and the topological properties of the configuration
space itself. It seems plausible that the modular parameter of the target space ${\mathbb T}^2$ is forced to be our
$\tau={\omega_3}/{\omega_1}$ and the sphere ${\mathbb S}^2$ is restricted to a 2:1 embedding of this ${\mathbb T}^2(\omega)$ in
${\mathbb CP}^1$. These and other issues are now under consideration, and we hope to report on them in the near future.

\section*{Acknowledgements}
This work is part of a collaboration which has been financed by the Brazilian and Spanish governments: VIA, DB and LL thank CAPES, CLAF, CNPq and PRONEX-CNPq-FAPESQ, and MAGL and JMG thank Ministerio de Educacion y Ciencia, under grant FIS2006-09417, for partial support.


\end{document}